 \documentclass{emulateapj}
 
\newcommand{\Msol}{\mbox{$M_{\odot}$}}
\newcommand{\logm}{\mbox{$\log M_{\star}$}}
\newcommand{\Mstar}{\mbox{$M_{\star}$}}
\newcommand{\logmu}{\mbox{$\log \mu_\star$}}

\newcommand{\ssfr}{\mbox{$SFR/M_{\star}$}}

\newcommand{\sfrkpc}{\mbox{$\Sigma_{SFR}$}}

\newcommand{\Msun}{\mbox{${\bf M_{\odot} }$}}

\def\deg      {{\ifmmode^\circ\else$^\circ$\fi}} 

\newcommand{\GALEX}{{\it GALEX}}
 
\defcitealias{Wyder2007}{Paper I} 
\defcitealias{Martin2007}{Paper III} 
\defcitealias{Salim2007}{S07}  
 
\slugcomment{Accepted to the Astrophysical Journal Supplement, GALEX Special Issue}
 
\begin{document}
 

\title{The UV-Optical Color Magnitude Diagram II: Physical Properties and Morphological Evolution on and off of a Star-Forming Sequence}


 

%
%
%
\author{
David Schiminovich\altaffilmark{1},  
Ted K. Wyder\altaffilmark{2}, 
D. Christopher Martin\altaffilmark{2}, 
Benjamin. D. Johnson\altaffilmark{1}, 
Samir Salim\altaffilmark{3,4}, 
Mark Seibert\altaffilmark{5}, 
Marie A. Treyer\altaffilmark{2,6}, 
Tamas Budavari\altaffilmark{7}, 
Charles Hoopes\altaffilmark{7}, 
Michel Zamojski\altaffilmark{1},
Tom A. Barlow\altaffilmark{2}, 
Karl G. Forster\altaffilmark{2},  
Peter G. Friedman\altaffilmark{2}, 
Patrick Morrissey\altaffilmark{2}, 
Susan G. Neff\altaffilmark{8}, 
Todd A. Small\altaffilmark{2},
Luciana Bianchi\altaffilmark{9}, 
Jose Donas\altaffilmark{6}, 
Timothy M. Heckman\altaffilmark{7}, 
Young-Wook Lee\altaffilmark{10}, 
Barry F. Madore\altaffilmark{5}, 
Bruno Milliard\altaffilmark{6}, 
R. Michael Rich\altaffilmark{4}, 
Alex. S. Szalay\altaffilmark{7},
Barry Y. Welsh\altaffilmark{11},
Suk Young Yi\altaffilmark{10}}

\altaffiltext{1}{Department of Astronomy, Columbia University, 550 West 120th Street, New York, NY 10027; ds@astro.columbia.edu}
\altaffiltext{2}{California Institute of Technology, MC 405-47, 1200 East California Boulevard, Pasadena, CA 91125}
\altaffiltext{3}{NOAO, 950 North Cherry Ave., Tucson, AZ 85719}
\altaffiltext{4}{Department of Physics and Astronomy, University of California, Los Angeles, CA 90095}
\altaffiltext{4}{Observatories of the Carnegie Institution of Washington, 813 Santa Barbara St., Pasadena, CA 91101}
\altaffiltext{5}{Laboratoire d'Astrophysique de Marseille, BP8, Traverse du Siphon, F-13376 Marseille, France}
\altaffiltext{6}{Department of Physics and Astronomy, The Johns Hopkins University, Homewood Campus, Baltimore, MD 21218}
\altaffiltext{7}{Laboratory for Astronomy and Solar Physics, NASA Goddard Space Flight Center, Greenbelt, MD 20771}
\altaffiltext{8}{Center for Astrophysical Sciences, The Johns Hopkins` University, 3400 N. Charles St., Baltimore, MD 21218}
\altaffiltext{9}{Center for Space Astrophysics, Yonsei University, Seoul 120-749, Korea}
\altaffiltext{11}{Space Sciences Laboratory, University of California at Berkeley, 601 Campbell Hall, Berkeley CA 94720 }

 
%

%
 
  
\begin{abstract}

We use the UV-optical color magnitude diagram in combination with spectroscopic and photometric measurements derived from the SDSS spectroscopic sample to measure the distribution of galaxies in the local universe (z$<$0.25) and their physical properties as a function of specific star formation rate (\ssfr) and stellar mass ($M_\star$).  Throughout this study our emphasis is on the properties of galaxies on and off of a local ``star-forming sequence.''  We discuss how the physical characteristics of galaxies along this sequence are related to scaling relations typically derived for galaxies of different morphological types.   We find, among other trends that our measure of the star formation rate surface density, \sfrkpc ~is nearly constant along this sequence.  We discuss this result and implications for galaxies at higher redshift.   For the first time, we report on measurements of the local UV luminosity function versus galaxy structural parameters as well as inclination.   We also split our sample into disk-dominated and bulge-dominated subsamples using the $i$-band Sersic index and find that disk-dominated galaxies occupy a very tight locus in \ssfr~ vs. \Mstar~ space while bulge-dominated galaxies display a much larger spread of \ssfr~ at fixed stellar mass.  A significant fraction of galaxies with \ssfr~ and \sfrkpc~ above those on the ``star-forming sequence'' are bulge-dominated.  We can use our derived distribution functions to ask whether a significant fraction of these galaxies may be experiencing a final episode of star formation (possibly induced by a merger or other burst), soon to be quenched, by determining whether this population can explain the growth rate of the non-star-forming galaxies on the ``red sequence.''  We find that this is a plausible scenario for bulge-dominated galaxies near the characteristic transition mass under reasonable assumptions regarding quenching timescales.  Similarly we use this technique to estimate the rate of mergers/starbursts that take galaxies off of the star-forming sequence and show that the implied merger rates are consistent with local measurements.

\end{abstract}
 
 
 


 \section{Introduction}

What determines the star formation rate (SFR) of a galaxy?   Ample evidence suggests that it is the quantity and distribution of cold gas \citep{Schmidt1959, Kennicutt1998a} and the gas-dynamical processes responsible for triggering, regulating or quenching new star formation.  In the context of a hierarchical clustering scenario for galaxy formation \citep[e.g.][]{Kauffmann1993} these mechanisms are necessarily linked to the flow of dark and baryonic matter over a wide range of scales, densities and temperatures \citep{Keres2005}.   Given this complexity it is intriguing that the integrated light from many galaxies can be explained using simple star formation histories (SFH)  \citep[][although see also Kauffmann et al. 2006b]{Tinsley1968,Searle1973,Bruzual2003}.  Such work has led to an apparent understanding of star formation on cosmological scales \citep{HopkinsA2006}, although accurate physical models embedded within a realistic framework for galaxy assembly \citep[e.g. ][]{Stringer2007} are required to understand star formation in individual galaxies.

Measurements of the colors and structure of a galaxy should guide these models by providing insight into the connection between star formation and assembly.   The fact that galaxies in the local universe appear to show a remarkable correlation between their star formation history and their structure---disk-dominated galaxies show higher present to past-averaged star formation rates than bulge-dominated galaxies \citep{Kennicutt1998b}---would appear to suggest a straightforward link, but we now know that the explanation must be non-trivial \citep{De Lucia2006}.  A crucial component of these analyses is a quantitative and representative description of the galaxy population.  In this regard color-magnitude distributions and their derivatives have emerged as useful tools because they can be easily interpreted in terms of the star formation history and the stellar mass content and therefore are easily connected to models of galaxy assembly  and the build up of massive galaxies along the ``red sequence'' \citep{Faber2007}.

In this paper, the second in a series, we explore how the UV-optical properties of a large sample of galaxies in the local universe can be used to understand the distribution of SFR and the connection with assembly history across the population.  We accomplish this by expanding on the analysis of the UV-color magnitude diagram \citep[][hereafter Paper I]{Wyder2007} using observations of 26241 galaxies from the {\it Galaxy Evolution Explorer (GALEX)} Medium Imaging Survey (MIS), combined with the Sloan Digital Sky Survey (SDSS) primary spectroscopic survey and  incorporating additional ``value-added'' data related to the morphology/structure, star formation history and dust attenuation in each galaxy.  Our analysis has many similarities to recent studies conducted using SDSS \citep{Blanton2003c,Kauffmann2003a, Kauffmann2003b, Shen2003}, local surveys \citep{Driver2006, Jansen2001} and high-z investigations \citep{Bell2005, Noeske2007a, Noeske2007b}, although it extends those studies in several ways, described below.  

As discussed in \citetalias{Wyder2007}, a notable feature of the UV-optical color diagram is the very wide separation between the peaks of the blue and the red galaxy populations.  Of central importance is the strongly peaked locus of star-forming blue galaxies that has been variously called the ``blue sequence'' \citet{Blanton2006} and the ``main sequence'' \citet{Noeske2007a}.  \citetalias{Wyder2007}, \citet{Noeske2007a} and \citet[][hereafter S07]{Salim2007} show that the majority of star-forming galaxies of a given stellar mass possess a narrow range of SFR, a result already noted by \citet{Brinchmann2004}, \citet{Feulner2006} and \citet{Cattaneo2007}.  This stands in marked contrast to the optical color-based view, which emphasizes a tight ``red sequence'' and a scattered ``blue cloud'' \citep{Bell2005}.    It suggests strong similarities among star-forming galaxies and a greater diversity of SFR for those galaxies that optically appear  ``red-and-dead.''  In many respects, this alternate view is reminiscent of the progression from the Hubble classification scheme, with a rich description of spirals and only a few elliptical classes, towards work in recent years that revealed that ellipticals possess a greater range of structure (at low and high spatial frequency) than originally thought \citep{de Zeeuw1991}.

 \begin{figure*}[p]
\plotone{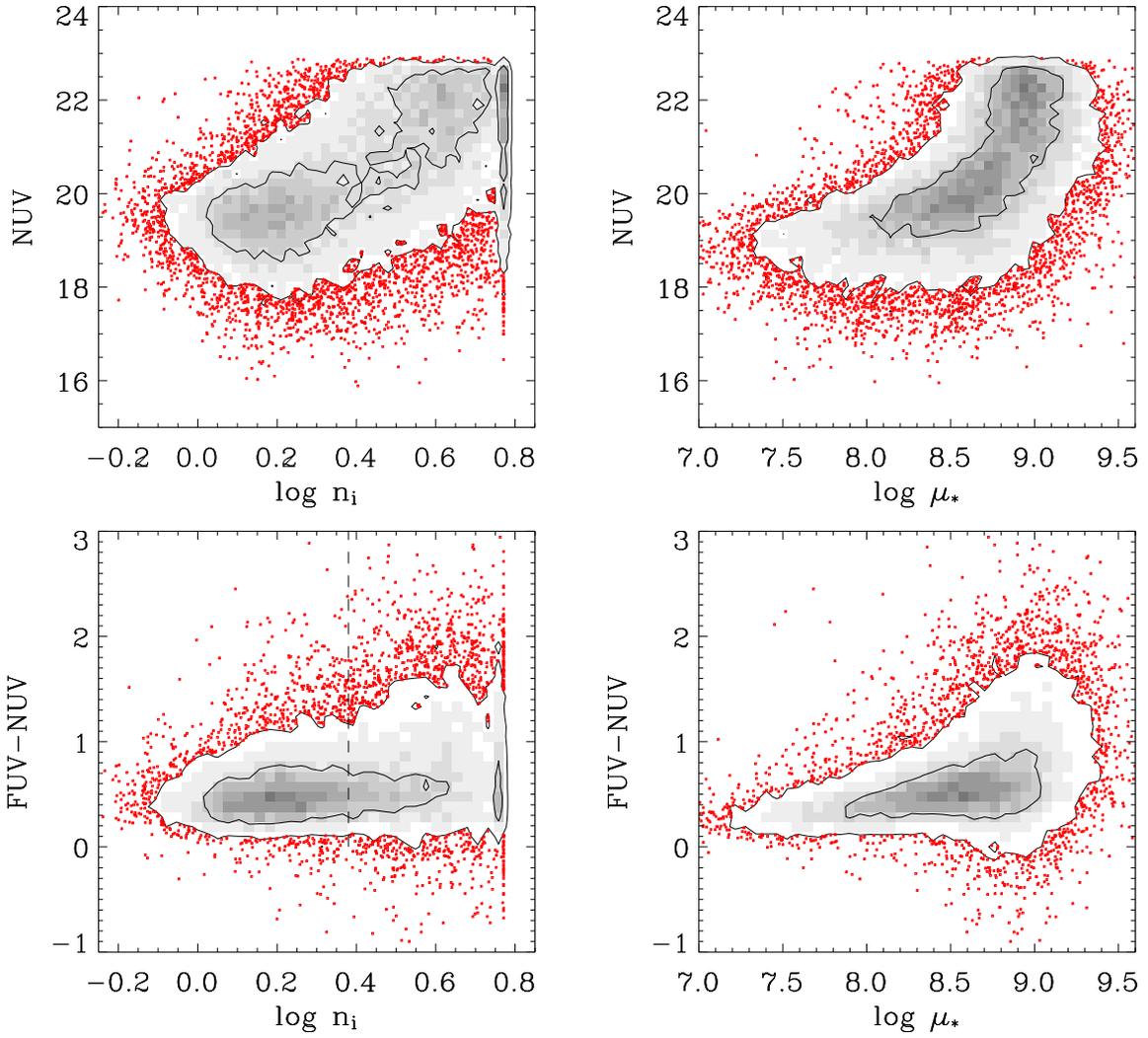}
\caption{Main galaxy sample bivariate number density distribution: {\it Top}: $NUV$ magnitude and $FUV-NUV$ color vs. log $n_i$. {\it Bottom}:  $NUV$ and $FUV-NUV$ vs. log $\mu_\star$.  Contours enclose 50\% and 90\% of the distribution, with outliers plotted individually.}
\label{fig:morph_magnitude}
\end{figure*}

 \begin{figure*}
\plotone{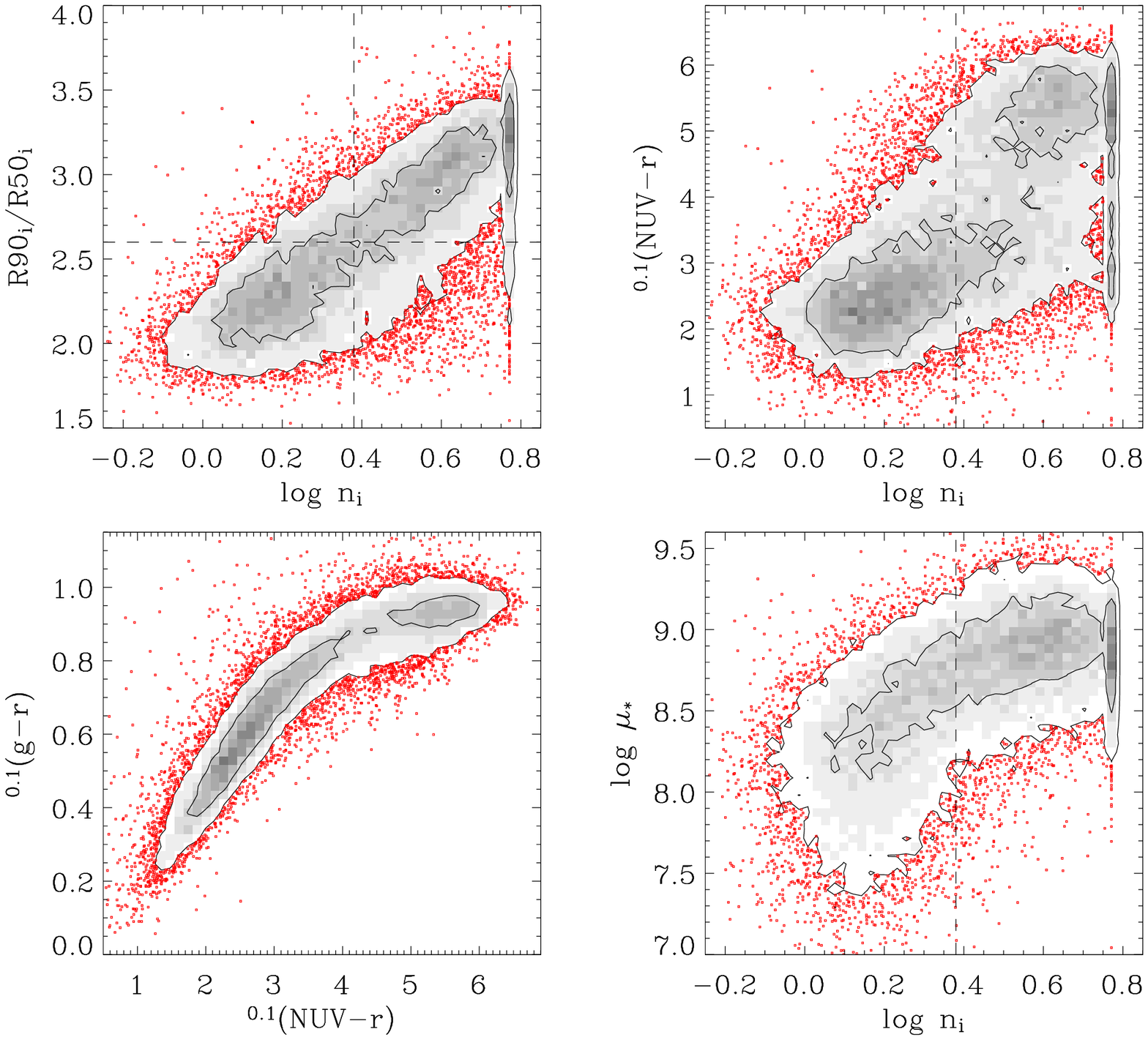}
\caption{Main galaxy sample bivariate number density distribution. {\it Top}: $i$-band concentration and $NUV-r$ color vs. log $n_i$. {\it Bottom left}: $g-r$ color vs. $NUV-r$. {\it Bottom right}: log $\mu_\star$ vs. log $n_i$.  Contours enclose 50\% and 90\% of the distribution, with outliers plotted individually.} 
\label{fig:morph_params}
\end{figure*}

Here we quantify the structural and physical properties of a local ``star-forming sequence'' (SF sequence) defined by a relationship between stellar mass and star formation rate and use it to understand the characteristics of the dominant galaxy population including the slope of the sequence itself.  Some of this analysis is quite complementary to the work of \citetalias{Salim2007}.  We also focus on the distribution disk-dominated and bulge-dominated galaxies across the full range of \ssfr~and \Mstar~and investigate the structure of outliers to the SF sequence.  While in \citet[][hereafter Paper III]{Martin2007}, specific attention has been given to galaxies in the intermediate region between the blue and red galaxy population (the ``green valley''), our ultimate focus will be on galaxies with specific star formation rates higher than those on the SF sequence.  This is the population in the \ssfr~vs. \Mstar~diagram that has evolved most dramatically since z$\sim$1.   

We briefly describe the outline of this paper.  After presenting the data in section 2 we investigate in section 3 the physical properties of galaxies on and off of the SF sequence. In section 4 we investigate the relationship between star formation history and structure and its connection with the evolution of galaxies on and off of the SF sequence.   Throughout this paper, we make use of the flat $\Lambda$CDM cosmology with $H_0$ = 70 km s$^{-1}$ Mpc$^{-1}$ and $\Omega_\Lambda =0.7$.

 \section{Data}
 
 \subsection{GALEX DR4-MIS Cross-Match}
 
{\it GALEX} data were obtained as part of the {\it GALEX} Medium Imaging Survey  \citep[MIS;][]{Martin2005}, cataloged as part of an internal data release 1.1 (IR1.1) and processed using standard {\it GALEX} pipeline processing \citep{Morrissey2005, Morrissey2007}.  The MIS reaches a limiting UV magnitude of $\sim$23 through single or multiple eclipse exposures that are typically 1 ks or greater in duration.  MIS targets were initially selected to overlap the SDSS Data Release 2 footprint, although some additional overlap with SDSS Data Release 4 \citep[DR4;][]{Adelman-McCarthy2006} made it advantageous to use the latter release for the cross-match.

A total of 67,883 SDSS DR4 spectroscopic objects were within 0.6$^\circ$ of the field centers of \GALEX~ MIS observations.  For each of these objects we searched for the closest \GALEX~ detection within a 4\arcsec~radius.  Objects with no match were considered \GALEX~ non-detections.  To produce a complete statistical sample, further cuts were applied.     SDSS objects were selected from the main galaxy sample with $r$-band magnitudes $14.5<r<17.6$, magnitude error $\sigma_r<0.2$ mag, redshift in the range $0.01<z<0.25$, and redshift confidence $z_{conf} > 0.67$.  The sample was limited to regions of sky with UV exposure times greater than 800 s, location on detector within 0.55$^{\circ}$ of field center, an NUV magnitude cut (16$<$NUV$<$23) and non-artifacts using ${\tt nuv\_artifact}$ $\le$ 1.  These cuts, source matching, and completeness are all described in more detail in \citetalias{Wyder2007} and in \citet{Bianchi2006}.   The main sample used in this paper contains a total of 26,241/18,091 galaxies detected with NUV/FUV $<$23 over an area of 485.321/411.266 deg$^2$.  For most of our analysis we will use the NUV-detected sample (``main galaxy sample''), noting exceptions where appropriate.

We combined our \GALEX/SDSS matched photometric catalog with extra derived parameters obtained from the  MPA/JHU \citep{Kauffmann2003a, Kauffmann2003b, Brinchmann2004, Tremonti2004} and NYU  \citep{Blanton2005b} value-added catalogs. We briefly discuss the parameters we have used below.

\begin{figure*}
\plotone{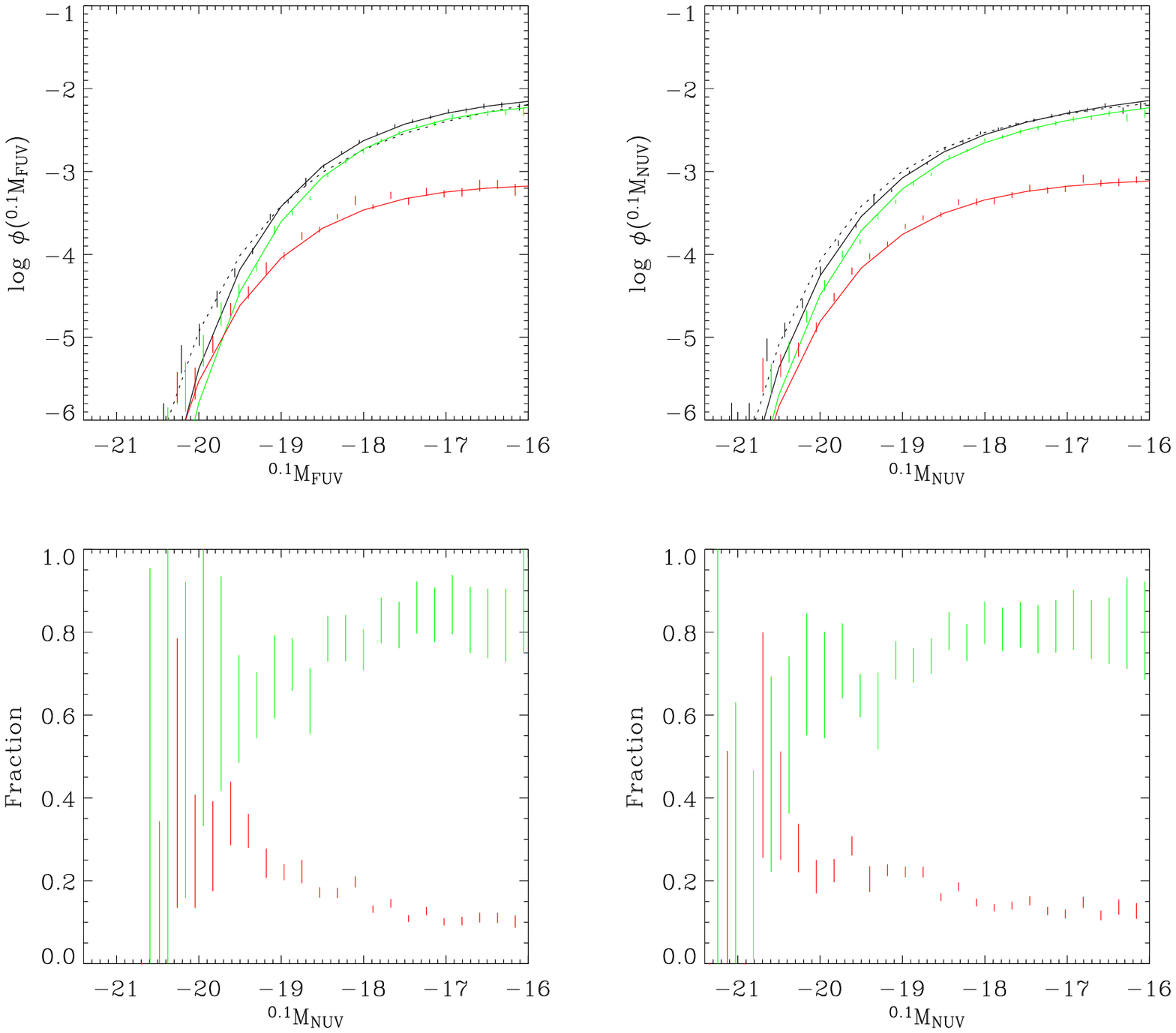}
\caption{{\it Top}: FUV and NUV luminosity function for complete sample ($black$) and disk and bulge-dominated subsamples split by $i$-band Sersic index $n$.  $n_i<2.4$ (log $n_i < 0.38$) ({\it green}) and  $n_i>2.4$ (log $n_i > 0.38$) ({\it red}).  Units of $\phi$ are in Mpc$^{-3}$ mag$^{-1}$. The dotted curve is from the \citet{Wyder2005} and \citet{Treyer2005} LF. {\it Bottom}: Relative fraction (1/Vmax-weighted) of disk vs. total and bulge-dominated vs. total in each magnitude bin.} 
\label{fig:lf_mis_n}
\end{figure*}

\begin{figure*}[t]
\plotone{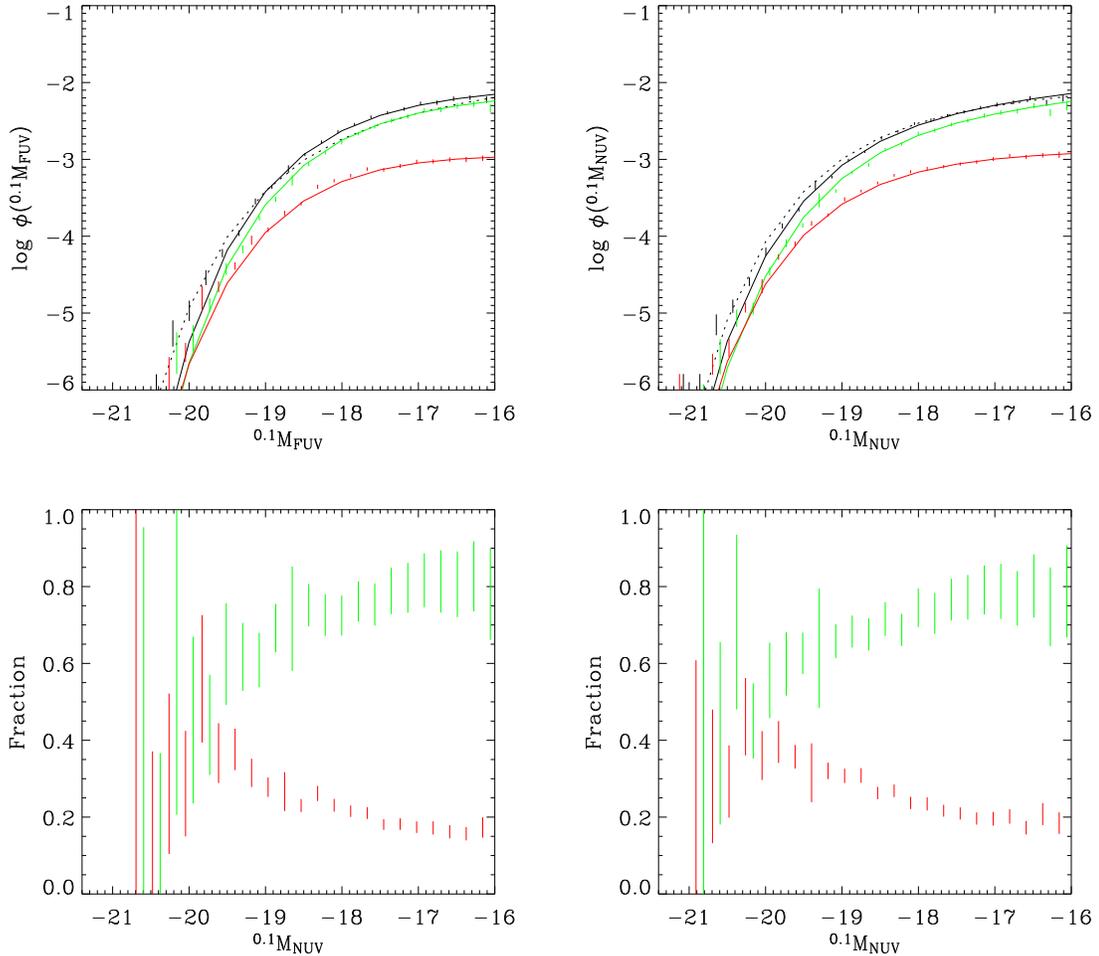}
\caption{{\it Top}: FUV and NUV luminosity function for complete sample ($black$) and low and high stellar mass surface density subsamples split by log $\mu_\star$:  log $\mu_\star<8.5$  ({\it green}) and   log $\mu_\star > 8.5$ ({\it red}).   Units of $\phi$ are in Mpc$^{-3}$ mag$^{-1}$.  The dotted curve is from the \citet{Wyder2005} and  \citet{Treyer2005} LF. {\it Bottom}: Relative fraction (1/Vmax-weighted) of low and high stellar mass surface density vs. total.} 
\label{fig:lf_mis_mu}
\end{figure*}

\begin{figure*}[t]
\plotone{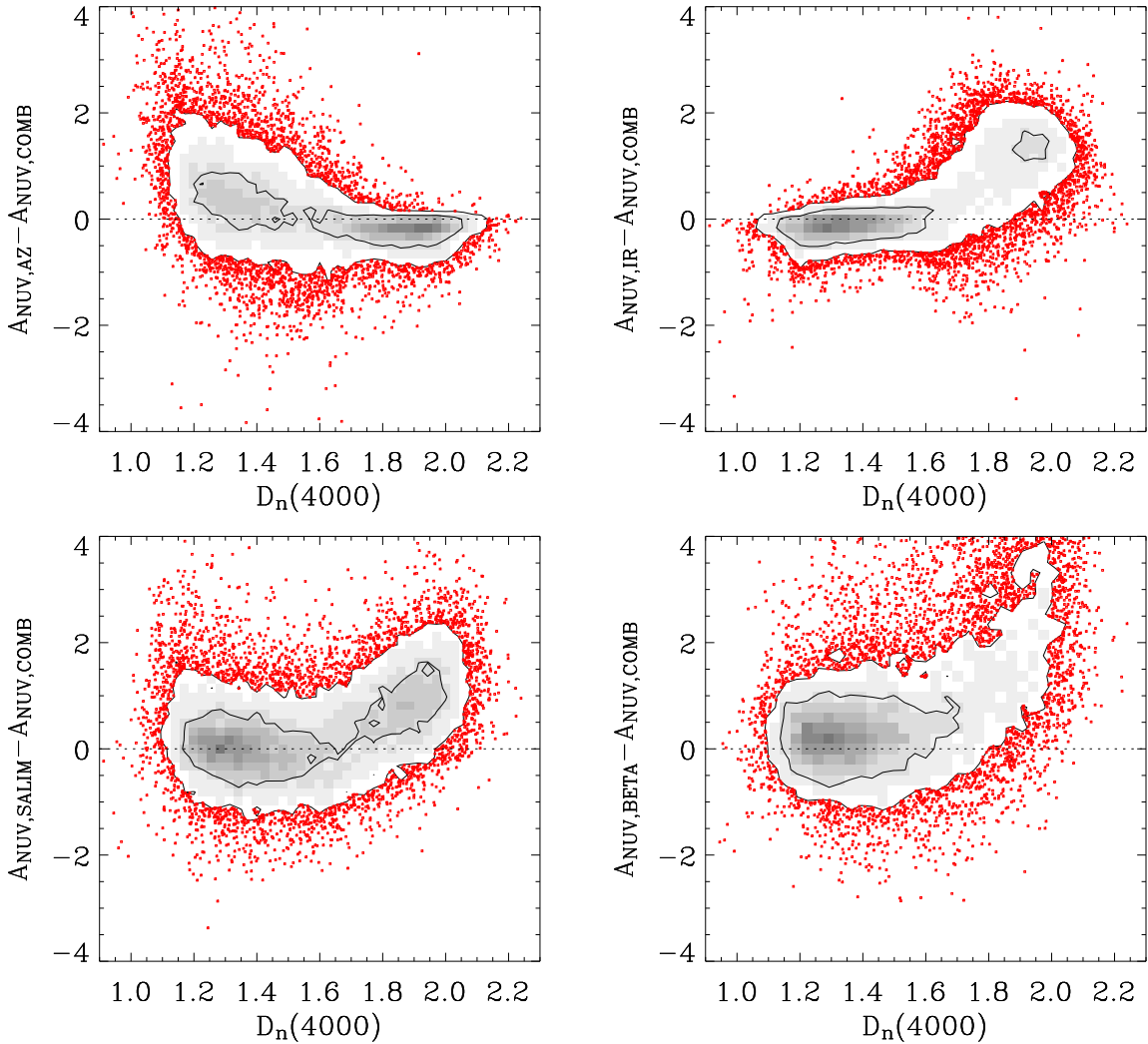}
\caption{Comparison of different dust attenuation measures (A$_{NUV}$) with the ``combined'' measure used in this paper. $Top~left:$ A$_{NUV}$ derived using $A_z$ from Kauffmann et al (2003).  $Top~right:$ A$_{NUV}$ derived using empirically derived attenuation measure from \citet{Johnson2007}.  $Bottom~left:$  A$_{NUV}$ derived by \citetalias{Salim2007}.  $Bottom~right:$ A$_{NUV}$ using the \citet{Seibert2005} IRX-$\beta$ relation.} 
\label{fig:anuvcomp}
\end{figure*}

{\it Photometry, redshifts.---} For GALEX $FUV$ and $NUV$ photometry we used Kron magnitudes and errors\footnote{Early versions of the GALEX pipeline occasionally underestimated the true error, corrected in this and later releases.} generated by the IR11 pipeline.  These magnitudes were dereddened for Galactic extinction as described in \citetalias{Wyder2007}. We used  $ugriz$ Petrosian magnitudes and dereddening values obtained from the SDSS DR4 pipeline.
Redshifts and redshift errors were obtained from the  Princeton reductions (http://spectro.princeton.edu) which have been subsequently incorporated into the MPA/JHU files.

{\it Galaxy size, light profile shape and model fit.---}  We used the DR4 pipeline Petrosian 50\% and 90\% radii.  Redshifts and assumed cosmology were used to convert these to physical sizes in kpc.      These values were used to derive secondary quantities, such as surface densities and concentration. The SDSS pipeline also performs model fits to the galaxy light profile using an exponential model commonly used for fitting spiral disks and a de Vaucouleurs model used for fitting bulges and ellipticals.  The pipeline quantity FracDeV\ (erroneously labelled FracPSF in SDSS output catalogs)  provides an estimate for how much light from the galaxy is coming from the bulge-like component.   For the bulk of our analysis we made use of the improved seeing-deconvolved, axisymmetric Sersic profile fits from the NYU VAGC \citep[described in the appendix of][]{Blanton2005b}.  The Sersic profile has the form
$$I(r)=A \exp \bigg[-(r/r_0)^{1/n}\bigg]$$
where the Sersic index $n$ is 1 for an exponential light profile and 4 for a de Vaucouleurs profile.  As reported in \citet{Blanton2005b}, these fits slightly underestimate high-$n$ galaxies (measuring 3.5 for galaxies with $n$=4), but are sufficient for our purposes.  We refer the reader to \citet{Blanton2003c} and \citet{Blanton2005b} for further discussion regarding the use of the Sersic index over a similar redshift range.  We use only the $i$-band fit, using a longer wavelength band that will be less sensitive to recent star formation, and typically express the Sersic index in logarithmic form (log $n_i$).   In later sections we also divide our sample into disk-dominated and bulge-dominated galaxies at $n_i=2.4$ (log $n_i$ = 0.38).   Note that this dividing line is similar to or slightly higher than that used in other analyses.  \citet{Vincent2005} separate disk and bulge-dominated galaxies using $n$=2.0 which yields a cut very close to FracDeV=0.5.  We chose our limit to conservatively restrict the number of disk-dominated galaxies identified as bulge-dominated.  We obtain 12,835 disk-dominated and 13,406 bulge-dominated galaxies in our main galaxy sample using this coarse classification. 

{\it Stellar mass and stellar mass surface density.---} Stellar masses were obtained from the MPA/JHU catalogs using the values discussed in \citet{Kauffmann2003a}.  Stellar masses and the $z$-band half-light radius were used to derive stellar mass surface densities:
$$\mu_\star = \frac{0.5 M_\star}{\pi R_{50,z}^2}$$

in $\Msun$ kpc$^{-2}$. In Figure \ref{fig:morph_magnitude} we plot the distribution of our sample as a function of measured magnitude, UV color ($FUV-NUV$), Sersic index, and $\log \mu_\star$.  Most disk and low stellar mass surface density galaxies (\logmu $<$ 8.5) in our sample are brighter than our magnitude limit, while some bulge-dominated and \logmu $>$ 8.5 galaxies are fainter than our limit.  In the next section we make use of the ($1/V_{max}$) method in order to correct for this incompleteness.  

\subsection{K-corrections and V$_{max}$ calculation}

Our sample has a median redshift of 0.086 (0.078/0.10 for disk-dominated/bulge-dominated). 
Using the {\tt kcorrect} code \citep[v4.1.4;][]{Blanton2007} we derived rest-frame absolute magnitudes in the bands ~$^{0.1}FUV$, $~^{0.1}NUV$, $~^{0.1}u$, $~^{0.1}g$, $~^{0.1}r$, $~^{0.1}i$, and $~^{0.1}z$, generated by shifting the observed bandpasses blueward by a factor in wavelength of $1/(1+z)$ ($ =1/1.1$ for $z=0.1$).  This approach, described in \citet{Blanton2003a}, minimizes the amplitude of the $K$-correction (beyond a trivial constant $-2.5 \log_{10}(1.1)$) determined for the typical galaxy in our sample.   We applied these for a given band $b$, using the equation:
 $$M_{b,0.1} = m_b-DM-K_{b,0.1}(z)+(z-0.1)Q$$  where DM is the distance modulus and $Q=1.6$ \citepalias[see][for description]{Wyder2007} is added to account for luminosity  evolution over the redshift range being considered \citep{Blanton2003b}.

\begin{deluxetable*}{l c c c c c  l}
\tablecaption{FUV and NUV LF Schechter Fit for Various Subsamples}
\tablewidth{0pt}
\tablehead{
\colhead{Subsample} & \colhead{$\log \phi_\star(^{0.1}M^\star)$}& \colhead{$~^{0.1}M^\star$}  & \colhead{$\alpha$} & \colhead{$\log \rho_\nu$}\\
\colhead{} & \colhead{(Mpc$^{-3}$)}&  \colhead{ }  & \colhead{ } & \colhead{(erg s$^{-1}$ Mpc$^{-3}$) }\\
}
\startdata
 & & FUV & &  \\
\hline
Total: & $ -2.08\pm  0.02$  & $-17.83\pm  0.04$  & $ -1.06\pm  0.04$  &  25.75 \\
 $n < 2.4$  & $ -2.12\pm  0.02$  & $-17.70\pm  0.04$  & $ -1.02\pm  0.05$  &  25.65 \\
 $n > 2.4$  & $ -3.03\pm  0.03$  & $-18.09\pm  0.06$  & $ -0.94\pm  0.07$  &  24.88 \\
 $\log \mu_\star < 8.5$  & $ -2.17\pm  0.02$  & $-17.78\pm  0.04$  & $ -1.08\pm  0.05$  &  25.64 \\
 $\log \mu_\star > 8.5$   & $ -2.79\pm  0.02$  & $-17.94\pm  0.05$  & $ -0.91\pm  0.05$  &  25.04 \\
 $q_{25} < 0.6$   & $ -2.31\pm  0.04$  & $-17.37\pm  0.07$  & $ -1.06\pm  0.09$  &  25.34 \\
 $q_{25} > 0.6$   & $ -2.27\pm  0.02$  & $-17.86\pm  0.04$  & $ -0.87\pm  0.05$  &  25.53 \\
 \hline
 & & NUV & &  \\
\hline
Total:  & $ -2.27\pm  0.02$  & $-18.44\pm  0.03$  & $ -1.21\pm  0.03$  &  25.88 \\
 $n < 2.4$  & $ -2.33\pm  0.02$  & $-18.34\pm  0.04$  & $ -1.20\pm  0.04$  &  25.77 \\
 $n > 2.4$ & $ -3.03\pm  0.03$  & $-18.49\pm  0.04$  & $ -0.99\pm  0.04$  &  25.06 \\
 $\log \mu_\star < 8.5$  & $ -2.37\pm  0.02$  & $-18.36\pm  0.04$  & $ -1.23\pm  0.04$  &  25.75 \\
 $\log \mu_\star > 8.5$  & $ -2.86\pm  0.02$  & $-18.51\pm  0.03$  & $ -1.01\pm  0.03$  &  25.24 \\
 $q_{25} < 0.6$ & $ -2.47\pm  0.03$  & $-17.91\pm  0.06$  & $ -1.18\pm  0.06$  &  25.45 \\
 $q_{25} > 0.6$  & $ -2.42\pm  0.02$  & $-18.43\pm  0.03$  & $ -1.00\pm  0.04$  &  25.65 \\
\enddata
\label{table:schechter}
\end{deluxetable*}

Figure \ref{fig:morph_params} shows the distribution of galaxies in our sample as a function of rest-frame color and structural parameters.  The $~^{0.1}(g-r)$ vs. $~^{0.1}(NUV-r)$ color-color diagram demonstrates quite clearly how the $NUV-r$ color covers a much wider magnitude range than $g-r$.  In addition, the $g-r$ color starts to saturate for red galaxies, while $NUV-r$ varies by more than 2 magnitudes.    We see indications from these plots that rest-frame UV-optical colors correlate with Sersic index (and concentration), although as with the color-color plot, \logmu~is nearly constant for bulge-dominated galaxies.

Using our $K$-corrected magnitudes we determined $V_{max}$, the maximum volume over which the galaxy would have been included in our sample.  This was calculated using our adopted cosmology for three bands individually ($FUV$, $NUV$, $r$), as well as the combination of any two (or all three).  For most of our analysis below we use $V_{max, NUV, r}$, which is the intersection of the detection volume in each individual band.  For any analysis using $FUV$ data, we use all three bands to determine the appropriate $V_{max}$.  Again, the reader is referred to \citetalias{Wyder2007} for a more detailed discussion.

\begin{figure*}[t]
\plotone{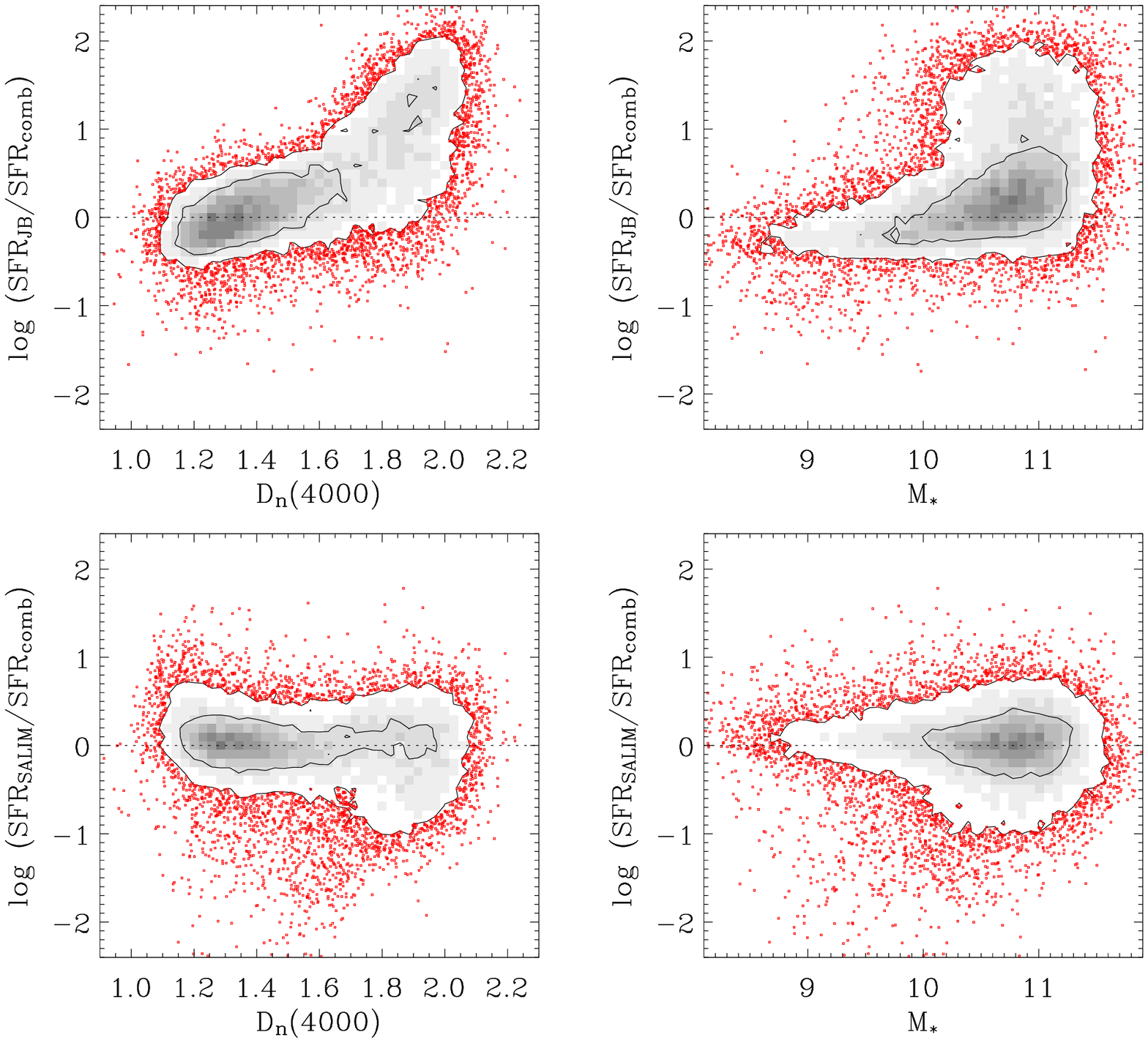}
\caption{Comparison of different SFR measures with the one used in this paper, plotted vs. $D_n$(4000) and \Mstar. {\it Top}: SFRs from \citet{Brinchmann2004}.  {\it Bottom}: SFRs from \citetalias{Salim2007}.} 
\label{fig:sfrcomp}
\end{figure*}

 \begin{figure*}[t]
 \epsscale{1.2}
 \plotone{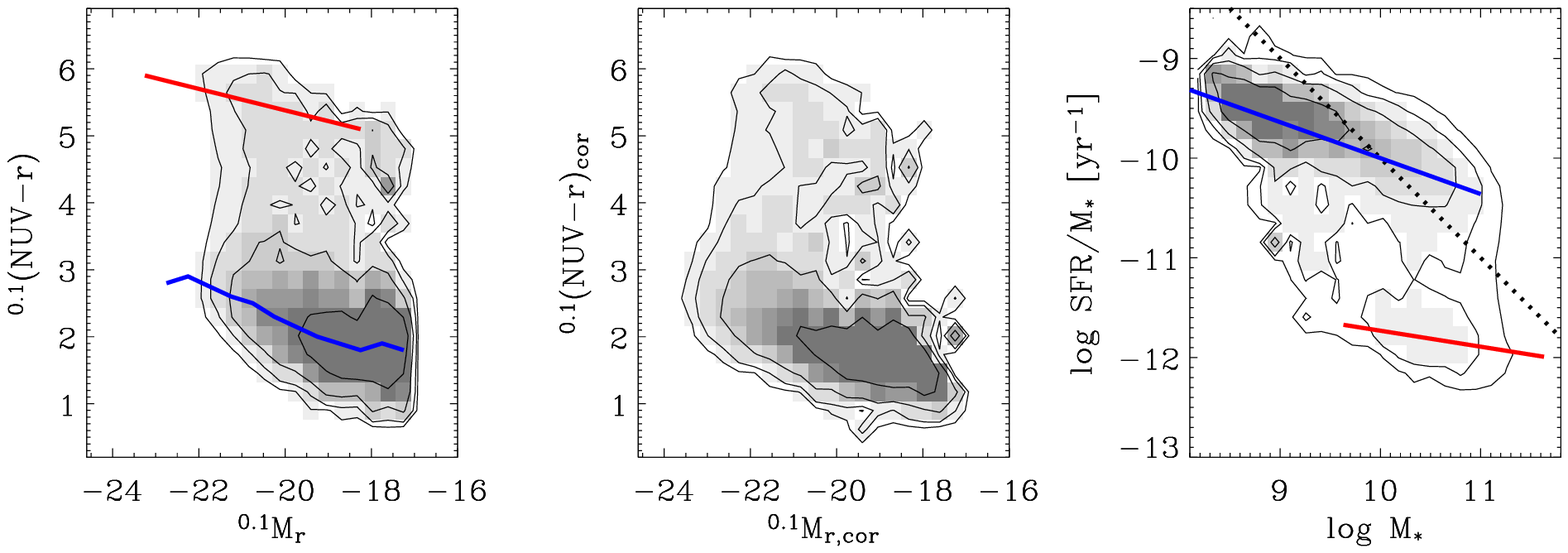}

\caption{Bivariate distribution (1/V$_{max}$-weighted) of galaxies as a function of color vs. magnitude and \ssfr~ vs. \Mstar.  {\it Left}: ~$^{0.1}(NUV-r)$ color vs. ~$^{0.1}M_r$ using $K$-corrected absolute magnitudes.  No dust-attenuation correction has been applied. Blue and red solid lines trace the ridge line of the blue and red sequences, taken from \citetalias{Wyder2007}.  {\it Center}: Dust attenuation corrections have been applied to both the $NUV$ and $r$ absolute magnitudes.  {\it Right}: \ssfr~ vs. \Mstar~.  The blue solid line shows the star-forming sequence fit from \citetalias{Salim2007}, and the red line shows approximate position of non-star-forming sequence on this diagram. Contours enclose 38\%, 68\%, 87\%, and 95\% of the distribution.  (See text for details).    {\it Dotted line}: $SFR = 1~M_\odot$ yr$^{-1}$}
\label{fig:massnrall}
\end{figure*}

\section{Physical Properties Across the Galaxy Population}

\subsection{A Starting Point: UV Luminosity Function vs. Structural Parameters}

We begin with an example that highlights the issues we will be considering in this paper.  In Figure \ref{fig:lf_mis_n} we plot the UV luminosity function calculated from our sample using the 1/$V_{max}$ method.  We show the FUV and NUV luminosity functions split by Sersic index $n_i$ where we use the separation described above to define disk-dominated and bulge-dominated samples.  We also fit Schechter parameters, which we include in Table \ref{table:schechter}.  The total luminosity function is consistent with the one determined by \citet{Wyder2005} and \citet{Treyer2005}, although the Schechter fit does undershoot the most luminous points, some of which might be active galactic nuclei (AGNs).

In Figure \ref{fig:lf_mis_mu} we plot the luminosity function split by stellar mass surface density, \logmu.  Below each plot we show the fractional abundance of bulge-dominated (or high \logmu) galaxies vs. UV magnitude.   We find in both sets of plots (and for both FUV and NUV) that the fraction of bulge-dominated (or high \logmu) galaxies increases with increasing UV luminosity.    These observations are consistent with those of \citet{Dahlen2007} based on data at higher redshift from GOODS, as well as with\citet{Menanteau2005}.  

We can nevertheless ask, How might we physically interpret this result?  If the SFR is considered to be proportional to the UV (i.e., we assume no dust correction is needed) then the luminosity functions suggest that galaxies with the highest SFRs show a higher prevalence of bulge-dominated galaxies.  This result is in agreement with \citet{Brinchmann2004} and \citet{Salim2005}, which both reported a population of high star formation rate, high concentration galaxies. However, some bulge-dominated systems are likely to be massive, so while SFR might be high, \ssfr~ may vary considerably.   If many UV-luminous galaxies are dusty, then they are also likely to span a wide range of (dust-corrected) SFRs.  In fact, as \citet{Hoopes2006} report, the most UV luminous galaxies are known to be a diverse population, containing disks and compact systems of a wide range of stellar masses and dust attenuations.  It is unclear from our luminosity functions how these luminous disk and bulge-dominated galaxies will be distributed once a dust correction is made.  
 
We have highlighted two crucial pieces of information that are needed to interpret the UV luminosity plots: the dependence on stellar mass, and the application of a dust correction that would allow us to interpret our results in terms of the galaxy's star formation history. (A third factor, inclination, is discussed further in the Appendix). To achieve this, we require reliable dust-attenuation corrections and SFR and $M_\star/L$ conversions.   Below we describe our dust-attenuation corrections and SFR calibration, which we then apply to the color-magnitude distribution to construct a  a \ssfr~ vs. \Mstar~diagram.  We use this more complete description of the galaxy population to study various scaling relations.  Later in the paper we use this description to study the disk and bulge-dominated populations separately.

\subsection{Star formation rates, dust attenuation, \ssfr, and $\Sigma_{SFR}$}

In order to derive star formation rates from our UV-optical measures, we need to account for any dust attenuation which may cause us to underestimate the intrinsic luminosity of the galaxy.  The dust-attenuation correction at any wavelength is nontrivial, since it is not simply a line-of-sight extinction correction but reflects assumptions about the geometry of the emitting regions and the surrounding dust.  Several recent investigations have explored how this dust geometry might be linked to galaxy morphology\citep[e.g.,][]{Pierini2004,Dale2007,Zheng2007}.

We hope to obtain reliable star formation rates for disk and bulge-dominated galaxies, and the latter population, with its lower \ssfr~and intrinsically red spectrum presents a considerable challenge in this regard.  Dust-attenuation measures derived from the UV using the relation determined for starburst or star-forming galaxies are unlikely to provide an accurate value \citep[e.g.,][]{Bell2002, Kong2004}.  Fiber emission-line measures (e.g. Balmer decrement) face similar problems in addition to requiring aperture corrections, and are often not available at high S/N for bulge-dominated galaxies \citep{Brinchmann2004}.  UV-optical SED fitting, such as that adopted by \citetalias{Salim2007}, relies heavily on the calibration and accuracy of the models of young and old stellar populations and the dust-attenuation law, which are highly uncertain for evolved systems.  Even the IR-to-UV flux ratio (unavailable for this work), often hailed as the most reliable dust-attenuation measure, remains difficult to interpret for evolved galaxies because of the uncertainty in determining which population (young or old stars, AGN) is contributing to the heating of the dust \citep[e.g.,][]{Johnson2007}.

Here we adopt a hybrid approach which attempts to combine measures that work effectively on different subsets of the galaxy distribution.  \citet{Johnson2006, Johnson2007} present an IR-calibrated measure of the FUV attenuation ($A_{IRX}$) for a sample of 1000 SDSS galaxies,  based on UV-optical colors and D$_n$(4000) which correlates with star formation history,
$$A_{IRX} = 0.81 -1.32x+1.07 y-0.81 x y$$
where $x=~^{0.1}(NUV-z)-2$ and $y= D_n(4000)-1.25$ using coefficients taken from Table 2 in \citet{Johnson2007}.    The derivation of $A_{IRX}$ is most accurate for galaxies with D$_n$(4000)$<$1.7.  
\citet{Kauffmann2003a} provide an attenuation measure $A_z$ that is based on detailed model fits to the SDSS absorption-line spectrum and broadband SED, likely to be accurate for galaxies with higher  D$_n$(4000).  Our combined fit provides a weighted mean of these two measures, using the measurement errors and published scatter for $A_{IRX}$ and the 1 $\sigma$ confidence intervals for $A_z$.  Specifically we combine them as:
$$A_{NUV,comb}=\frac{A_{FUV,IRX}/ ( k_{FUV} \sigma_{A_{FUV,IRX}}^2)+A_z/(k_{z} \sigma_{A_z}^2)}{1/(k_{FUV} \sigma_{A_{FUV,IRX}})^2+1/(k_{z} \sigma_{A_z})^2} $$
where we have used the \citet{Calzetti2000} attenuation curve to derive $k_{FUV} = A_{NUV,c}/A_{FUV,c} = 0.81$ and $k_{z} = A_{NUV,c}/A_{z,c} = 3.21$.

We show in Figure \ref{fig:anuvcomp} a comparison between our derived dust attenuation and other measures, all converted to $A_{NUV}$ using the \citet{Calzetti2000} attenuation curve and plotted as a function of D$_n$(4000): $A_{NUV, IR}$ from  \citet{Johnson2007}, $A_{NUV,z}$ from \citet{Kauffmann2003a}, $A_{NUV, SALIM}$ from \citetalias{Salim2007}, and $A_{NUV,BETA}$ derived using the IRX-$\beta$ relation obtained by \citet{Seibert2005}.    The first two plots compare quantities used to derive $A_{NUV, COMB}$ and illustrate the range over which each attenuation measure is given higher weight, with A$_z$ being the dominant measure for older galaxies and A$_{NUV, IR}$ for younger galaxies.  As expected, A$_{NUV,IR}$ is systematically higher for older galaxies \citep{Johnson2007}.   Our hybrid attenuation measure shows relatively good agreement ($|\overline{\Delta A_{NUV}}|  < 0.5$) with \citetalias{Salim2007} and \citet{Seibert2005}, albeit with large scatter throughout and lower values for the oldest galaxies.  \citet{Johnson2007} explore the scatter and systematic differences between these measures and others including the Balmer decrement (see also \citetalias{Wyder2007}), which are beyond the scope of this work.  We have repeated almost all analyses in this paper using each of the different measures, and although there are notable systematic differences, our overall results and conclusions do not change significantly.

We applied this dust-attenuation correction to our $NUV$ luminosities to obtain an ``intrinsic'' $NUV$ luminosity.  We then used these values to determine a star formation rate using the formula:
$$ SFR {\rm (M_\odot/yr)} = 10^{-28.165} L_{\nu} {\rm (erg ~ s^{-1} Hz^{-1})}$$
derived by \citetalias{Salim2007} assuming a Kroupa IMF and a continuous recent (100-300 Myr) star formation history, which makes these star formation rates directly comparable to those in that work and most other recent determinations.  Note that for a standard Salpeter IMF (between 0.1 and 100 $M_\odot$), star formation rates would be a factor of $\sim1.5$ higher.

Figure \ref{fig:sfrcomp} compares our derived SFRs with those of \citetalias{Salim2007} and \citet{Brinchmann2004}.  Again, we find relatively good agreement with \citetalias{Salim2007} except at the highest stellar masses and oldest galaxies, where our star formation rates tend to be slightly higher.  A wider systematic trend is observed with the H$\alpha$ [and color-D$_n$(4000)] derived star formation rates from \citet{Brinchmann2004}, with our values significantly lower for the most massive and oldest galaxies.  Galaxies not classified as ``star-forming'' in \citet{Brinchmann2004} have SFRs derived using D$_n$(4000), optical colors, and aperture corrections.  UV-derived SFRs have been shown to possess a higher dynamic range and are therefore applicable to a broader range of galaxy types \citepalias[see][for an extensive discussion of these differences]{Salim2007}.  As with the attenuation measures discussed above, we find that despite the differences, our overall conclusions are largely independent of the choice of SFR measure used.

All specific star formation rates (\ssfr) in this paper are calculated using the $NUV$-derived star formation rate, $SFR_{comb}$ and the MPA/JHU stellar mass, \Mstar.    Global star formation rate surface densities ($\Sigma_{SFR}$) were calculated using $SFR_{comb}$ and the $u$-band half light radius, R$_{u,50}$,
$$\Sigma_{SFR} = \frac{0.5SFR}{\pi R_{u,50}^2}$$   Our definition differs somewhat from other studies which calculate a global $\Sigma_{SFR}$, out to the edge of the optical disk \citep[e.g.][]{Martin2001}.  For the purposes of our analysis, our definition is sufficiently similar that we can neglect this difference, but will return to it in future work.

\subsection{The Color-Magnitude  and \ssfr -- M$_{\star}$~distribution}

Figure \ref{fig:massnrall} replicates the final result from \citetalias{Wyder2007}, which we reproduce here in slightly modified form using the analysis described above.  In the left panel of Figure \ref{fig:massnrall} we plot the 1/V$_{max}$-weighted distribution of galaxies as a function of~$^{0.1}(NUV-r)$ vs.~$^{0.1}M_r$  calculated as described in the previous section.    In the central panel we plot~$^{0.1}(NUV-r)_{cor}$ vs.~$^{0.1}M_{r,cor}$, where we have applied our derived dust attenuation corrections A$_{NUV,COMB}$ to the measurement in each band.  Finally, in the right panel we plot the specific star formation rate \ssfr~vs. \Mstar~ based on the quantities derived from our dust-corrected luminosities.  Note that in this paper we adhere to an adopted terminology in which ``blue sequence'' refers to the locus of blue galaxies in the uncorrected or dust-corrected color magnitude diagram, and ``SF sequence'' refers to the locus of star-forming galaxies in the \ssfr~vs. \Mstar~diagram.

As described in \citetalias{Wyder2007}, the blue sequence follows a shallow slope in the UV-optical color-magnitude diagram, with a reddening trend towards higher stellar mass galaxies.  Above a stellar mass of $10^{10}$ \Msol~  the sharpness of the sequence decreases and a wide spread of colors are seen, extending up to the red limit of our sample.  We have overplotted the curve following the peak of the blue sequence derived in \citetalias{Wyder2007}.  In the middle plot, the corrected blue sequence displays a much shallower slope with stellar mass, and a more sharply defined ``peak''.   On the right this sequence tilts again, a result of the conversion of dust-corrected r-band luminosity to stellar mass which varies across the sequence (with $M_\star/L$ increasing with stellar mass).  We overplot the SF sequence with a line defined by

$$\log SFR/M_\star = -0.36 \log M_\star - 6.4$$
taken from \citetalias{Salim2007}\footnote{Taken from an earlier analysis from S07 prior to final published result ($\log SFR/M_\star = -0.35 \log M_\star - 6.33$).}.  We show below that this line closely follows the ridge line (peak) of the star forming distribution, even to high stellar masses: therefore we do not attempt to provide a more rigorous definition.  The line is also consistent with a z$\sim$0.1 extrapolation of the one derived by Noeske et al (2007a, 2007b) over a range of redshifts (0.3$<z<$1.1), in both slope and normalization.  

We overplot a trend line for non-star-forming (non-SF) galaxies in the \ssfr~vs. \Mstar~diagram, but caution the reader that unlike the SF sequence, this locus is likely to misrepresent the true distribution of non-SF galaxies.   In fact, the \ssfr~of galaxies in this part of the diagram is best taken as an upper limit; in practice it remains extremely difficult to probe star formation at levels below log \ssfr $\sim$ -12. Many evolved galaxies possess weak UV emission that is probably not associated with recent star formation \citep[e.g.,][]{Rich2005, Yi2005}.  This UV light from evolved stars is most likely responsible for the red sequence locus in the color-magnitude diagram.    While some authors have suggested that AGN may contribute a fraction of the NUV luminosity in galaxies with reduced levels of star formation \citep[e.g.,][]{Agueros2005, Tremonti2007}, emission-line diagnostics suggest that this is unlikely to be significant for the majority of galaxies in our sample.

In the next series of figures we also highlight the location of intermediate (``green valley'') galaxies with a line defined by the geometric mean of the blue and the red.  The green valley is meant to identify galaxies that lie between the blue and red sequences on the color magnitude diagram, although as we discuss in the next section this population is quite heterogeneous \citep[e.g.,][]{Johnson2007}.   Considering the above caveats for the (log-log) \ssfr~vs. \Mstar~diagram, describing this region as a ``valley'' may not be accurate---galaxies may not actually show a double peaked distribution in \ssfr~at fixed stellar mass.\footnote{However, it is possible that galaxies of a given \Mstar~might show a low level of star formation fed by gas associated with stellar mass loss \citep{Knapp1992} and/or cooling flows \citep{Fabian1982}, or other phenomena \citep{Mathews2003}.}  The ``green valley'' is most physically meaningful in  describing a residual star-forming population \citep[e.g.,][]{Yi2005} in the dust-corrected color-magnitude diagram shown as the middle plot of Figure \ref{fig:massnrall} and discussed extensively in \citetalias{Martin2007}.  In reference to the \ssfr~ vs. \Mstar~ diagram, we refer to these as ``residual-SF'' galaxies with log \ssfr $\sim$ -11.  At their present rate, residual-SF galaxies will form an additional $\sim$1\%--10\% in stellar mass over a gigayear.

\subsection{Trends along the SF, residual-SF and non-SF ``sequences''}

\begin{figure}[t]
\epsscale{1.2}
\plotone{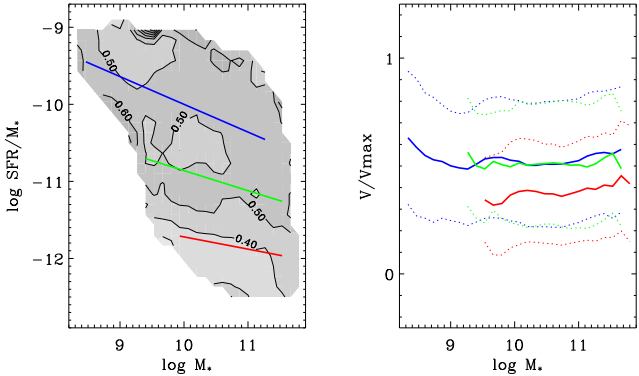} 
\caption{{\it Left}: Mean of $V/V_{max}$  as a function of \ssfr~ vs. log $M_\star$.  The blue solid line shows SF sequence ridge line from \citetalias{Salim2007}, and the red solid line for non-SF galaxies is based on red sequence fit from \citetalias{Wyder2007}.  The green solid line represents residual-SF galaxies (``green valley''), following the geometric mean of SF and non-SF sequences.  {\it Right}: Weighted mean and $\pm1~\sigma$ distribution width on $V/V_{max}$ along blue, green, and red lines.}
\label{fig:mass_ssfr_vmax}
\end{figure}

\begin{figure}[t]
\epsscale{1.2}
\plotone{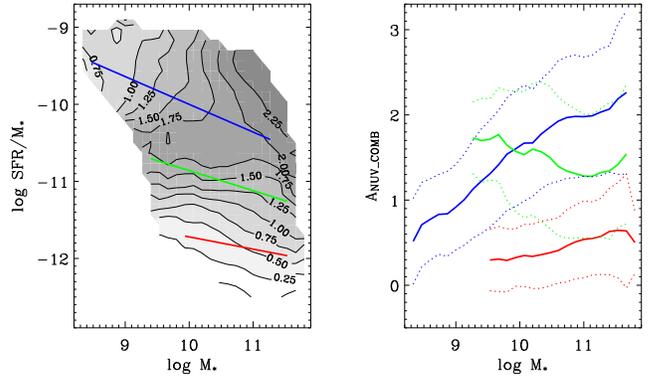} 
\caption{{\it Left}: Weighted mean of NUV-band atttenuation, $A_{NUV,comb}$.    {\it Right}: Weighted mean and $\pm1 \sigma$ distribution width for $A_{NUV,comb}$ along similarly colored curves in above plots. (See caption of Fig. \ref{fig:mass_ssfr_vmax} for explanation.)} 
\label{fig:mass_ssfr_airx}
\end{figure}

\begin{figure}[b]
\epsscale{1.2}
\plotone{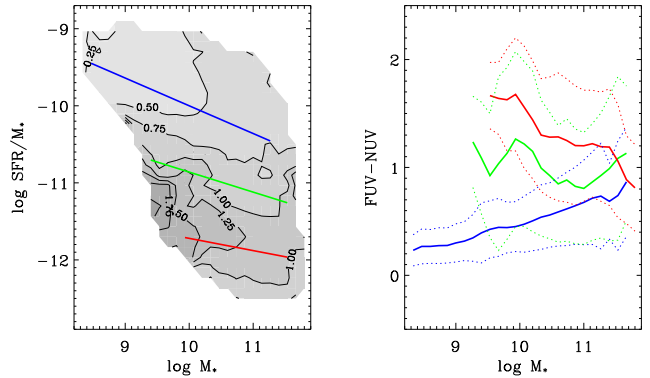} 
\caption{{\it Left}: Weighted mean of rest-frame $FUV-NUV$ color.    {\it Right}: Weighted mean and $\pm1~\sigma$ distribution width for $FUV-NUV$ along similarly colored curves in above plots. (See caption of Fig. \ref{fig:mass_ssfr_vmax} for explanation.)} 
\label{fig:mass_ssfr_fn} 
\end{figure}

\begin{figure}[t]
\epsscale{1.2} 
\plotone{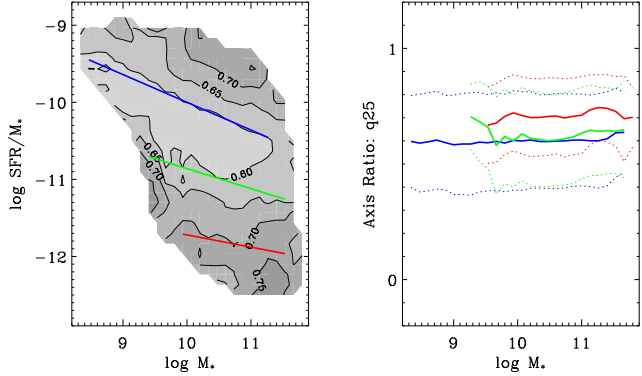} 
\caption{{\it Left}: Weighted mean of axis ratio $q_{25}$, used as a measure of inclination.    {\it Right}: Weighted mean and $\pm1~\sigma$ distribution width for axis ratio $q_{25}$ along similarly colored curves in above plots. (See caption of Fig. \ref{fig:mass_ssfr_vmax} for explanation.)} 
\label{fig:mass_ssfr_q25} 
\end{figure}

\begin{figure}[t]
\epsscale{1.2} 
\plotone{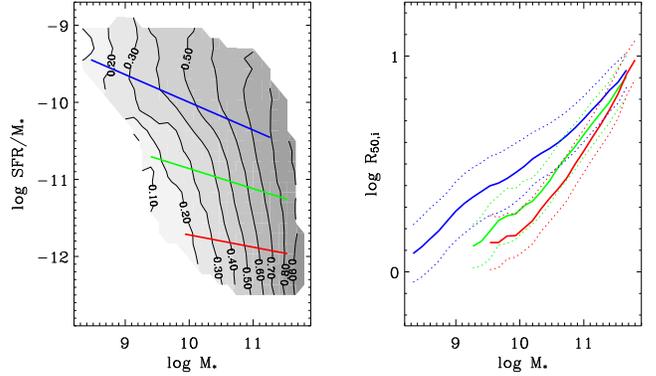} 
\caption{{\it Left}: Weighted mean of logarithm of the $i$-band half-light (50\%) radius, $\log r_{50,i}$ (physical, in kpc ).    {\it Right}: Weighted mean and $\pm1~\sigma$ distribution width for $\log r_{50,i}$ along similarly colored curves in above plots. (See caption of Fig. \ref{fig:mass_ssfr_vmax} for explanation.)} 
\label{fig:mass_ssfr_r50}
\end{figure}

\begin{figure}[b]
\epsscale{1.2} 
\plotone{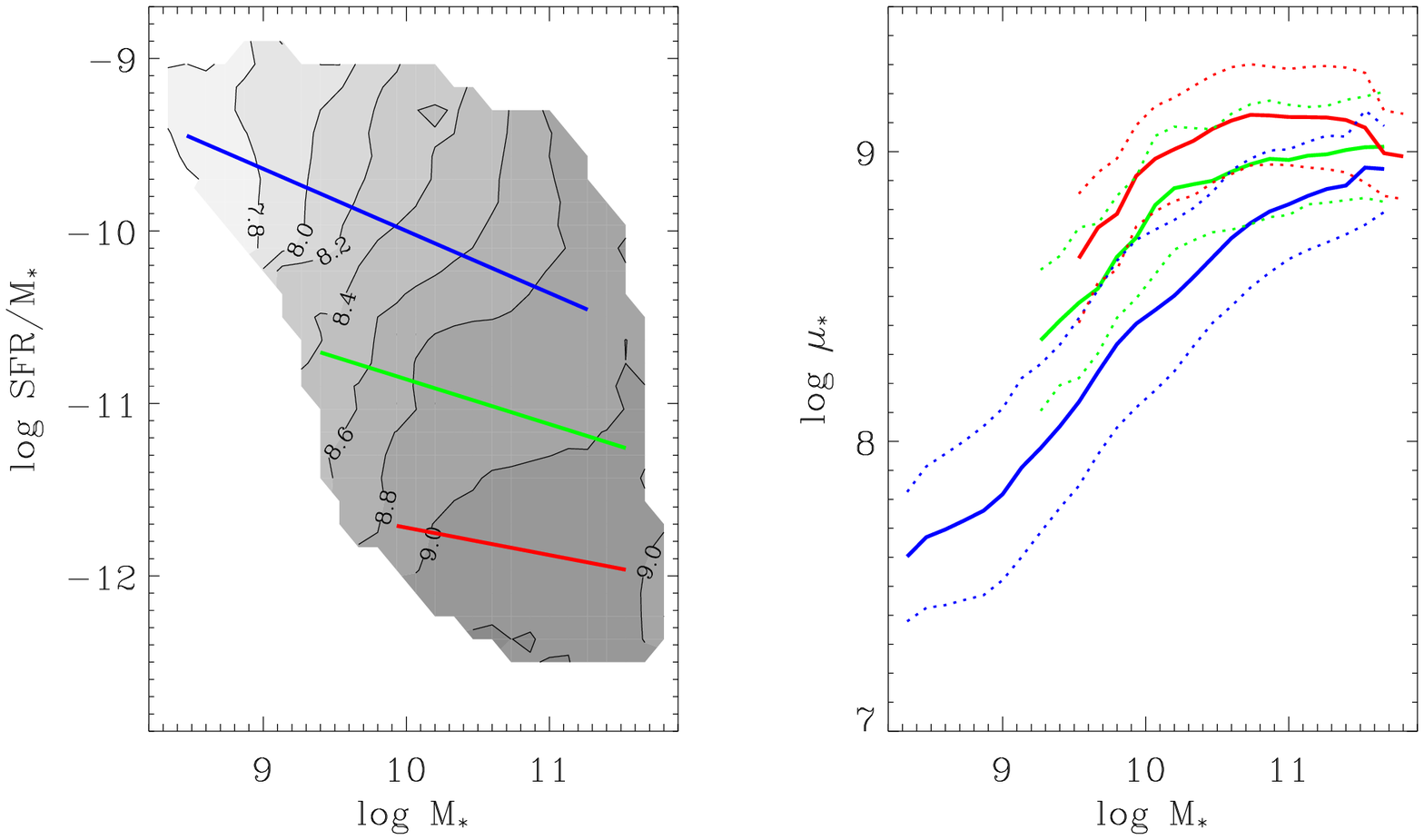} 
\caption{{\it Left}: Weighted mean of logarithm of the stellar mass surface density \logmu.    {\it Right}: Weighted mean and $\pm1!\sigma$ distribution width for \logmu~along similarly colored curves in above plots. (See caption of Fig. \ref{fig:mass_ssfr_vmax} for explanation.)} 
\label{fig:mass_ssfr_logmu}
\end{figure}

\begin{figure}[t]
\epsscale{1.2} 
\plotone{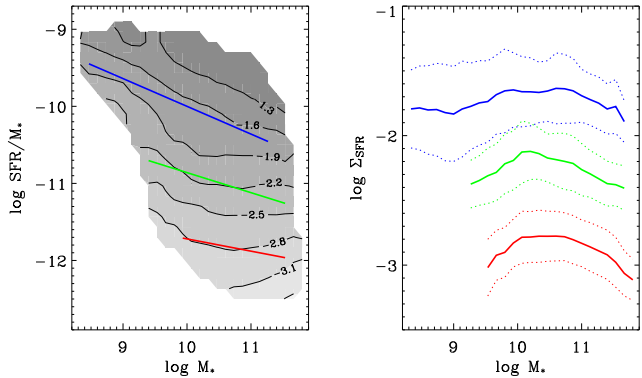} 
\caption{{\it Left}: Weighted mean of logarithm of the star formation rate surface density, $\log \Sigma_{SFR}$.    {\it Right}: Weighted mean and $\pm1~\sigma$ distribution width for $\log \Sigma_{SFR}$ along similarly colored curves in above plots. (See caption of Fig. \ref{fig:mass_ssfr_vmax} for explanation.)} 
\label{fig:mass_ssfr_sfrkpc2_nuv}
\end{figure}

\begin{figure}[t]
\epsscale{1.2} 
\plotone{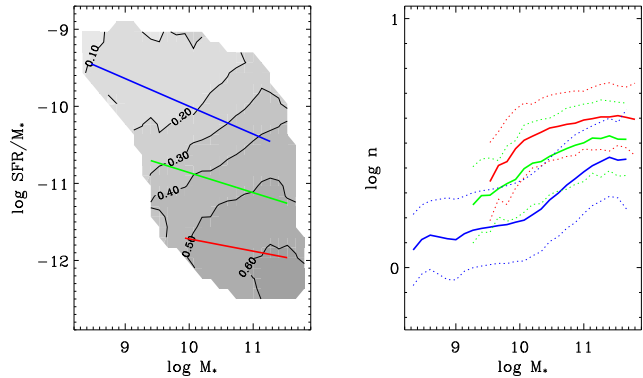} 
\caption{{\it Left}: Weighted mean of logarithm of the Sersic index, $\log n_i$.    {\it Right}: Weighted mean and $\pm1~\sigma$ distribution width for $\log n_i$ along similarly colored curves in above plots. (See caption of Fig. \ref{fig:mass_ssfr_vmax} for explanation.)} 
\label{fig:mass_ssfr_logn}
\end{figure}

\begin{figure}[b]
\epsscale{1.2}
\plotone{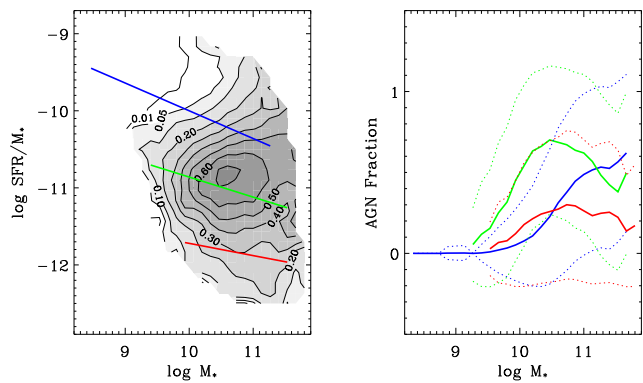}  
\caption{{\it Left}: Weighted mean of AGN fraction.    {\it Right}: Weighted mean and $\pm1~\sigma$ distribution width for AGN fraction along similarly colored curves in above plots. (See caption of Fig. \ref{fig:mass_ssfr_vmax} for explanation.)} 
\label{fig:mass_ssfr_agnfrac}
\end{figure}

The identification of the SF sequence presents a unique opportunity to study the properties of the dominant star-forming population independently of any color, spectroscopic, or morphologically defined selection criteria.   In this section we focus on trends along the SF sequence, and we also compare the properties of SF, residual-SF and non-SF galaxies.  To accomplish this, we calculated the 1/V$_{max}$-weighted mean and standard deviation of a given measure (linear or logarithmic, as indicated) in bins of \ssfr~vs. \Mstar.   Bins were spaced by 0.13 dex in both \ssfr~and \Mstar~and we only considered bins containing more than 20 galaxies.   For each two-dimensional distribution we then measured weighted means (and standard deviations) vs. \Mstar~along the SF, residual-SF, and non-SF sequences defined by the linear relations described above.  Results along the SF sequence for a number of measured and derived properties are given in Table \ref{table:sfsequence}.  Figures \ref{fig:mass_ssfr_vmax}-\ref{fig:mass_ssfr_agnfrac} show the two-dimensional and one-dimensional distribution of weighted means of several key measurements and physical properties.

Since we are considering trends across the entire galaxy population, we first investigate whether any region of parameter space suffers from sample incompleteness which might influence our subsequent analysis.  Figure \ref{fig:mass_ssfr_vmax} shows the (unweighted) mean distribution of $V/V_{max}$ for which we expect a value of 0.5.  We find deviations from this value only for the lowest stellar masses and lowest \ssfr, neither of which will significantly impact our results.    Another crucial component of our analysis is the dust-attenuation correction.  We plot A$_{NUV,COMB}$ in Figure \ref{fig:mass_ssfr_airx} which shows a clear trend towards increasing A$_{NUV,COMB}$ along the SF sequence towards higher \Mstar.  The highest dust attenuations are found for galaxies with the highest \ssfr~and \Mstar~ in the top right of the diagram.  Residual-SF and non-SF galaxies show lower attenuations for the same \Mstar.  We can compare these results with the average $FUV-NUV$ colors shown in Figure \ref{fig:mass_ssfr_fn} which has been shown by \citet{Meurer1999}, \citet{Seibert2005}, and \citet{Johnson2007} to correlate with dust-attenuation in galaxies with ongoing star formation.  Not surprisingly, we find a similar trend along the SF sequence with $FUV-NUV \simeq A_{NUV,COMB}/3.1$, in almost exact agreement with our expectation based on \citet{Seibert2005} and \citet{Calzetti2000}.

Observed axis ratios are known to correlate with galaxy inclination \citep{Binney1981}.  In Figure \ref{fig:mass_ssfr_q25} we use the isophotal axis ratio $q_{25} = b_{25}/a_{25}$ to study the distribution of inclination across the galaxy population.  While the total variation is small, we do find a minimum at \ssfr~ slightly below the SF sequence.   A color-derived attenuation correction will fail to fully remove
inclination effects at optical depths $\tau>$1, e.g. saturating in edge-on galaxies.  One possible interpretation is that our dust-attenuation correction has not fully removed inclination-dependent effects.  However, an alternative is that there is a prevalence of disk-dominated galaxies (which show a wider distribution of axis ratios than irregular or bulge-dominated galaxies) in that part of the \ssfr~vs. \Mstar ~ diagram.   We consider this point further in the Appendix.

The next set of plots shows the variation of structural parameters along the SF sequence, highlighting four parameters: $r_{50}$ and \logmu, which are related to the underlying stellar mass distribution,  \sfrkpc, a measure of the star formation rate surface density, and the Sersic index $n_{i}$, which provides an indication of the ratio of bulge-to-total light.  Figure \ref{fig:mass_ssfr_r50} shows that  the $i$-band half-light radius predominantly increases with \Mstar.  \citet{Shen2003} derived the stellar mass-radius relation for galaxies from SDSS and found for disk galaxies a dependence of r~$\propto $~\Mstar$^{0.15}$ for $\logm < 10.5$ and r~$ \propto $~\Mstar$^{0.4}$ for $\logm > 10.5$.  A fit to the relation along the SF sequence results in r~$ \propto $~\Mstar$^{0.22}$, with a dependence that steepens to r~$\propto$~\Mstar$^{0.3}$ at higher stellar masses.  Residual-SF and non-SF sequence galaxies display an even steeper slope for the \Mstar-$r_{50}$ relation.  Complementary trends are found for \logmu, shown in Figure \ref{fig:mass_ssfr_logmu}.  Surface mass density \logmu varies smoothly along the SF sequence with $\mu \propto \Mstar^{0.52}$ for 9~$<$~\logm~$<$~11.  Over most of the stellar mass range, the residual-SF and non-SF galaxies show much less variation, saturating to nearly a constant value near \logmu~=~9.   These results are similar to those found by \citet{Kauffmann2003b} for young, disk-dominated and old, bulge-dominated systems.

Unlike \logmu, the trend in the star formation rate surface density, \sfrkpc~ plotted in Figure \ref{fig:mass_ssfr_sfrkpc2_nuv}, shows almost no variation along the SF sequence, with \sfrkpc $\simeq 0.02$ $M_\odot$ yr$^{-1}$ kpc$^{-2}$, similar to the nearly constant FUV surface brightness, $I_{FUV}$ vs. \Mstar~noted in \citet{Hoopes2006}.  This value lies approximately at the middle of the range identified for ``normal spirals'' and ``central regions of normal disks'' by \citet{Kennicutt1998b}, although it is approximately 2 orders of magnitude below the typical \sfrkpc~for starbursts \citep{Kennicutt1998b} and star-forming galaxies at z=2 \citep{Erb2006}.   Given the scaling relations identified above, it is straightforward to derive a link between the ridge line of the SF sequence and the near constancy of \sfrkpc~for these galaxies.  If we assume that the ridge line has the approximate functional form
					$$ SFR \propto M_\star^{2/3}$$
then combined with a stellar mass-radius relation
					$$ r \propto M_\star^{1/3-\epsilon}$$
we find that
					$$\Sigma_{SFR} \propto M_\star^{2\epsilon}$$
or nearly constant for small $\epsilon$.  This intriguing result would appear to suggest that the vast majority of star-forming galaxies are distributed such that their scaling relations are consistent with a single value for \sfrkpc.   It may also suggest a possible connection (via the global Schmidt law) to observations that show that disk-dominated and star-forming galaxies possess a narrow range in gas mass surface densities despite displaying a much wider range of gas masses and radii \citep{Haynes1984, Roberts1994, Kennicutt1998a}.  Low surface brightness and starbursting galaxies will lie off of the SF sequence, but they are not a dominant population at $z\sim0.1$.

This result also has interesting implications at higher redshift.  Noeske et al. (2007a, 2007b) suggest that the SF sequence exists up to $z\sim1$ with a similar slope vs.  \Mstar but an evolving SFR intercept (normalization).  Combined with the weak evolution of the stellar mass-radius relation out to z$\sim$1 \citep[e.g.,][]{Somerville2006}, the fundamental change between the local and z=1 SF sequence is an increase in \sfrkpc.  In light of this, it is intriguing that a significant fraction of LIRGs at higher redshift appear to be star-forming disks with elevated SFRs \citep{Zheng2007, Melbourne2005}.

The highest \sfrkpc, about 5 times higher (in the mean), are found in the high-\Mstar, high-\ssfr~ region which also shows the highest mean attenuations.  This extreme part of the diagram is occupied by ultralumininous infrared galaxies and ultraviolet luminous galaxies \citep[UVLGs;][]{Heckman2005,Hoopes2006}, although these populations are rare and will not contribute significantly to the weighted mean.  The residual-SF and non-SF sequences show little variation of \sfrkpc~with stellar mass, with levels of \sfrkpc~ that are at most 1/3 and 1/10 that of the SF sequence.

For the logarithm of the Sersic index, $\log n_i$ shown in Figure \ref{fig:mass_ssfr_logn},  the mean profile along the SF sequence is relatively constant until stellar masses close to the ``transition mass'' (\logm $\simeq 10.5$) identified in \citet{Kauffmann2003a}, where the Sersic index increases sharply.  At any given \Mstar, the mean Sersic index for residual-SF and non-SF galaxies is higher than on the SF sequence, with both showing a trend of increasing $\log n_i$ with increasing \Mstar, particularly below the transition mass.     It is clear from this plot that the SF sequence is not uniform in its structural properties, with a steady increase in bulge-dominated galaxies with increasing stellar mass, and significant scatter over the entire stellar mass range.

The detectability of an AGN has been shown to be correlated with the luminosity of the bulge component and the presence of gas within a galaxy \citep{Kauffmann2003c}.  The fraction of AGN [with log L(OIII)$_{extcor} > 5$] as a function of \ssfr~ and \Mstar~ is plotted in Figure \ref{fig:mass_ssfr_agnfrac}.  The trend along the SF sequence mimics the trend in Sersic index.  In addition, along the residual-SF line the AGN fraction remains high, suggesting that galaxies with small amounts of SF may still show the AGN phenomenon.  The AGN fraction and its connection to residual-SF in ``green valley'' galaxies is discussed in more detail in  \citetalias{Martin2007} and in \citetalias{Salim2007}.

We conclude this section by restating some key results: 

1)  In almost all respects, we find that the physical properties of galaxies vary smoothly (vs. \Mstar) along the SF sequence.  Although we find known scaling relations for star-forming, disk and/or bulge-dominated galaxies we do recover some new and notable trends.

2) We find that \sfrkpc~remains nearly constant vs. \Mstar~along the SF sequence, with $\sfrkpc$ for non-SF galaxies least a factor of 10 lower on average.

3) At fixed \Mstar, $r_{50}$ is higher and $\logmu$ and $\log n_i$ lower for the SF sequence compared to the residual-SF and non-SF sequences. This suggests that non-SF galaxies are not simply SF galaxies of the same stellar mass that have stopped forming stars---a result consistent with the observed correlation between SFH and structure.   However, these differences between the SF and non-SF sequences are most profound near the transition mass.  Above this stellar mass the structural properties of the population show considerably less scatter, despite the significant differences in \ssfr~and \sfrkpc.

\begin{deluxetable*}{l c c c c c c c c l}
\small
\tablecaption{Measured and Derived Properties along the Star-Forming Sequence}
\tablewidth{0pt}
\tablehead{
\colhead{} & \multispan{7}{\hfill log (M$_\star$/M$_\odot$) \hfill}\\
\colhead{Measurement} & \colhead{8.5}& \colhead{9.0} & \colhead{9.5}  & \colhead{10.0} & \colhead{10.5}& \colhead{11.0} & \colhead{11.5}\\
}
\startdata
$V/V_{max}$ &   0.58 &   0.50 &   0.53 &   0.52 &   0.51 &   0.52 &   0.56 \\
\hfill $\pm$  &   0.31 &   0.25 &   0.27 &   0.28 &   0.29 &   0.29 &   0.30 \\
\hline
$~^{0.1}(FUV-NUV)$ &   0.27 &   0.30 &   0.42 &   0.46 &   0.57 &   0.68 &   0.73 \\
\hfill $\pm$ &   0.16 &   0.18 &   0.27 &   0.25 &   0.30 &   0.39 &   0.47 \\
$~^{0.1}(NUV-r)$ &   1.77 &   1.98 &   2.32 &   2.65 &   2.97 &   3.24 &   3.52 \\
\hfill $\pm$  &   0.36 &   0.39 &   0.49 &   0.46 &   0.48 &   0.50 &   0.54 \\
$A_{NUV,COMB}$ &   0.73 &   0.92 &   1.29 &   1.63 &   1.87 &   1.98 &   2.16 \\
\hfill $\pm$ &   0.49 &   0.50 &   0.63 &   0.68 &   0.67 &   0.70 &   0.85 \\
$A_{z}$ &   0.26 &   0.32 &   0.45 &   0.61 &   0.70 &   0.73 &   0.81 \\
\hfill $\pm$ &   0.29 &   0.28 &   0.31 &   0.38 &   0.35 &   0.35 &   0.43 \\
$A_{NUV,Salim}$ &   0.99 &   1.04 &   1.19 &   1.45 &   1.68 &   1.89 &   2.16 \\
\hfill $\pm$ &   0.62 &   0.68 &   0.59 &   0.67 &   0.79 &   0.99 &   1.23 \\
\hline
$q_{25}$ &   0.59 &   0.58 &   0.59 &   0.60 &   0.60 &   0.60 &   0.63 \\
\hfill $\pm$  &   0.21 &   0.21 &   0.21 &   0.21 &   0.20 &   0.20 &   0.20 \\
\hline
$\log SFR_{JB}$ &  -1.13 &  -0.80 &  -0.37 &   0.04 &   0.47 &   0.83 &   1.04 \\
\hfill $\pm$  &   0.23 &   0.26 &   0.25 &   0.23 &   0.22 &   0.21 &   0.22 \\
$\log SFR_{SALIM}$ &  -0.71 &  -0.49 &  -0.20 &   0.06 &   0.34 &   0.69 &   1.03 \\
\hfill $\pm$  &   0.32 &   0.38 &   0.39 &   0.38 &   0.27 &   0.36 &   0.49 \\
\hline
$\log r_{i,50}$ &   0.13 &   0.28 &   0.39 &   0.47 &   0.58 &   0.71 &   0.87 \\
\hfill $\pm$  &   0.14 &   0.14 &   0.14 &   0.13 &   0.12 &   0.11 &   0.10 \\
$\log \mu_\star$ &   7.68 &   7.82 &   8.12 &   8.43 &   8.65 &   8.82 &   8.93 \\
\hfill $\pm$  &   0.25 &   0.30 &   0.29 &   0.28 &   0.23 &   0.19 &   0.19 \\
\hline
$\log n_i$ &   0.12 &   0.11 &   0.16 &   0.19 &   0.28 &   0.38 &   0.43 \\
\hfill $\pm$  &   0.14 &   0.16 &   0.15 &   0.16 &   0.17 &   0.17 &   0.15 \\
$f_{DeV}$ &   0.14 &   0.14 &   0.18 &   0.21 &   0.33 &   0.54 &   0.62 \\
\hfill $\pm$  &   0.23 &   0.22 &   0.24 &   0.27 &   0.31 &   0.32 &   0.30 \\
\hline
$\log \Sigma_{SFR}$ &  -1.79 &  -1.83 &  -1.74 &  -1.65 &  -1.65 &  -1.69 &  -1.78 \\
\hfill $\pm$  &   0.33 &   0.37 &   0.32 &   0.28 &   0.24 &   0.20 &   0.20 \\
$\log f_{gas}$ &  -0.47 &  -0.56 &  -0.71 &  -0.91 &  -1.13 &  -1.32 &  -1.49 \\
\hfill $\pm$  &   0.09 &   0.10 &   0.11 &   0.12 &   0.11 &   0.11 &   0.11 \\
f$_{AGN}$ &   0.00 &   0.00 &   0.01 &   0.06 &   0.25 &   0.49 &   0.56 \\
\hfill $\pm$  &   0.00 &   0.04 &   0.08 &   0.24 &   0.43 &   0.50 &   0.50 \\
\enddata
\tablecomments{Values for 1$\sigma$ confidence interval halfwidth are given on second line of each entry.  Units for SFR in M$_\odot$ yr$^{-1}$, r$_{i,50}$ in kpc, $\mu$ in M$_\odot$ kpc$^{-2}$ and $\Sigma_{SFR}$ in M$_\odot$ yr$^{-1}$ kpc$^{-2}$.} 
\label{table:sfsequence}
\end{deluxetable*}

\section{On the relationship between star formation history and structure}

\subsection{\ssfr~vs. $\log n_i$ and the SFR excess, $\Delta_{SFR}$ ~vs. $\log n_i$}

\begin{figure*}[t]
\plotone{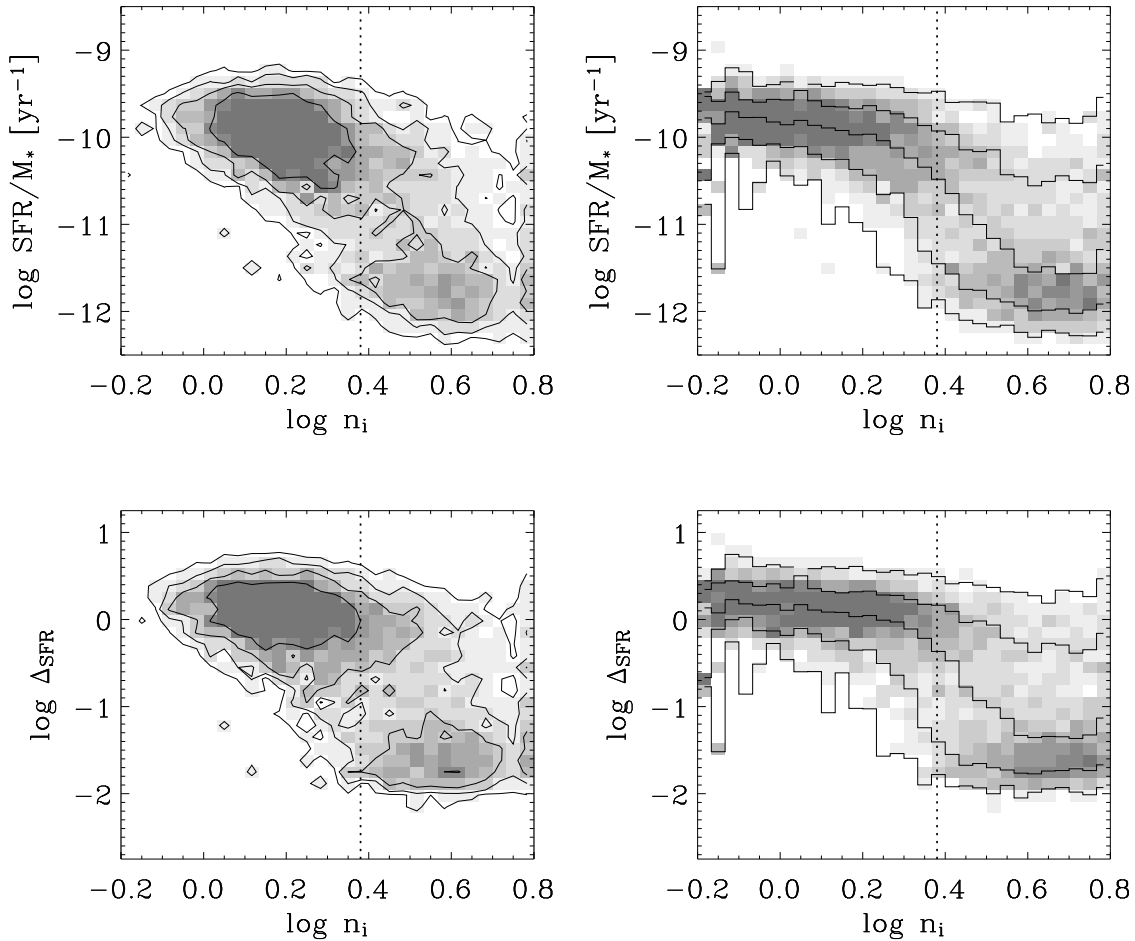} 
\caption{Specific star formation rate and ``SFR excess'' vs. galaxy light profile shape, plotting the   bivariate distribution (1/V$_{max}$-weighted) of galaxies as a function of \ssfr~ ($top$) and $\Delta_{SFR}$ ($bottom$) vs. log of $i$-band Sersic index $n_i$.   {\it Left}:  Contours enclose 38\%, 68\%, 87\%, 95\% of the distribution. {Right:} Conditional distribution in each $\log n_i$ bin. Quintiles represent 10\%, 25\%, 50\%, 75\%, 90\% of the distribution in that bin.  } 
\label{fig:n_nuvr_all}
\end{figure*}

As discussed in $\S$ 1, galaxies in the local universe exhibit a strong correlation between their structure and their star formation history.  This section explores this connection using our local galaxy sample.  The upper plots in Figure \ref{fig:n_nuvr_all} show the distribution of \ssfr~vs. $\log n_i$ for the full galaxy population, where both the galaxy $1/V_{max}$ weighted density distribution and the conditional density distribution are shown.  Broadly speaking there is a clear trend with $\log n_i$, with \ssfr~ decreasing as one moves towards bulge-dominated systems.  However, the quintiles clearly demonstrate that the star formation history of disk-dominated galaxies falls within a narrow range of values, whereas bulge-dominated galaxies show a much wider range of \ssfr.  This demonstrates (1) a dearth of passive disk-dominated galaxies, (2) a significant population of bulge-dominated galaxies with significant $\ssfr$, and (3) a fairly sharp transition between these two zones at $\log n_i \sim 0.38$.   We also reiterate that the ``peak'' in the distribution of no-SF galaxies at $\log \ssfr \sim -12$ may simply result from the fact that we are unable to measure values of \ssfr~significantly below this value---the true distribution is likely to display an even wider dispersion.

One limitation of this plot is that it necessarily combines galaxies with a wide range of \Mstar~which means that differences between low and high stellar mass galaxies may introduce scatter, thereby masking a tighter relation within a single \Mstar~bin.  We have found that using narrower ranges in stellar mass does reduce scatter although in general the overall trends we have identified remain.

We can effectively remove the stellar mass dependence by measuring a quantity that we call the SFR excess,  $\Delta_{SFR} (M_\star)$.  We define this using the equation for the ridge line of the SF sequence
$$ \log \Delta_{SFR} = \log SFR - (0.64) \log M_\star + 6.4 $$
which provides a measure of the excess or deficiency of SFR as a function of stellar mass.

In certain respects this residual is similar to the gas deficiency parameter defined by \citet{Haynes1984}, although it is different in two important ways (other than the trivial sign difference and the fact that we are using SFR rather than gas content to describe an excess).  First of all, we correct for a dependence on stellar mass rather than morphological type or the combination of morphological type and radius, as in \citet{Solanes2001}.  Our method has the advantage that stellar mass is much more reliably measured than morphological type.  In addition,  we have a basis for physically interpreting the SFR excess since it can be related (on average) to the SFR intensity, \sfrkpc, rather than to a median or mean of the full galaxy population in a given stellar mass/luminosity bin.

The SFR excess, $\Delta_{SFR}$ ~vs. $\log n_i$ is plotted in the lower panels of Figure \ref{fig:n_nuvr_all}.  The distribution of $\Delta_{SFR}$ for disk-dominated galaxies is quite narrow below the disk/bulge transition value.  While bulge-dominated galaxies do show a considerable spread in $\Delta_{SFR}$, there is little variation of $\Delta_{SFR}$ vs. Sersic index within the separate subpopulations, suggesting an additional controlling parameter for $\ssfr$.  See, for example, \citet{Blanton2005a} for an exploration into the connection between Sersic index, color and evironment.

\subsection{\ssfr~vs. M$_{\star}$  and \logmu~for disk and bulge-dominated subsamples and derived gas fractions}

\begin{figure*}[t]
 \plotone{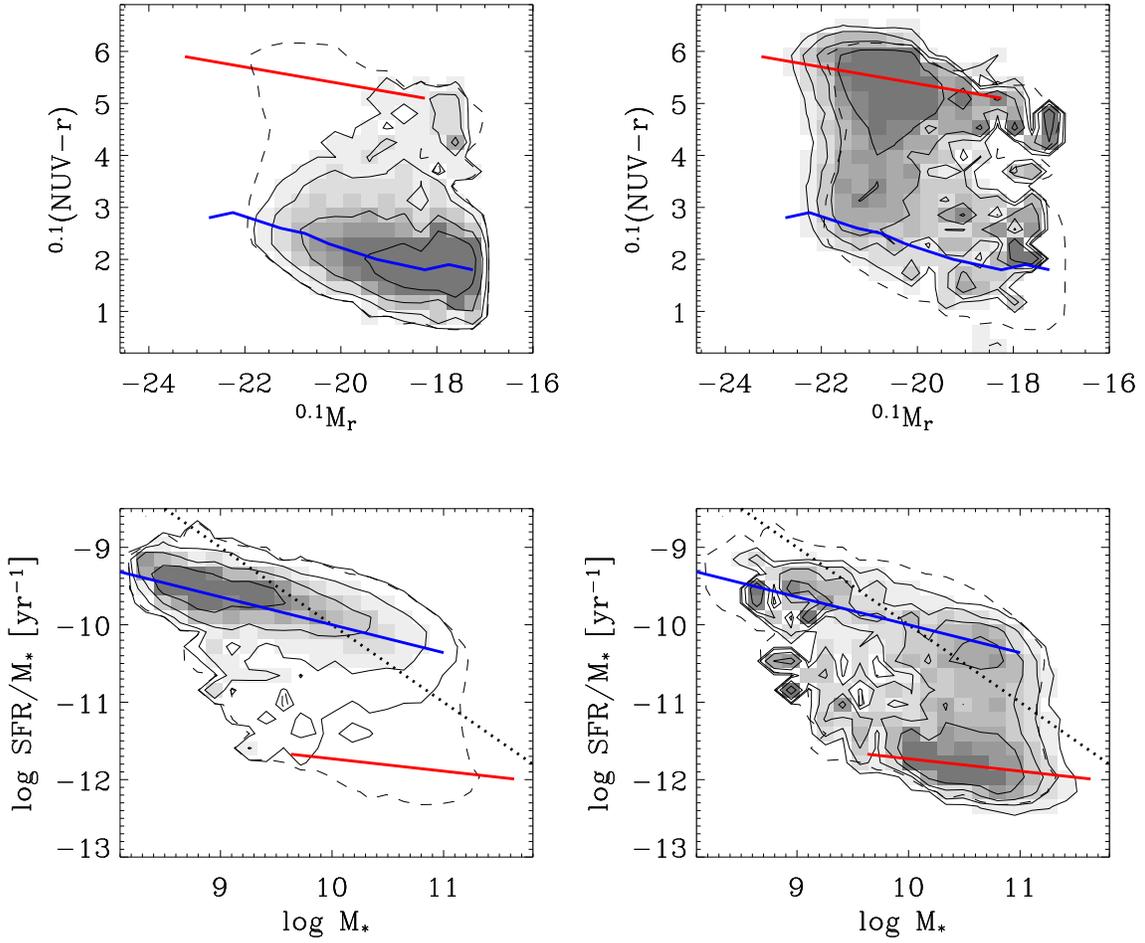} 
\caption{Bivariate distribution (1/V$_{max}$-weighted) of galaxies as a function of $NUV-r$ vs. $M_r$ ({\it top}) and \ssfr~ vs. \Mstar~({\it bottom}) for subsamples split by Sersic index: $n_i$.  $n_i<2.4$ (log $n_i < 0.38$) ({\it left}) and $n_i>2.4$ (log $n_i > 0.38$) ({\it right}).  Solid lines and contours are as in Fig. \ref{fig:massnrall}. Dashed contour follows the outermost contour from Fig. \ref{fig:massnrall}} 
\label{fig:colmag_subsample}
\end{figure*}

In this section we use the Sersic index $n_i$ to split our main sample into disk and bulge-dominated subsamples in order to investigate the relation between structural properties and the star formation history (and other physical properties) of galaxies in the local universe. Our disk/bulge-dominated cut is similar to the one made in \citet{Vincent2005} and \citet{Blanton2003c}.  In Figure \ref{fig:colmag_subsample} we present the same plots shown in Figure \ref{fig:massnrall} for each of our subsamples.  Disks remain on the blue (and SF) sequence, with a weak tail towards decreasing star formation activity.  Most of this tail is at lower stellar masses.  We also find that the relative width of the distribution of disk galaxies is wider in uncorrected color than in the \ssfr~vs. \Mstar~plot, indicating that much of the scatter in color spread was removed when the dust-attenuation correction was applied.  Inclined disks are largely responsible for this spread.    Bulge-dominated galaxies, on the other hand, while predominantly populating the no-SF region, show a much greater diversity in color-magnitude and \ssfr~vs. \Mstar~ space.   Some bulge-dominated galaxies appear to have specific star formation rates comparable to disk galaxies, and also populate the SF sequence.

 \begin{figure*}[p]
 \epsscale{1.15}
 \plotone{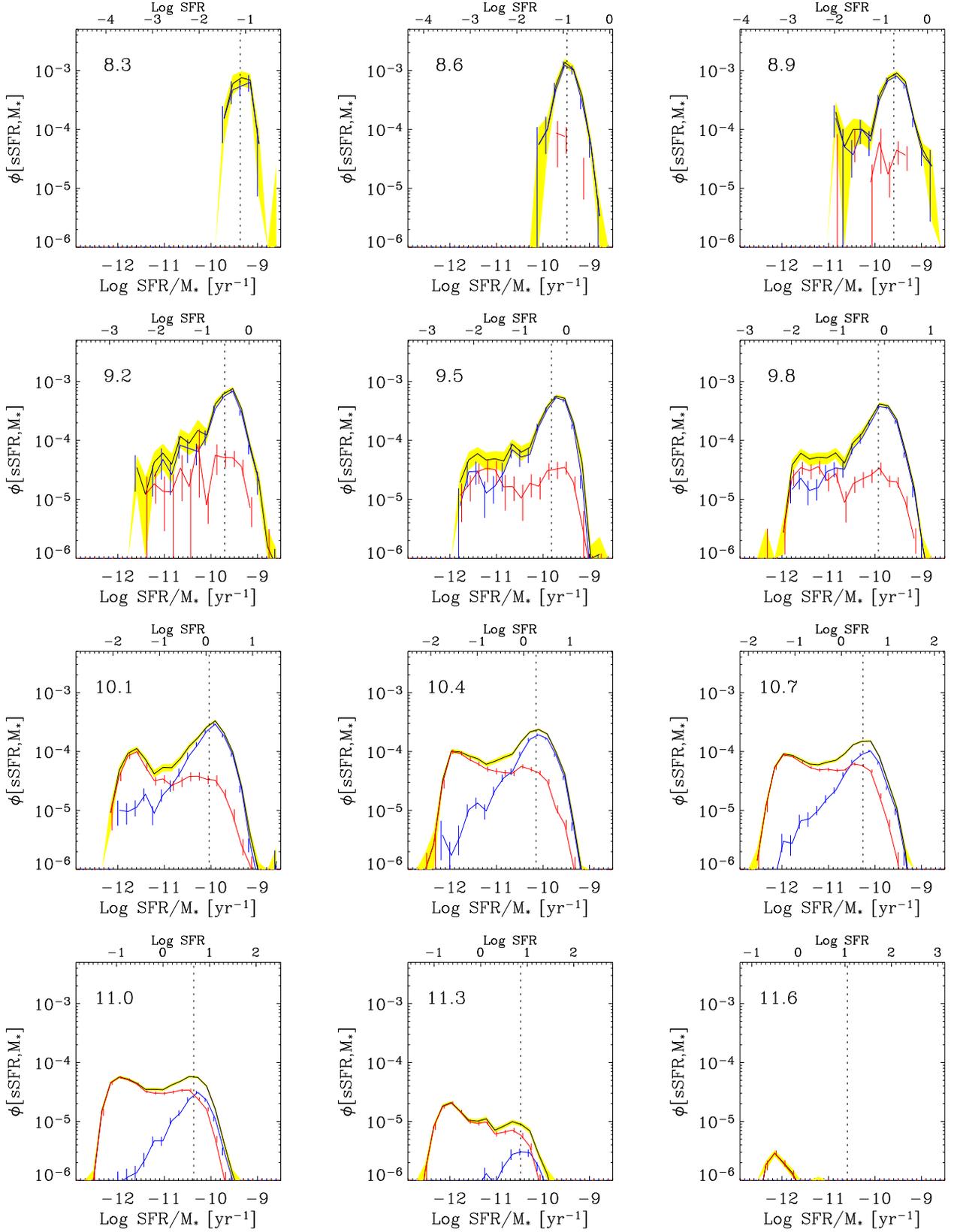} 
\caption{Galaxy density distribution (1/V$_{max}$-weighted) vs. \ssfr~ in bins of log \Mstar~ for total galaxy sample and in disk/bulge-dominated subsamples split by Sersic index $n_i$ ({it blue/red}; split at $n_i= 2.4$ or  log $n_i = 0.38$).  Error bars ({\it yellow filled regions and vertical lines}) have been determined using bootstrap resampling.  Upper axis denotes log SFR using average $< \log \Mstar >$ ({\it upper left corner of each frame}), with the vertical dotted line showing the SFR of the SF sequence ridge for corresponding $<\log \Mstar>$.   Units of $\phi(\ssfr, \Mstar)$ are in Mpc$^{-3}$ bin$^{-1}$, where each bin is 0.3 dex wide in \Mstar~and 0.1875 dex wide in \ssfr.} 
\label{fig:ssfr_mass_cut}
\end{figure*}

 \begin{figure*}[p]
 \plotone{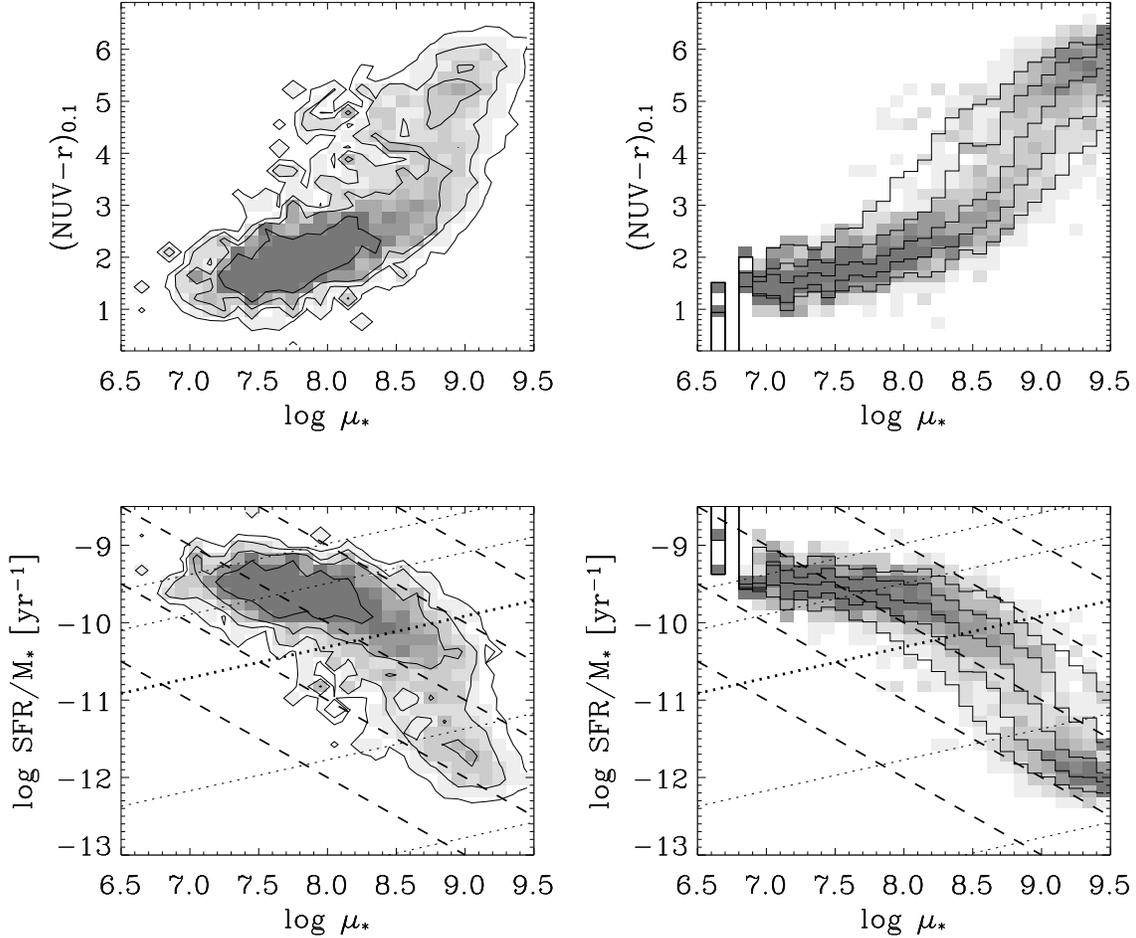} 
\caption{UV-optical color and specific star formation rate vs. stellar mass surface density, showing the bivariate distribution (1/V$_{max}$-weighted) of galaxies as a function of rest-frame $NUV-r$ color ($top$) and \ssfr~($bottom$) vs. \logmu.  For $NUV-r$ vs. \logmu~ no dust-attenuation correction has been applied to the color. {\it Left}:  Contours enclose 38\%, 68\%, 87\%, 95\% of the distribution.  Diagonal dashed lines are isopleths in \sfrkpc~spaced by 1 dex.  Dotted lines show constant gas fractions of (decreasing) 0.5, 0.3, 0.1, 0.01, 0.001.  The bold dotted line plots $f_{gas}=0.1$. {\it Right}: Conditional distribution in each \logmu~bin. Quintiles represent 10\%, 25\%, 50\%, 75\%, 90\% of the distribution in that bin. } 
\label{fig:munuvrall}
\end{figure*}

 \begin{figure*}[t]
 \plotone{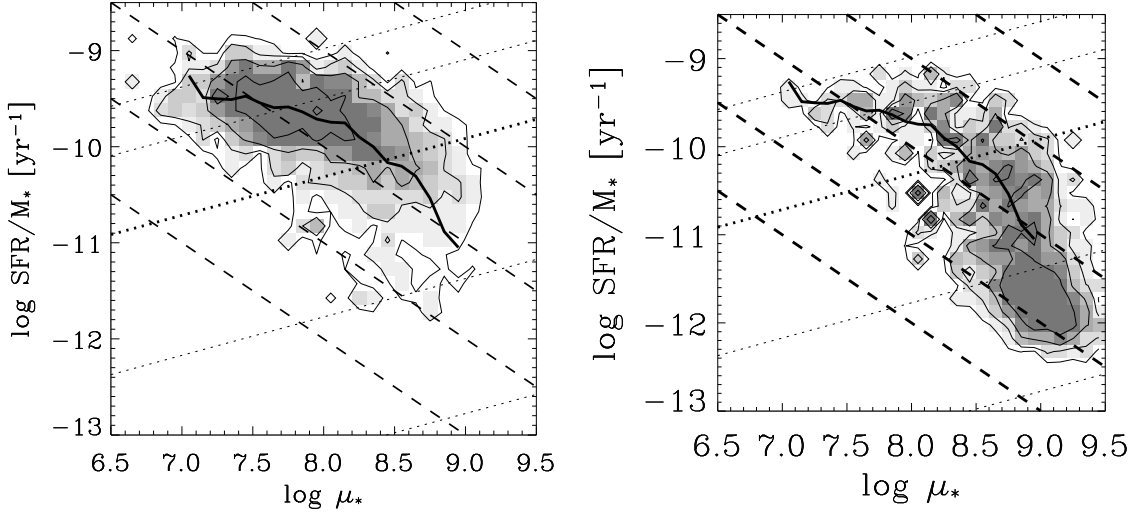} 
\caption{UV-optical color and specific star formation rate vs. stellar mass surface density for disks and bulge-dominated galaxies, showing the bivariate distribution (1/V$_{max}$-weighted) of galaxies as a function of \ssfr~ vs. \logmu~and split by Sersic index $n_i$:   $n_i<2.4$ (log $n_i < 0.38$) ({\it left}) and, $n_i>2.4$ (log $n_i > 0.38$) ({\it right}).  Contours enclose 38\%, 68\%, 87\%, 95\% of the distribution.  Diagonal dashed lines are isopleths in \sfrkpc~spaced by 1 dex,  dotted lines show constant gas fractions of (decreasing) 0.5, 0.3, 0.1, 0.01, 0.001, bold dotted line plots $f_{gas}=0.1$, and the solid line shows the median \ssfr~ vs. \logmu~ for full sample.} 
\label{fig:logmu_nuvr_subsample}
\end{figure*}

\begin{figure*}[t]
\epsscale{1.15}
\plotone{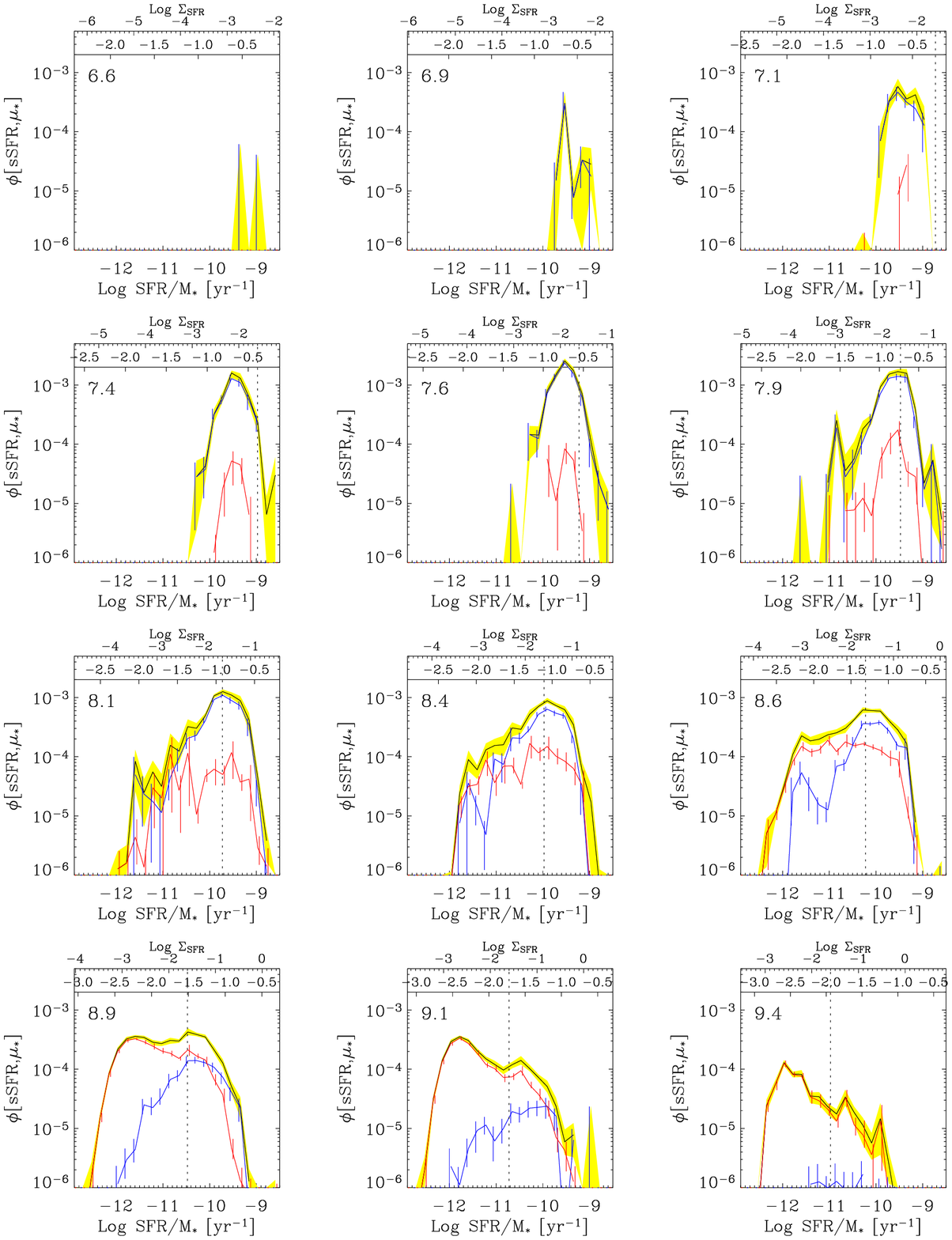} 
\caption{ \ssfr~ distribution (1/V$_{max}$-weighted) in bins of fixed \logmu~ for total galaxy sample and in disk/bulge-dominated subsamples split by Sersic index $n_i$ ({\it blue/red}; split at $n_i= 2.4$ or  log $n_i = 0.38$).  Error bars ({\it yellow filled regions and vertical lines}) have been determined using bootstrap resampling.  Upper axes denote star formation rate surface density and gas fraction (log $\Sigma_{SFR}$ and $\log f_{gas}$) using average $< \logmu >$ ({\it upper left corner of each box}), with the vertical dotted line corresponding to log $\Sigma_{SFR}=-1.7$. Units of $\phi(\ssfr, \Mstar)$ are in Mpc$^{-3}$ bin$^{-1}$ where each bin is 0.25 dex wide in \logmu~and 0.1875 dex wide in \ssfr. } 
\label{fig:ssfr_logmu_cut}
\end{figure*}

Numerous studies have detected star formation in bulge-dominated galaxies, including several key studies in the local universe in the UV \citep{Rich2005, Yi2005}.  Nevertheless, we re-iterate the essential $qualitative$ difference between the plots in the lower half of Figure \ref{fig:colmag_subsample} and those from other studies based on the color-magnitude diagram.  While disks have been known to occupy the diffuse ``blue cloud'' of the optical color magnitude diagram \citep{Bell2005}, it is clear that the majority of disks fall on the very tight SF sequence.  In fact, it is the bulges, which dominate the red sequence in color-magnitude diagrams, that show a great diversity of \ssfr.  The majority of bulge-dominated galaxies are not undergoing major episodes of stellar mass build-up.  Nevertheless, the spread suggests that bulge-dominated galaxies show a rich distribution in SFR despite homogeneity in optical colors and, to some extent, structure.

To quantify this picture, we extracted the one-dimensional (1/V$_{max}$) distribution of \ssfr~in 12 different stellar mass bins, shown in Figure \ref{fig:ssfr_mass_cut}.  In each plot we show the total distribution, as well as the distribution for disk and bulge-dominated subsamples.  For all the samples and subsamples, the 1 $\sigma$ confidence interval on the distribution was calculated using bootstrap resampling.  We indicate on the top horizontal axis the SFR that corresponds to the equivalent range of \ssfr~for that particular \Mstar bin.  A vertical dotted line indicates the location of the SF sequence defined by the relation derived in \citetalias{Salim2007}.  As expected, these lines intercept the mode of the star-forming peak (SF sequence ridge line) in our derived distributions.  

These plots show the relative contribution of disk and bulge-dominated galaxies to any part of the \ssfr~vs. \Mstar~ diagram and further confirm the trends identified above.   At low stellar masses the galaxies that lie on the SF sequence are disk-dominated.  At higher stellar masses residual and no-SF galaxies are bulge-dominated.  The transition between the prevalence of star-forming and ``dead'' galaxies appears to take place at a stellar mass of \logm=10.4.   At nearly all stellar masses, bulge-dominated galaxies show a spread in \ssfr.  Disk galaxies are predominantly located within the SF sequence, although there is evidence for a relative increase in passive disks at low stellar masses.   While the relationship between \ssfr~ and structure is not uniform across all stellar masses, for the $10<$\logm$<11$ range we do find that disk galaxies dominate the SF sequence and bulge-dominated galaxies dominate the residual-SF and no-SF populations.  In this stellar mass range, given \ssfr, we can predict with high likelihood whether the galaxy is disk or bulge-dominated.
Figure \ref{fig:ssfr_mass_cut} also shows that at stellar masses above \logm=10.4, a significant fraction of galaxies in the SF sequence are bulge-dominated galaxies, with these systems becoming the majority at stellar masses above \logm=11.0.  These galaxies are among those with the highest SFRs in the local universe.  Extreme versions of these would be the massive, compact UVLGs and/or rare ULIRGs.

  \begin{figure*}
  \plotone{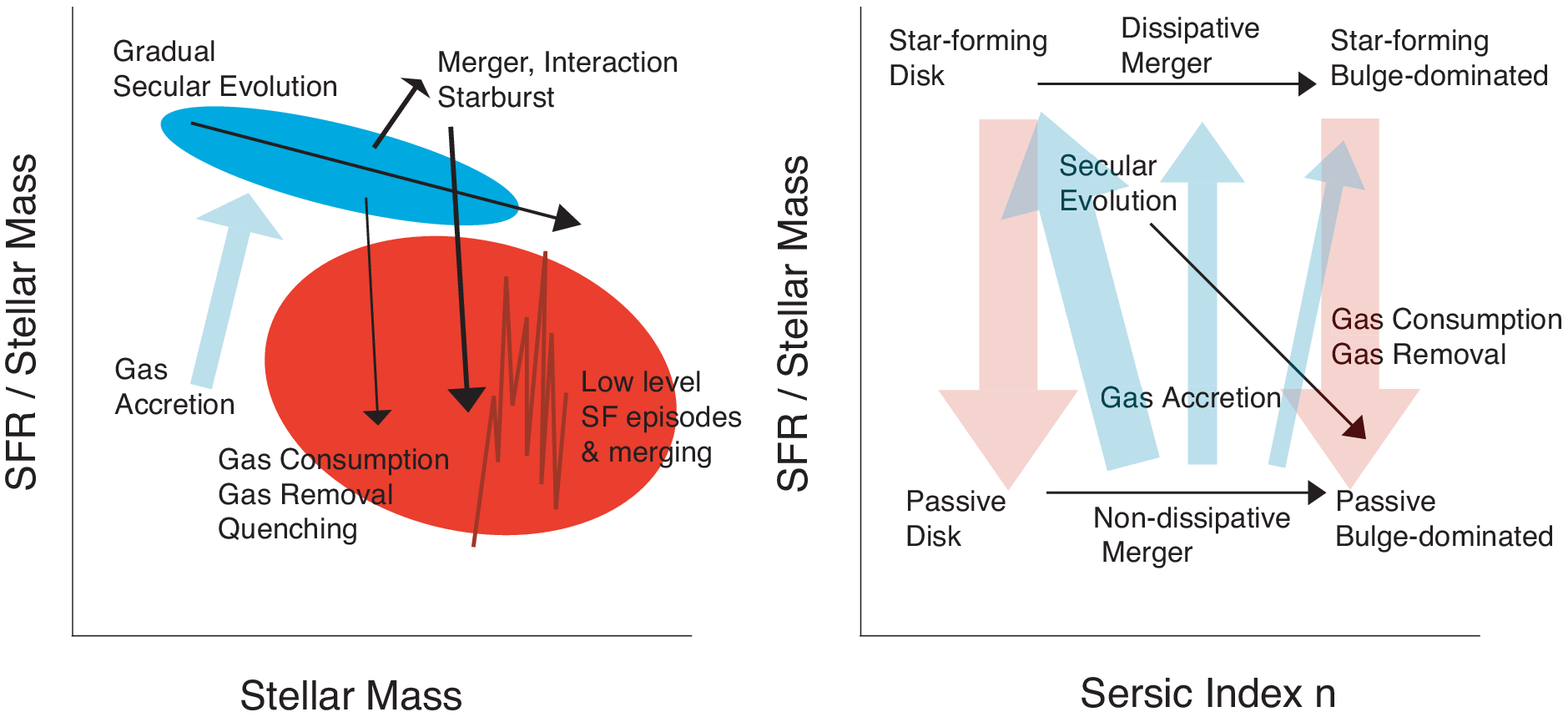} 
\caption{Schematic views of star formation history vs. stellar mass ({\it left}) and morphology ({\it right}, as measured by Sersic index {\it n}) and its evolution.  See text for details.} 
\label{fig:morphevolution}
\end{figure*}

 \begin{figure*}[p]
 \epsscale{1.15}
 \plotone{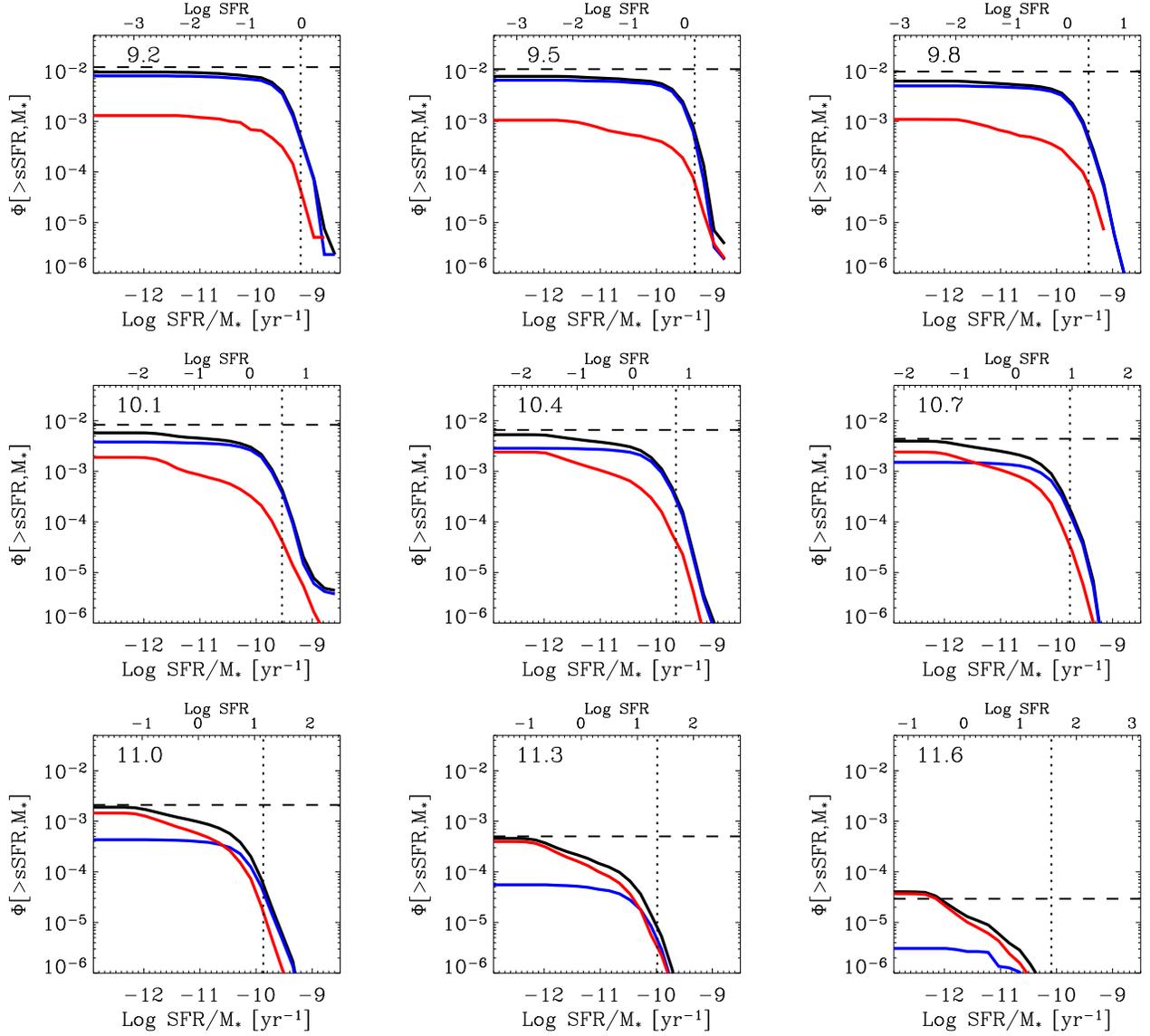} 
\caption{Integrated galaxy density distribution (1/V$_{max}$-weighted)  for all galaxies that exceed \ssfr~ in a given bin of fixed log \Mstar~  for subsamples plotted in Fig. \ref{fig:ssfr_mass_cut} (see caption).  Upper axis denotes log SFR using average $< \log \Mstar >$ ({\it upper left corner of each box}), with the vertical dotted line showing the log $\Delta_{SFR}=0.5$ for corresponding $<\log \Mstar>$.  {\it Black horizontal dashed line}: Total galaxy density in stellar mass bin taken from local stellar mass function of \citet{Bell2003b} and \citet{Borch2006}.   } 
\label{fig:ssfr_massint_cut}
\end{figure*}

 \begin{figure*}[t]
 \plotone{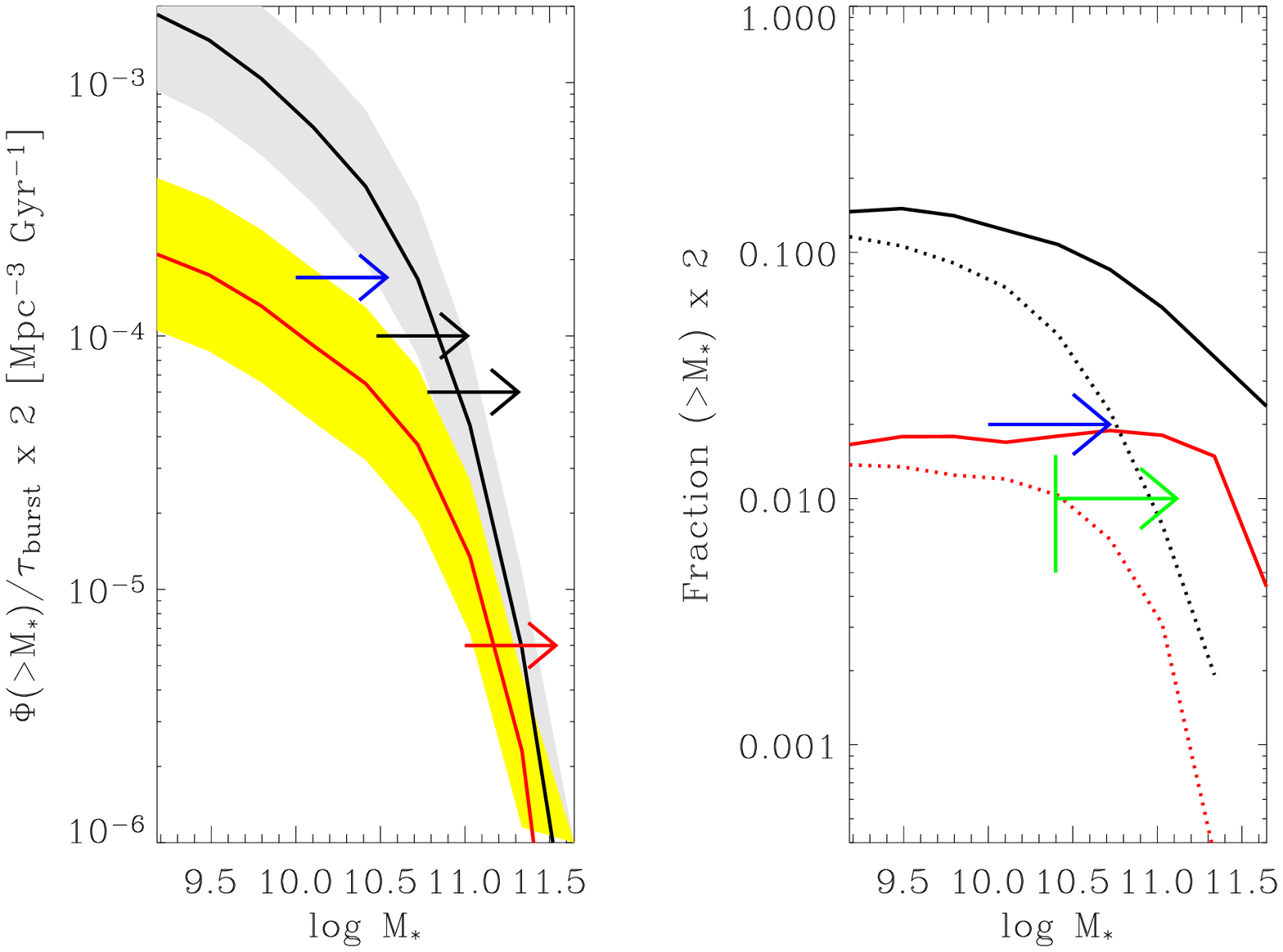} 
\caption{{\it Left}: Estimated ``burst'' rate density  ($\times 2$) for galaxies above a given \Mstar, compared to various merger rates from the literature.  {\it Solid red line}:  Rate density using bulge-dominated galaxies with $\log \Delta_{SFR} > 0.5$. {\it Solid black line}: Total rate density for all galaxies with $\log \Delta_{SFR} > 0.5$.  {\it Arrows}:  Local measurements from \citet{De Propris2005} ({\it blue}), \citet{Masjedi2006} ({\it red}) and model prediction from \citet{Maller2006} ({\it black}).  {\it Right}: Fraction of galaxies ($\times 2$) experiencing $\log \Delta_{SFR} > 0.5$ compared with local merger fraction from literature. Colors as above.  Fraction is calculated using integrated number density of galaxies with  $>$\Mstar~({\it solid line}) and $>$\Mstar/2 ({\it dashed line}) to account for possible range in merger mass ratios.  {\it Arrows}:  Local measurements from \citet{De Propris2005} ({\it blue}), and \citet{Bell2006} ({\it green}).  }
\label{fig:intflow}
\end{figure*}

Previous work indicates that \ssfr~might be better correlated with \logmu~than with \Mstar~\citep{Bell2000, Kauffmann2003b, Brinchmann2004}.  In  Figure \ref{fig:munuvrall} we show the distribution of \ssfr~ vs. \logmu.  On this plot isopleths in \sfrkpc~ lie along straight lines of negative unit slope (under the simplistic assumption that r$_{50,SFR}$=r$_{50,r}$).  The trends seen in Figure \ref{fig:munuvrall} closely match the results from \citet{Brinchmann2004}.  The most striking feature of these plots is the fact that over a significant range in \logmu, \ssfr~is slowly varying (although nearly constant in \sfrkpc) until the transition point \logmu $\sim 8.5$ beyond which \ssfr~varies by several orders of magnitude and \logmu nearly constant.  A simple interpretation of this result is that stellar mass surface density increases for star forming galaxies up to a threshold $\logmu$, above which it remains nearly constant.  It is intriguing that the highest SFR intensities are observed for galaxies in the range of \logmu~where this transition is observed. 

We can use our derived physical properties (\ssfr~ and \logmu) to attempt to connect our results to the gas content of galaxies.  Recently, a number of investigations \citep{Boselli2001, Bell2003a, Kannappan2004, Reddy2006, Erb2006} have made use of the correlation between specific star formation rate and/or specific UV luminosities and gas mass  fractions.  \citet{Reddy2006} obtain a simple expression $\ssfr = C f_{gas} / (1-f_{gas})$ where the constant $C$ depends both on the constant in the Schmidt law, and the quantity $M_{gas,initial}=M_{gas}+M_{\star}$.   We simply make use of the global Schmidt law \citep{Kennicutt1998a} to derive:
$$\Sigma_{SFR}  \propto  \Sigma_{gas}^{\alpha}  \propto  \bigg( \frac{M_{gas,SFR}}{\pi r_{SFR}^2} \bigg)^\alpha  \propto \bigg( \frac{f_{gas} \mu_{\star}}{1-f_{gas}} \bigg)^\alpha $$
where the gas fraction\footnote{We have adopted the notation $f_{gas}$ rather than $\mu$ to prevent confusion with $\mu_\star$. Some authors define gas fraction as $M_{HI}/M_\star$.} $f_{gas} = M_{gas}/(M_\star + M_{gas})$, and where we have assumed that $r_{SFR} \sim r_\star$ and that
all of the gas in the galaxy is involved in recent star formation.  For an assumed star formation law (e.g. $\alpha=1.4$), we can use $\Sigma_{SFR}$ and \logmu~ to estimate the gas fraction in a galaxy.  Adopting the star formation law from \citet{Kennicutt1998b} we have:
$$\log \Sigma_{SFR} = \log (\frac{2.5\times10^{-4}}{1.5})$$
$$ + 1.4(\log \mu_\star-6) + 1.4 \log (\frac{ f_{gas}}{1-f_{gas}}) $$
and 
$$\log \ssfr = \log \Sigma_{SFR} - \log \mu $$
$$= \log (\frac{2.5\times10^{-4}}{1.5}) + 0.4(\log \mu_\star-6) + 1.4 \log (\frac{ f_{gas}}{1-f_{gas}}) $$

Figure \ref{fig:munuvrall} shows where our estimated gas mass fraction isopleths fall.  The transition point which delineates star-forming from non-star-forming galaxies occurs at $f_{gas} \sim 0.1$.  Most star-forming galaxies have estimated gas mass fractions in the range $0.1<f_{gas}<0.5$.  Residual-SF and non-SF galaxies have $f_{gas}<0.1$.  These estimates suggest that a diminishing gas mass fraction is strongly correlated with a decrease in SF activity in galaxies.  We reiterate the  caveats that (1) f$_{gas}$ only includes gas that is traced by star formation, (2) a simple empirical relationship has been used to connect star formation to gas content,  and (3) we have employed the optical half-light radius to estimate the half-light radius of the star-forming disk.  The first caveat is likely to lead to an underestimate of total gas content of the galaxy, although the amount of gas involved in recent star formation (vs. the gas also being accreted) is itself a physically useful quantity.  The effect of the latter two caveats could cause our estimate to be off in either direction. Clearly, these new estimates would benefit from empirical constraints based on resolved and unresolved measurements of cold gas in galaxies. Work along these lines is in progress.

We show in Figure \ref{fig:logmu_nuvr_subsample} a similar plot, where we have split the population into disks and bulge-dominated galaxies.  This split emphasizes the transition described above; disk galaxies show very little variation in \ssfr~over a wide range in \logmu, and bulge-dominated galaxies show little variation in \logmu~over a wide range of \ssfr.   We see that most disk galaxies fall within a narrow range of \sfrkpc.  However, both disks and bulge-dominated galaxies have the highest \sfrkpc~near the transition \logmu.  These results are quantitatively illustrated in Figure \ref{fig:ssfr_logmu_cut} which divides the complete sample into bins of $\logmu$ in a similar manner as Figure \ref{fig:ssfr_mass_cut}.  On each plot, the horizontal axis corresponds to \ssfr, but at fixed \logmu~ also corresponds to an axis of \sfrkpc.    We indicate on each plot the average value  for the star forming sequence (log \sfrkpc $\sim$-1.7), which helps to identify those galaxies that are forming stars at an intensity higher than typical for their measured \Mstar.

\subsection{On the evolution of star-forming bulge-dominated galaxies off of the SF sequence}

Are star-forming bulge-dominated galaxies plausible candidates for galaxies that are soon to leave the environs of the star-forming sequence?  In a study of normal and peculiar galaxies (drawn from the Arp atlas), \citet{Larson1978} found that peculiar galaxies show an increased scatter in their color when compared with normal galaxies, providing evidence for recent bursts.   Similarly, we would expect mergers and interactions to trigger star formation activity, resulting in elevated SFRs and/or \sfrkpc.  Could heavily star-forming bulge-dominated galaxies be a short-lived phase in the evolution of a galaxy off of the SF sequence?    

Figure \ref{fig:morphevolution} ({\it left}) provides a schematic view of the SF sequence and possible evolutionary scenarios. Steady star formation can carry a galaxy along the SF sequence, while merging and quenching may push a galaxy off of the SF sequence.  The SF sequence itself and physical properties along the sequence are also expected to evolve with time.  While gaseous accretion, in some cases via merging, might move a galaxy stochastically through the residual-SF zone and occasionally back onto the SF sequence itself, a plausible track for major and minor mergers is one which takes a galaxy above and below the SF sequence.   These and other scenarios may lead to morphological transformations which might be reflected in the relationship between star formation and structure (Fig.  \ref{fig:morphevolution}, {\it right}).

The galaxy distribution function presented in Figure \ref{fig:ssfr_mass_cut} and Table \ref{table:ssfr_mass_cut}, combined with very simple assumptions regarding the evolution of star-forming bulge-dominated galaxies, can be used to quantify the rate at which galaxies might be leaving the SF sequence.  We can compare our estimates with other measurements to test whether or not these bulge-dominated galaxies might consititute a significant fraction of the population being quenched.    One strength of this approach is that it uses measurements of the galaxies with the highest SFR (and $L_{bol}$) a fact that could facilitate its application at higher redshift.  This approach does rely on accurately characterizing the distributions of galaxies along the SF sequence \citep[as has been done by ][]{Noeske2007a, Labbe2007}.

 Figure \ref{fig:ssfr_massint_cut} shows the integrated number of galaxies,
$$\Phi(>\ssfr,\Mstar) =  $$ 
 $$ \int_{\ssfr}^\infty  \phi(\ssfr,\Mstar) d(\ssfr) $$ 
in each stellar mass bin with \ssfr~greater than a specified value. In the limit $\ssfr \rightarrow 0$, $ \Phi(>\ssfr,\Mstar) \rightarrow \phi(M_\star)$ which shows good agreement with the local stellar mass function derived by \citet{Borch2006} \citep[taken from ][]{Bell2003b}. This function will be used to determine the total volume density of (e.g. bulge-dominated) galaxies of a particular \Mstar~forming stars at a rate higher than galaxies on the SF sequence.  

We then make three simple assumptions:
\begin{itemize}
\item[1.] {\it The end state of a galaxy leaving the SF sequence (``quenched galaxy'') is a bulge-dominated galaxy.}  As discussed above, passive disks are rare.
\item[2.] {\it All galaxies leaving the SF sequence  experience a brief period of elevated SFR for their given \Mstar.}  Models and observations suggest that most processes that lead to morphological transformation and quenching (interactions, mergers w/ possible feedback) experience periods of significantly enhanced star formation. Our method will not identify galaxies which do not experience this phase.
\item[3.] {\it All bulge-dominated galaxies with elevated SFR will be quenched.}  For our simple analysis we assume elevated SFR occurs at $\log \Delta_{SFR}>0.5$ and that $\tau_{burst} = 0.5$ Gyr.  We discuss the implications of this third assumption below.
\end{itemize}

As a check on our method we can compare the number rate density for the production of star-forming bulge-dominated galaxies to the merger rate by assuming that bursts result from mergers and that we have relative steady state over short timescales (e.g. $\dot{n}_{SF,\downarrow} \simeq \dot{n}_{merger}$).  We calculate the number rate density,
$$ \dot{n}_{SF,\downarrow} (>M_\star) $$
$$=  \frac{ \int_{\Mstar}^\infty \int_{\ssfr}^\infty  \phi(\ssfr,\Mstar) d(\ssfr)d(\Mstar)}   {\tau_{burst}}$$ 
and the bursting fraction:
$$ {Frac}_{SF,\downarrow} (>M_\star)$$
$$= \frac{ \int_{\Mstar}^\infty \int_{\ssfr}^\infty  \phi(\ssfr,\Mstar) d(\ssfr)d(\Mstar)}   { \int_{\Mstar}^\infty   \phi(\Mstar) d(\Mstar)}$$ 
which are shown in Figure \ref{fig:intflow} (in plotting both quantities we have multiplied by a factor of 2 in order to compare to merger rates and fractions in the literature, which are based on pre-merger galaxy counts).  Comparison merger rate densities are  taken from  the Millenium Survey \citep{De Propris2005}, SDSS LRGs \citep{Masjedi2006} and merger fractions are taken from \citet{De Propris2005} and \citet{Bell2006}.  For \citet{De Propris2005} and \citet{Masjedi2006}, we estimated a stellar mass limit based on their luminosity cut.  We also provide comparison points from a recent theoretical prediction \citep{Maller2006}.  We find reasonably good agreement with the numbers that result from our scenario.

 \begin{figure*}[t]
 \plotone{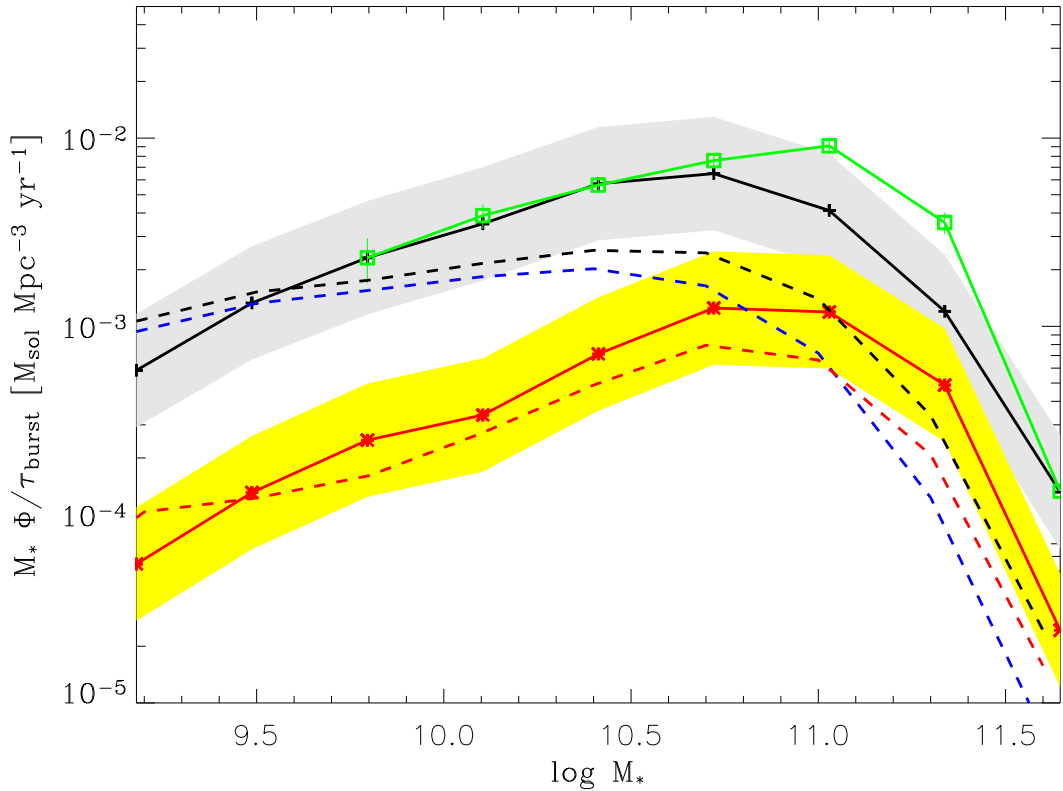} 
\caption{Estimated stellar mass flux density off of the SF sequence and comparison with the SFR density  vs. \Mstar.  {\it Solid lines:} Total stellar mass flux density for all galaxies ({\it black}) and bulge-dominated galaxies ({\it red}) with $\log \Delta SFR > 0.5$. Values are per 0.3 dex $M_\star$ bin.  {\it Green line:} Stellar mass flux rate for transition galaxies from \citetalias{Martin2007} (Table 3). {\it Dashed lines:} SFR density vs. \Mstar~for total ({\it black}), disk-dominated ({\it blue}), and bulge-dominated ({\it red}) subsamples.} 
\label{fig:massflow}
\end{figure*}

The total flux of stellar mass transitioning off of the SF sequence can be defined as
$$\dot{\rho}_{\star,\downarrow} (\Mstar)d\Mstar = \frac{\eta_{\downarrow}}{\eta_{burst}} \frac{\Mstar \Phi(>\Delta_{SFR},\Mstar) }{\tau_{burst}}d\Mstar$$ 
where $\eta_{\downarrow}$ denotes the fraction of galaxies that will not return to the SF sequence and $\eta_{burst}$ is a completeness factor that takes into account the flux from galaxies missed by this method (we adopt $\eta_\downarrow \sim 1$ and $\eta_{burst} \sim 1$). 

In Figure \ref{fig:massflow} we plot (for each \Mstar~ bin) the total derived stellar mass flux  
that results from bulge-dominated galaxies with enhanced star formation. Evidence suggests that the stellar mass function of the SF sequence has remained relatively constant with time \citep[][etc.]{Faber2007, Bell2007}  Therefore we would expect that star-forming galaxies should not be moving  from the SF sequence any faster than the rate at which new stars are being created.  Given the considerable assumptions, comparison of the two suggest good agreement, although the star-forming bulge-dominated stellar mass flux distribution is shifted to slightly higher \Mstar.       In addition, we compare these results with the stellar mass flux from the blue to red sequence (across the ``green valley'') determined in \citetalias{Martin2007}.   We find a considerably lower stellar mass flux, although the distributions agree quite well.   The offset between the stellar mass flux measurements coming from each side of the SF sequence could result from factors such as the time delay; ``green valley'' galaxies may have had their bursts several Gyr in the past, when they may have been more numerous. Also the differences between the two measurements may also provide some insight into the number of ``hidden'' (non-bursting or heavily obscured) quenched galaxies, or alternatively, the timescales over which transformations from disk to bulge-dominated might take place.  

An important difference between the present analysis and the one in \citetalias{Martin2007} is that the latter makes use of both the volume density and the timescales implied by a galaxy's position in the color-magnitude diagram and spectral indices D$_n(4000)$ and H$\delta_A$ whereas this work only   {\it assumes} the relevant timescale.   Our value of 0.5 Gyr is based on burst timescales calculated for mergers \citep[e.g.,][]{Hopkins2006}.  Some studies suggest that the period of elevated SFR might be shorter ($\sim$ 200 Myr) or longer, which would raise or lower our estimate.   It would not be hard to incorporate additional morphological (e.g. asymmetry, M20), dynamical (velocity dispersions, mass-dependent quantities), or recent star formation history information in order to generate an improved estimate of the relevant \Mstar-dependent timescale (as well as  $\eta_\downarrow, \eta_{burst}$); however this is beyond the scope of the present study.    

Both this work and \citetalias{Martin2007} demonstrate a number of new applications for the UV-optical color magnitude diagram and associated physical properties and distributions.  Ultimately, it is a combination of approaches that (1) compare the evolution of the \ssfr~ vs. \Mstar~distribution at different redshifts, and (2) use physically motivated timescales to predict the rate of change within a given time slice, which will lead to significant progress in modeling and measuring a complete history of star formation and morphological transformation.

\section{Conclusions}

We have generated a catalog of galaxies in the local (z$<$0.25) universe with a combination of UV-optical photometry, spectroscopic measures, structural parameters, and value-added and physical quantities, and have used it for an investigation into the distribution of star formation across galaxies of different morphologies and stellar masses.  Our chief results are as follows,

1.  We have derived a new set of physical properties of the galaxies in our samples, including star formation and stellar mass rates and surface densities, dust attenuations, and gas fractions.  Our measurements incorporate a slightly modified prescription for dust attenuation, designed to use the best available data to derive star formation rates across the whole galaxy sample.  In general, this follows most closely the approaches described in \citet{Johnson2007} and \citet{Kauffmann2003a} but ultimately we hope to develop it as a refinement over current methodology.

2.  For the first time we have measured the local UV luminosity function against galaxy structural parameters as well as inclination.  Among our key results is that we have shown that the fraction of intermediate and early-type galaxies is highest for the most UV luminous galaxies, dropping off to low fractions for the least luminous galaxies.

3.  Throughout this study our emphasis has been on the properties of galaxies on and off of a local ``star-forming sequence'' defined by $\log \ssfr = -0.36 \log M_\star - 6.4$.   We find, among other trends, that our measure of the star formation rate surface density, \sfrkpc~(measured within $r_{u,50}$) is nearly constant along this sequence.  

4.   We have split our sample into disk and bulge-dominated galaxies using the $i$-band Sersic index, and find that disk galaxies occupy a very tight locus in SFR vs. $M_\star$ space, while bulge-dominated galaxies display a much larger spread of SFRs at fixed stellar mass.  In particular, a significant fraction of galaxies with SFR and \sfrkpc~ above those on the ``star-forming sequence'' are bulge-dominated. 

5. We have used our derived distribution functions to ask whether a significant fraction of these galaxies may be experiencing a final episode of star formation (possibly induced by merger of other bursts), soon to be quenched, by determining whether this population can explain the growth rate of the non-star-forming population We find that this is a plausible scenario for bulge-dominated galaxies near the characteristic transition mass under reasonable assumptions regarding quenching timescales.  We use this technique to estimate the rate of mergers/starbursts that take galaxies off of the star-forming sequence and show that the implied merger rates are consistent with local measurements.


 \acknowledgments
 
D.S. gratefully acknowledges discussions with Eric Bell and Michael Blanton and the hospitality of the Max Planck Institut f$\ddot{u}$r Astronomie in Heidelberg and the Aspen Center for Physics.  This work has made extensive use of the \url[http://spectro.princeton.edu/]{{\tt idlutils}}, \url[]{{\tt kcorrect}} and \url[]{Goddard} IDL libraries as well as the \url[]{MPA/JHU} and the \url[]{NYU} SDSS value-added catalogs.   \url[]{\GALEX} ({\it Galaxy Evolution Explorer}) is a NASA Small Explorer, launched in April 2003. We gratefully acknowledge NASAÕs support for construction, operation, and science analysis for the \GALEX~ mission, developed in cooperation with the Centre National dÕEtudes Spatiales of France and the Korean Ministry of Science and Technology.

 

{\it Facilities:}  \facility{GALEX}. \facility{SDSS} 
 
 
\appendix
 
\section{DEPENDENCE ON VIEW ANGLE: INCLINATION, DUST ATTENUATION, AND GALAXY STRUCTURE}

In the absence of dust attenuation (scattering and absorption), the far-field integrated photometric properties of galaxies should have no angular dependence  [$F(\theta, \phi) = const$].   However even modest amounts of dust can have a considerable impact on the distribution of emitted flux.  A simple axisymmetric disk geometry for light and dust would result in axisymmetry for the light distribution [$F(\theta, \phi) = F(\phi)$].  In cases where galaxy axis ratios can be used to deduce the viewing angle, one can incorporate model assumptions to derive the emitted and intrinsic luminosity of a galaxy given a measurement along a single line of sight.  More complex dust geometries will naturally require more detailed modelling of the three-dimensional distribution of stars and dust and the resulting two-dimensional radiation field \citep[e.g.,][]{Jonsson2006}. We make no attempt here to consider the broad set of possible attenuation curves and dust geometries that might impact our measurements in the ultraviolet.  Instead, we choose to highlight  two results from our study and discuss possible implications.

\begin{figure*}[p]
\epsscale{1.0}
\plotone{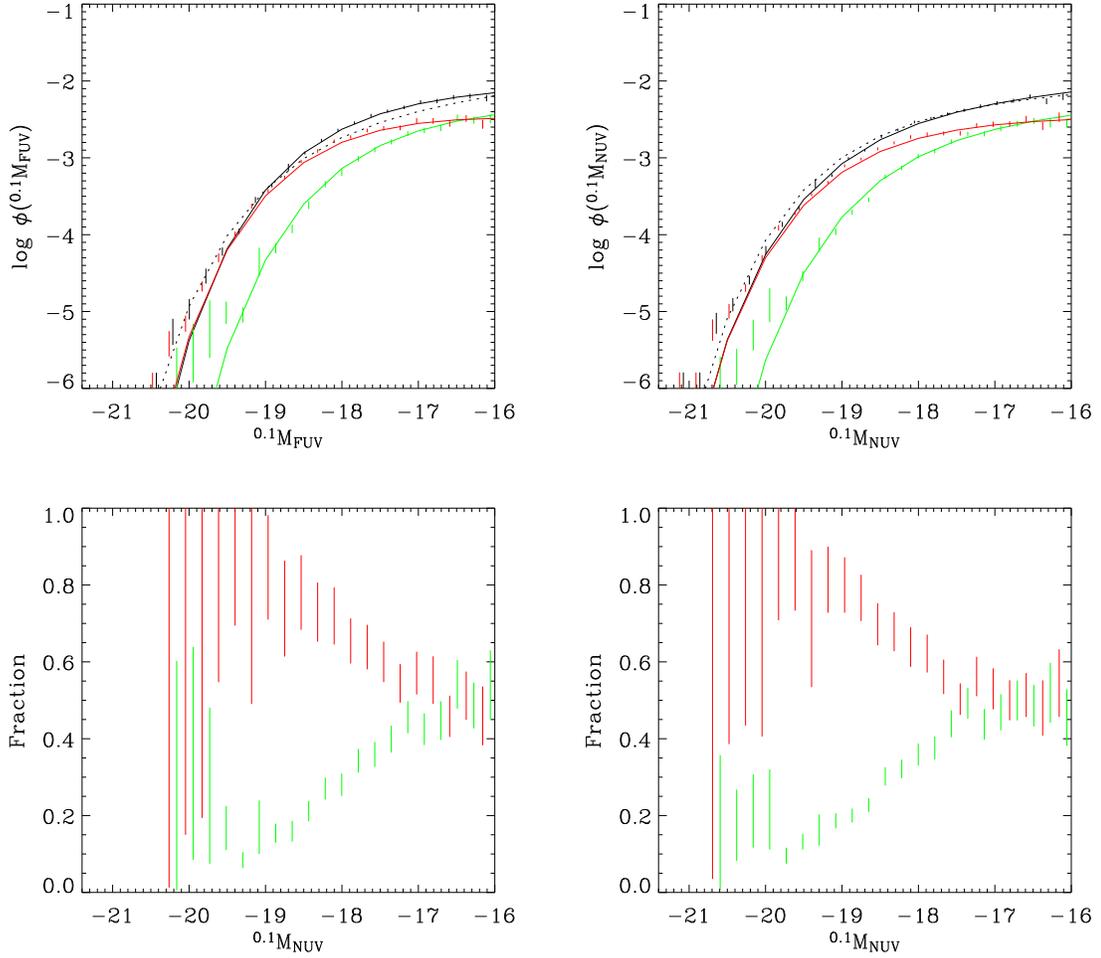}  
\caption{{\it Top}: FUV and NUV luminosity function for complete sample ($black$) and subsamples split by axis ratio:  $q_{25}=b_{25}/a_{25}<0.6$ ({\it green}) and $b_{25}/a_{25}> 0.6$ ({\it red}). Units of $\phi$ are in Mpc$^{-3}$ mag$^{-1}$. The dotted curve is from the \citet{Wyder2005} and the \citet{Treyer2005} LF.  $Bottom$: Relative fraction (1/V$_{max}$-weighted) of low and high axis ratio vs. total.} \label{fig:lf_mis_inclin} 

\end{figure*}

\begin{figure*}[p]
\epsscale{1.0}
\plotone{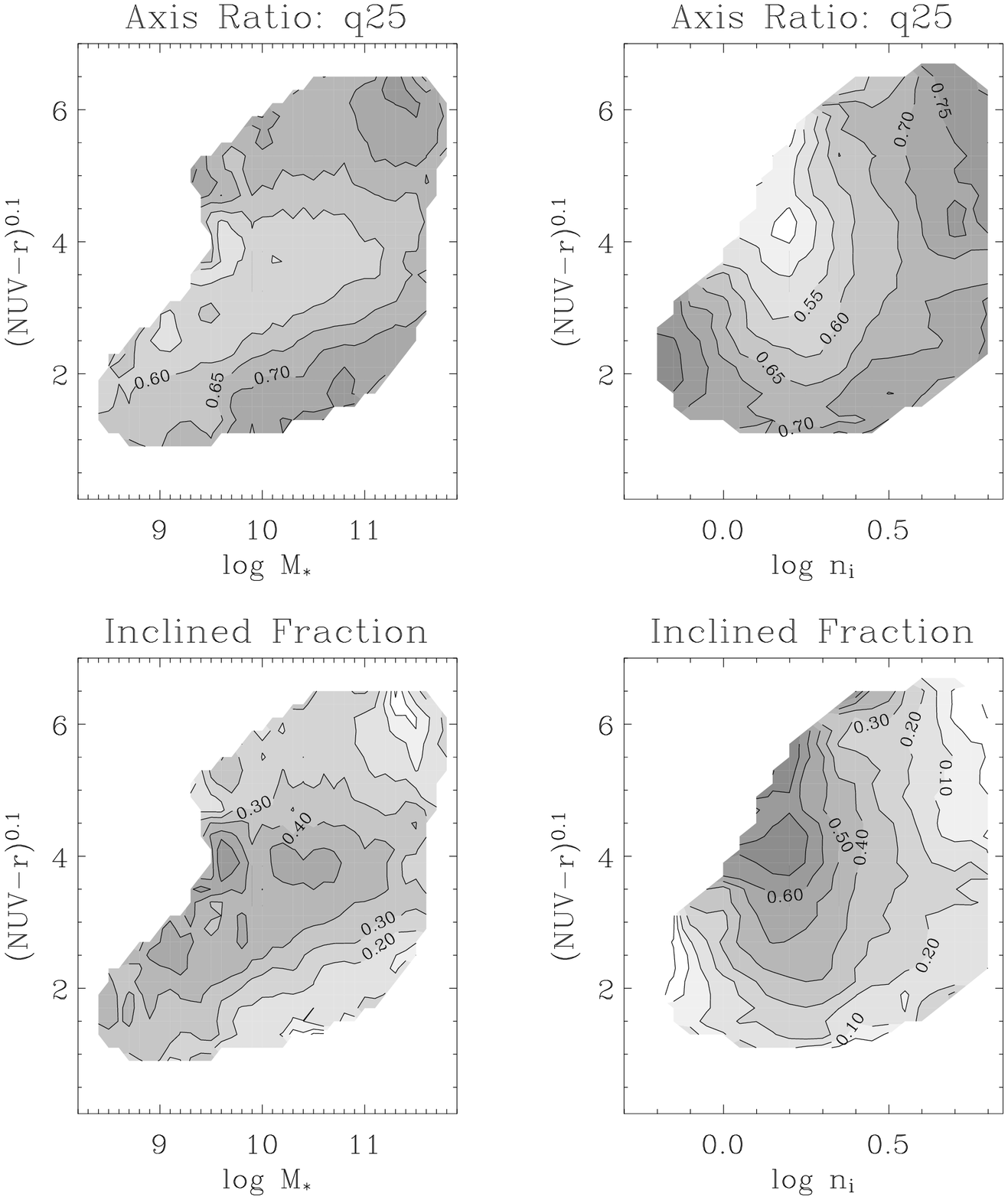}  
\caption{Weighted mean of axis ratio $q_{25}$ ($top$) and inclined fraction ($bottom$) in color vs. \Mstar ($left$) and color vs. log $n_i$ diagrams ($right$).} 
\label{fig:inclin_plots} 
\end{figure*}

In Figure \ref{fig:lf_mis_inclin} we show the luminosity function split by observed axis ratio.  We have split the sample at axis ratio $q_{25}=b_{25}/a_{25}=0.6$. Even though we have made no attempt to distinguish between disks and bulge-dominated galaxies, it is clear that the observed axis ratio has a considerable effect on the luminosity function.  While the shape of the LF is largely preserved, the low axis ratio (highly inclined) subset are 0.5 mag less luminous than the high axis ratio (low-inclination) subsample.  It is tempting to assume that a proper dust-attenuation correction should remove this discrepancy due to an apparent view angle effect.  However, this assumption is not valid in practice, because the distribution of axis ratios (observed or intrinsic) is known to correlate with galactic structure \citep{Binney1981}, and is not likely to be independent of intrinsic luminosity and/or other physical properties. 

Two aspects of this are demonstrated in Figure \ref{fig:inclin_plots}, where we show the mean axis ratio, $q_{25}$, and fraction of inclined galaxies with $q_{25}<0.6$ in $NUV-r$ vs. stellar mass and $NUV-r$ vs. Sersic index planes.  The plots on the left hand side shows that the distribution of inclinations is highly peaked in the observational ``green valley.''  A likely explanation is that most of these galaxies are SF-sequence galaxies which have a higher dust attenuation at our view angle and therefore are more highly reddened than their less inclined counterparts.  This would appear to be supported by Figure \ref{fig:mass_ssfr_q25}, which shows that this effect becomes less dramatic after application of a dust-attenuation correction.

The plots on the right of Figure \ref{fig:inclin_plots} show that the highest axis ratios are also found in those galaxies with intermediate Sersic indices, and that the distribution of axis ratios is clearly dependent on galaxy structure.  This may be due to the fact that at high (or low) Sersic index, galaxies tend to be bulge-dominated (or irregulars) with intrinsic axis ratios that differ from disks and disk/bulge composites with intermediate Sersic indexes.  A dust-attenuation correction may put most of the intermediate Sersic index galaxies on the SF sequence but it will not change the overall trend in axis ratio distribution in this plot.   This, combined with the fact that the dust-attenuation properties of galaxies are likely to vary with structure \citep{Pierini2004} suggests that conclusions drawn from the distribution of axis ratios alone should be treated with caution.  Since measures of inclination itself are strongly dependent on galaxy type, they should only be used in conjunction with a suitable structural quantity.

 
 


\vfill
\eject

\LongTables
{{\small

\begin{deluxetable}{l c c c c c c c c l}
\tablecaption{$\phi(\ssfr, \Mstar)$ for Total, Disk and Bulge-dominated subsamples}
\tablewidth{0pt}
\tablehead{
\colhead{log } & \colhead{log}&  &  \colhead{$\log \phi(\ssfr, \Mstar)\tablenotemark{a}$} & & $ \log $  & Fraction \\
\Mstar &  \ssfr &  Total & $n_i < 2.4$ & $n_i > 2.4$ & $\Delta$SFR & $n_i>2.4$\\
}
\startdata
   8.25 &   -9.72 &  $  -3.82 \pm    0.21$  &  $  -3.82 \pm    0.21$ & ... &   -0.35 & ... \\
   8.25 &   -9.53 &  $  -3.22 \pm    0.13$  &  $  -3.33 \pm    0.15$ & ... &   -0.16 & ... \\
   8.25 &   -9.34 &  $  -3.12 \pm    0.11$  &  $  -3.26 \pm    0.12$ & ... &    0.03 & ... \\
   8.25 &   -9.16 &  $  -3.16 \pm    0.11$  &  $  -3.20 \pm    0.12$ & ... &    0.22 & ... \\
   8.25 &   -8.97 &  $  -4.25 \pm    0.27$  &  $  -4.25 \pm    0.27$ & ... &    0.40 & ... \\
   8.25 &   -8.59 &  $  -4.90 \pm    0.30$  & ...& ... &    0.78 & ... \\
\hline
   8.56 &  -10.09 &  $  -4.26 \pm    0.30$  &  $  -4.26 \pm    0.30$ & ... &   -0.61 & ... \\
   8.56 &   -9.91 &  $  -4.00 \pm    0.22$  &  $  -4.00 \pm    0.21$ & ... &   -0.42 & ... \\
   8.56 &   -9.72 &  $  -3.25 \pm    0.10$  &  $  -3.32 \pm    0.11$ &  $  -4.06 \pm    0.22$  &   -0.24 &   0.15 \\
   8.56 &   -9.53 &  $  -2.86 \pm    0.06$  &  $  -2.91 \pm    0.06$ &  $  -4.12 \pm    0.17$  &   -0.05 &   0.06 \\
   8.56 &   -9.34 &  $  -2.98 \pm    0.07$  &  $  -2.99 \pm    0.07$ & ... &    0.14 & ... \\
   8.56 &   -9.16 &  $  -3.44 \pm    0.08$  &  $  -3.55 \pm    0.08$ &  $  -4.71 \pm    0.22$  &    0.33 &   0.05 \\
   8.56 &   -8.97 &  $  -4.26 \pm    0.15$  &  $  -4.26 \pm    0.15$ & ... &    0.51 & ... \\
   8.56 &   -8.78 &  $  -5.47 \pm    0.30$  &  $  -5.47 \pm    0.30$ & ... &    0.70 & ... \\
\hline
   8.87 &  -10.84 &  $  -3.70 \pm    0.17$  &  $  -3.81 \pm    0.18$ &  $  -4.38 \pm    0.30$  &   -1.25 &   0.21 \\
   8.87 &  -10.66 &  $  -4.29 \pm    0.30$  &  $  -4.29 \pm    0.30$ & ... &   -1.06 & ... \\
   8.87 &  -10.47 &  $  -4.00 \pm    0.17$  &  $  -4.43 \pm    0.23$ &  $  -4.20 \pm    0.21$  &   -0.88 &   0.63 \\
   8.87 &  -10.28 &  $  -4.00 \pm    0.16$  &  $  -4.00 \pm    0.16$ & ... &   -0.69 & ... \\
   8.87 &  -10.09 &  $  -4.12 \pm    0.15$  &  $  -4.19 \pm    0.16$ &  $  -4.90 \pm    0.30$  &   -0.50 &   0.17 \\
   8.87 &   -9.91 &  $  -3.43 \pm    0.09$  &  $  -3.51 \pm    0.09$ &  $  -4.22 \pm    0.23$  &   -0.31 &   0.16 \\
   8.87 &   -9.72 &  $  -3.13 \pm    0.05$  &  $  -3.17 \pm    0.05$ &  $  -4.76 \pm    0.20$  &   -0.13 &   0.02 \\
   8.87 &   -9.53 &  $  -3.04 \pm    0.04$  &  $  -3.10 \pm    0.04$ &  $  -4.35 \pm    0.15$  &    0.06 &   0.05 \\
   8.87 &   -9.34 &  $  -3.19 \pm    0.04$  &  $  -3.26 \pm    0.05$ &  $  -4.44 \pm    0.16$  &    0.25 &   0.06 \\
   8.87 &   -9.16 &  $  -3.85 \pm    0.07$  &  $  -3.87 \pm    0.07$ & ... &    0.44 & ... \\
   8.87 &   -8.97 &  $  -4.35 \pm    0.12$  &  $  -4.44 \pm    0.13$ & ... &    0.62 & ... \\
   8.87 &   -8.78 &  $  -4.62 \pm    0.28$  &  $  -4.62 \pm    0.27$ & ... &    0.81 & ... \\
\hline
   9.18 &  -11.59 &  $  -4.46 \pm    0.21$  &  $  -4.46 \pm    0.21$ & ... &   -1.89 & ... \\
   9.18 &  -11.41 &  $  -4.92 \pm    0.30$  & ...&  $  -4.92 \pm    0.30$  &   -1.70 &   1.00 \\
   9.18 &  -11.22 &  $  -4.37 \pm    0.17$  &  $  -4.61 \pm    0.20$ &  $  -4.74 \pm    0.19$  &   -1.51 &   0.43 \\
   9.18 &  -11.03 &  $  -4.21 \pm    0.16$  &  $  -4.32 \pm    0.19$ &  $  -4.86 \pm    0.25$  &   -1.33 &   0.22 \\
   9.18 &  -10.84 &  $  -4.40 \pm    0.18$  &  $  -4.58 \pm    0.20$ &  $  -4.88 \pm    0.30$  &   -1.14 &   0.34 \\
   9.18 &  -10.66 &  $  -3.93 \pm    0.12$  &  $  -4.08 \pm    0.15$ &  $  -4.47 \pm    0.19$  &   -0.95 &   0.29 \\
   9.18 &  -10.47 &  $  -4.05 \pm    0.15$  &  $  -4.14 \pm    0.16$ &  $  -4.78 \pm    0.30$  &   -0.76 &   0.18 \\
   9.18 &  -10.28 &  $  -3.83 \pm    0.18$  &  $  -4.19 \pm    0.14$ &  $  -4.08 \pm    0.27$  &   -0.58 &   0.56 \\
   9.18 &  -10.09 &  $  -3.91 \pm    0.10$  &  $  -3.94 \pm    0.11$ &  $  -5.10 \pm    0.18$  &   -0.39 &   0.07 \\
   9.18 &   -9.91 &  $  -3.38 \pm    0.05$  &  $  -3.47 \pm    0.06$ &  $  -4.25 \pm    0.19$  &   -0.20 &   0.13 \\
   9.18 &   -9.72 &  $  -3.21 \pm    0.04$  &  $  -3.25 \pm    0.04$ &  $  -4.29 \pm    0.12$  &   -0.01 &   0.08 \\
   9.18 &   -9.53 &  $  -3.12 \pm    0.03$  &  $  -3.16 \pm    0.03$ &  $  -4.29 \pm    0.11$  &    0.17 &   0.07 \\
   9.18 &   -9.34 &  $  -3.46 \pm    0.04$  &  $  -3.53 \pm    0.04$ &  $  -4.44 \pm    0.12$  &    0.36 &   0.10 \\
   9.18 &   -9.16 &  $  -4.10 \pm    0.07$  &  $  -4.15 \pm    0.07$ &  $  -5.15 \pm    0.18$  &    0.55 &   0.09 \\
   9.18 &   -8.97 &  $  -4.69 \pm    0.15$  &  $  -4.69 \pm    0.15$ & ... &    0.74 & ... \\
   9.18 &   -8.78 &  $  -5.80 \pm    0.30$  & ...&  $  -5.80 \pm    0.30$  &    0.92 &   1.00 \\
   9.18 &   -8.59 &  $  -6.14 \pm    0.30$  &  $  -6.14 \pm    0.30$ & ... &    1.11 & ... \\
\hline
   9.49 &  -11.78 &  $  -4.71 \pm    0.20$  &  $  -5.11 \pm    0.30$ &  $  -4.93 \pm    0.22$  &   -1.97 &   0.60 \\
   9.49 &  -11.59 &  $  -4.33 \pm    0.13$  &  $  -4.53 \pm    0.16$ &  $  -4.76 \pm    0.17$  &   -1.78 &   0.37 \\
   9.49 &  -11.41 &  $  -4.22 \pm    0.11$  &  $  -4.52 \pm    0.16$ &  $  -4.53 \pm    0.13$  &   -1.59 &   0.50 \\
   9.49 &  -11.22 &  $  -4.33 \pm    0.13$  &  $  -4.89 \pm    0.21$ &  $  -4.47 \pm    0.15$  &   -1.40 &   0.72 \\
   9.49 &  -11.03 &  $  -4.31 \pm    0.12$  &  $  -4.77 \pm    0.18$ &  $  -4.49 \pm    0.15$  &   -1.22 &   0.66 \\
   9.49 &  -10.84 &  $  -4.35 \pm    0.14$  &  $  -4.55 \pm    0.17$ &  $  -4.79 \pm    0.20$  &   -1.03 &   0.36 \\
   9.49 &  -10.66 &  $  -4.07 \pm    0.10$  &  $  -4.16 \pm    0.11$ &  $  -4.78 \pm    0.16$  &   -0.84 &   0.19 \\
   9.49 &  -10.47 &  $  -4.20 \pm    0.11$  &  $  -4.28 \pm    0.13$ &  $  -4.98 \pm    0.22$  &   -0.65 &   0.17 \\
   9.49 &  -10.28 &  $  -4.12 \pm    0.09$  &  $  -4.24 \pm    0.11$ &  $  -4.73 \pm    0.15$  &   -0.47 &   0.25 \\
   9.49 &  -10.09 &  $  -3.72 \pm    0.06$  &  $  -3.77 \pm    0.06$ &  $  -4.78 \pm    0.14$  &   -0.28 &   0.09 \\
   9.49 &   -9.91 &  $  -3.42 \pm    0.04$  &  $  -3.48 \pm    0.04$ &  $  -4.51 \pm    0.12$  &   -0.09 &   0.08 \\
   9.49 &   -9.72 &  $  -3.25 \pm    0.03$  &  $  -3.28 \pm    0.03$ &  $  -4.49 \pm    0.13$  &    0.10 &   0.06 \\
   9.49 &   -9.53 &  $  -3.28 \pm    0.03$  &  $  -3.32 \pm    0.03$ &  $  -4.46 \pm    0.09$  &    0.28 &   0.07 \\
   9.49 &   -9.34 &  $  -3.69 \pm    0.04$  &  $  -3.76 \pm    0.04$ &  $  -4.71 \pm    0.12$  &    0.47 &   0.10 \\
   9.49 &   -9.16 &  $  -4.39 \pm    0.13$  &  $  -4.69 \pm    0.11$ &  $  -5.44 \pm    0.25$  &    0.66 &   0.09 \\
   9.49 &   -8.97 &  $  -6.01 \pm    0.21$  &  $  -6.36 \pm    0.30$ &  $  -6.26 \pm    0.30$  &    0.85 &   0.56 \\
   9.49 &   -8.78 &  $  -5.93 \pm    0.18$  &  $  -6.24 \pm    0.30$ &  $  -6.22 \pm    0.30$  &    1.03 &   0.51 \\
\hline
   9.80 &  -12.34 &  $  -5.80 \pm    0.30$  & ...&  $  -5.80 \pm    0.30$  &   -2.42 &   1.00 \\
   9.80 &  -11.97 &  $  -5.56 \pm    0.20$  & ...&  $  -5.56 \pm    0.20$  &   -2.04 &   1.00 \\
   9.80 &  -11.78 &  $  -4.42 \pm    0.10$  &  $  -4.83 \pm    0.16$ &  $  -4.63 \pm    0.11$  &   -1.85 &   0.62 \\
   9.80 &  -11.59 &  $  -4.22 \pm    0.08$  &  $  -4.64 \pm    0.13$ &  $  -4.46 \pm    0.09$  &   -1.67 &   0.57 \\
   9.80 &  -11.41 &  $  -4.32 \pm    0.09$  &  $  -4.85 \pm    0.16$ &  $  -4.51 \pm    0.10$  &   -1.48 &   0.63 \\
   9.80 &  -11.22 &  $  -4.29 \pm    0.09$  &  $  -4.80 \pm    0.14$ &  $  -4.44 \pm    0.10$  &   -1.29 &   0.69 \\
   9.80 &  -11.03 &  $  -4.30 \pm    0.08$  &  $  -4.53 \pm    0.11$ &  $  -4.68 \pm    0.12$  &   -1.10 &   0.41 \\
   9.80 &  -10.84 &  $  -4.21 \pm    0.08$  &  $  -4.47 \pm    0.12$ &  $  -4.56 \pm    0.12$  &   -0.92 &   0.45 \\
   9.80 &  -10.66 &  $  -4.38 \pm    0.11$  &  $  -4.49 \pm    0.12$ &  $  -5.05 \pm    0.18$  &   -0.73 &   0.21 \\
   9.80 &  -10.47 &  $  -4.04 \pm    0.07$  &  $  -4.13 \pm    0.08$ &  $  -4.73 \pm    0.12$  &   -0.54 &   0.20 \\
   9.80 &  -10.28 &  $  -3.88 \pm    0.07$  &  $  -3.97 \pm    0.08$ &  $  -4.66 \pm    0.11$  &   -0.35 &   0.17 \\
   9.80 &  -10.09 &  $  -3.63 \pm    0.04$  &  $  -3.70 \pm    0.04$ &  $  -4.59 \pm    0.10$  &   -0.17 &   0.11 \\
   9.80 &   -9.91 &  $  -3.38 \pm    0.02$  &  $  -3.43 \pm    0.03$ &  $  -4.47 \pm    0.10$  &    0.02 &   0.08 \\
   9.80 &   -9.72 &  $  -3.41 \pm    0.02$  &  $  -3.45 \pm    0.02$ &  $  -4.69 \pm    0.09$  &    0.21 &   0.05 \\
   9.80 &   -9.53 &  $  -3.64 \pm    0.03$  &  $  -3.69 \pm    0.03$ &  $  -4.70 \pm    0.08$  &    0.40 &   0.09 \\
   9.80 &   -9.34 &  $  -4.18 \pm    0.04$  &  $  -4.25 \pm    0.04$ &  $  -5.05 \pm    0.11$  &    0.58 &   0.13 \\
   9.80 &   -9.16 &  $  -4.80 \pm    0.07$  &  $  -4.86 \pm    0.08$ &  $  -5.66 \pm    0.16$  &    0.77 &   0.14 \\
   9.80 &   -8.97 &  $  -5.83 \pm    0.18$  &  $  -5.83 \pm    0.18$ & ... &    0.96 & ... \\
   9.80 &   -8.78 &  $  -6.58 \pm    0.30$  &  $  -6.58 \pm    0.30$ & ... &    1.15 & ... \\
\hline
  10.10 &  -12.16 &  $  -5.04 \pm    0.17$  & ...&  $  -5.04 \pm    0.16$  &   -2.12 &   1.00 \\
  10.10 &  -11.97 &  $  -4.32 \pm    0.06$  &  $  -5.00 \pm    0.16$ &  $  -4.42 \pm    0.07$  &   -1.93 &   0.79 \\
  10.10 &  -11.78 &  $  -4.03 \pm    0.05$  &  $  -5.02 \pm    0.14$ &  $  -4.08 \pm    0.05$  &   -1.74 &   0.90 \\
  10.10 &  -11.59 &  $  -3.95 \pm    0.04$  &  $  -4.94 \pm    0.12$ &  $  -4.01 \pm    0.04$  &   -1.56 &   0.88 \\
  10.10 &  -11.41 &  $  -4.13 \pm    0.05$  &  $  -4.72 \pm    0.09$ &  $  -4.26 \pm    0.06$  &   -1.37 &   0.74 \\
  10.10 &  -11.22 &  $  -4.38 \pm    0.06$  &  $  -5.05 \pm    0.13$ &  $  -4.49 \pm    0.07$  &   -1.18 &   0.77 \\
  10.10 &  -11.03 &  $  -4.27 \pm    0.05$  &  $  -4.74 \pm    0.10$ &  $  -4.47 \pm    0.07$  &   -0.99 &   0.64 \\
  10.10 &  -10.84 &  $  -4.28 \pm    0.06$  &  $  -4.58 \pm    0.09$ &  $  -4.58 \pm    0.07$  &   -0.81 &   0.50 \\
  10.10 &  -10.66 &  $  -4.12 \pm    0.05$  &  $  -4.35 \pm    0.07$ &  $  -4.51 \pm    0.07$  &   -0.62 &   0.41 \\
  10.10 &  -10.47 &  $  -3.91 \pm    0.04$  &  $  -4.07 \pm    0.05$ &  $  -4.42 \pm    0.07$  &   -0.43 &   0.31 \\
  10.10 &  -10.28 &  $  -3.77 \pm    0.03$  &  $  -3.88 \pm    0.04$ &  $  -4.42 \pm    0.07$  &   -0.24 &   0.22 \\
  10.10 &  -10.09 &  $  -3.59 \pm    0.02$  &  $  -3.66 \pm    0.03$ &  $  -4.47 \pm    0.06$  &   -0.06 &   0.13 \\
  10.10 &   -9.91 &  $  -3.48 \pm    0.02$  &  $  -3.54 \pm    0.02$ &  $  -4.49 \pm    0.08$  &    0.13 &   0.10 \\
  10.10 &   -9.72 &  $  -3.68 \pm    0.02$  &  $  -3.74 \pm    0.02$ &  $  -4.71 \pm    0.07$  &    0.32 &   0.09 \\
  10.10 &   -9.53 &  $  -4.01 \pm    0.03$  &  $  -4.06 \pm    0.03$ &  $  -5.08 \pm    0.09$  &    0.51 &   0.09 \\
  10.10 &   -9.34 &  $  -4.58 \pm    0.05$  &  $  -4.63 \pm    0.05$ &  $  -5.59 \pm    0.11$  &    0.69 &   0.10 \\
  10.10 &   -9.16 &  $  -5.41 \pm    0.09$  &  $  -5.57 \pm    0.10$ &  $  -5.93 \pm    0.16$  &    0.88 &   0.30 \\
  10.10 &   -8.97 &  $  -6.06 \pm    0.18$  &  $  -6.25 \pm    0.21$ &  $  -6.52 \pm    0.20$  &    1.07 &   0.35 \\
  10.10 &   -8.78 &  $  -6.86 \pm    0.30$  &  $  -6.86 \pm    0.30$ & ... &    1.26 & ... \\
  10.10 &   -8.59 &  $  -5.86 \pm    0.23$  &  $  -5.93 \pm    0.25$ &  $  -6.68 \pm    0.30$  &    1.44 &   0.15 \\
\hline
  10.41 &  -12.53 &  $  -6.02 \pm    0.30$  & ...&  $  -6.02 \pm    0.30$  &   -2.38 &   1.00 \\
  10.41 &  -12.34 &  $  -5.56 \pm    0.20$  & ...&  $  -5.56 \pm    0.22$  &   -2.20 &   1.00 \\
  10.41 &  -12.16 &  $  -4.49 \pm    0.07$  &  $  -5.42 \pm    0.21$ &  $  -4.54 \pm    0.07$  &   -2.01 &   0.88 \\
  10.41 &  -11.97 &  $  -3.99 \pm    0.03$  &  $  -5.76 \pm    0.21$ &  $  -4.00 \pm    0.03$  &   -1.82 &   0.98 \\
  10.41 &  -11.78 &  $  -4.01 \pm    0.03$  &  $  -5.44 \pm    0.17$ &  $  -4.03 \pm    0.03$  &   -1.63 &   0.95 \\
  10.41 &  -11.59 &  $  -4.07 \pm    0.03$  &  $  -4.99 \pm    0.09$ &  $  -4.13 \pm    0.03$  &   -1.45 &   0.87 \\
  10.41 &  -11.41 &  $  -4.12 \pm    0.03$  &  $  -4.87 \pm    0.08$ &  $  -4.21 \pm    0.03$  &   -1.26 &   0.82 \\
  10.41 &  -11.22 &  $  -4.22 \pm    0.04$  &  $  -5.01 \pm    0.11$ &  $  -4.30 \pm    0.04$  &   -1.07 &   0.83 \\
  10.41 &  -11.03 &  $  -4.17 \pm    0.04$  &  $  -4.66 \pm    0.07$ &  $  -4.33 \pm    0.05$  &   -0.88 &   0.68 \\
  10.41 &  -10.84 &  $  -4.10 \pm    0.04$  &  $  -4.45 \pm    0.06$ &  $  -4.36 \pm    0.05$  &   -0.70 &   0.55 \\
  10.41 &  -10.66 &  $  -4.05 \pm    0.04$  &  $  -4.34 \pm    0.05$ &  $  -4.37 \pm    0.04$  &   -0.51 &   0.47 \\
  10.41 &  -10.47 &  $  -3.83 \pm    0.03$  &  $  -4.04 \pm    0.03$ &  $  -4.25 \pm    0.04$  &   -0.32 &   0.38 \\
  10.41 &  -10.28 &  $  -3.67 \pm    0.02$  &  $  -3.79 \pm    0.02$ &  $  -4.30 \pm    0.04$  &   -0.13 &   0.23 \\
  10.41 &  -10.09 &  $  -3.63 \pm    0.02$  &  $  -3.72 \pm    0.02$ &  $  -4.36 \pm    0.04$  &    0.05 &   0.18 \\
  10.41 &   -9.91 &  $  -3.70 \pm    0.02$  &  $  -3.78 \pm    0.02$ &  $  -4.51 \pm    0.05$  &    0.24 &   0.16 \\
  10.41 &   -9.72 &  $  -4.01 \pm    0.02$  &  $  -4.07 \pm    0.03$ &  $  -4.97 \pm    0.06$  &    0.43 &   0.11 \\
  10.41 &   -9.53 &  $  -4.38 \pm    0.03$  &  $  -4.45 \pm    0.03$ &  $  -5.24 \pm    0.08$  &    0.62 &   0.14 \\
  10.41 &   -9.34 &  $  -5.19 \pm    0.06$  &  $  -5.27 \pm    0.07$ &  $  -5.95 \pm    0.15$  &    0.80 &   0.17 \\
  10.41 &   -9.16 &  $  -6.09 \pm    0.14$  &  $  -6.16 \pm    0.15$ &  $  -6.91 \pm    0.30$  &    0.99 &   0.15 \\
  10.41 &   -8.97 &  $  -6.95 \pm    0.30$  &  $  -6.95 \pm    0.30$ & ... &    1.18 & ... \\
  10.41 &   -8.78 &  $  -6.69 \pm    0.23$  &  $  -6.90 \pm    0.30$ &  $  -7.11 \pm    0.30$  &    1.37 &   0.38 \\
\hline
  10.72 &  -12.53 &  $  -5.85 \pm    0.21$  & ...&  $  -5.85 \pm    0.19$  &   -2.27 &   1.00 \\
  10.72 &  -12.34 &  $  -4.93 \pm    0.08$  & ...&  $  -4.93 \pm    0.08$  &   -2.08 &   1.00 \\
  10.72 &  -12.16 &  $  -4.27 \pm    0.04$  &  $  -6.18 \pm    0.22$ &  $  -4.27 \pm    0.03$  &   -1.90 &   0.98 \\
  10.72 &  -11.97 &  $  -4.04 \pm    0.03$  &  $  -5.53 \pm    0.13$ &  $  -4.06 \pm    0.03$  &   -1.71 &   0.96 \\
  10.72 &  -11.78 &  $  -4.07 \pm    0.02$  &  $  -5.56 \pm    0.14$ &  $  -4.09 \pm    0.02$  &   -1.52 &   0.96 \\
  10.72 &  -11.59 &  $  -4.14 \pm    0.03$  &  $  -5.18 \pm    0.10$ &  $  -4.18 \pm    0.03$  &   -1.33 &   0.90 \\
  10.72 &  -11.41 &  $  -4.21 \pm    0.03$  &  $  -5.14 \pm    0.11$ &  $  -4.27 \pm    0.03$  &   -1.15 &   0.88 \\
  10.72 &  -11.22 &  $  -4.23 \pm    0.03$  &  $  -5.00 \pm    0.07$ &  $  -4.31 \pm    0.03$  &   -0.96 &   0.82 \\
  10.72 &  -11.03 &  $  -4.18 \pm    0.03$  &  $  -4.80 \pm    0.06$ &  $  -4.30 \pm    0.03$  &   -0.77 &   0.76 \\
  10.72 &  -10.84 &  $  -4.15 \pm    0.03$  &  $  -4.66 \pm    0.05$ &  $  -4.32 \pm    0.03$  &   -0.58 &   0.68 \\
  10.72 &  -10.66 &  $  -4.04 \pm    0.02$  &  $  -4.39 \pm    0.04$ &  $  -4.32 \pm    0.03$  &   -0.40 &   0.53 \\
  10.72 &  -10.47 &  $  -3.89 \pm    0.02$  &  $  -4.17 \pm    0.03$ &  $  -4.21 \pm    0.02$  &   -0.21 &   0.47 \\
  10.72 &  -10.28 &  $  -3.83 \pm    0.02$  &  $  -4.05 \pm    0.02$ &  $  -4.24 \pm    0.03$  &   -0.02 &   0.39 \\
  10.72 &  -10.09 &  $  -3.82 \pm    0.02$  &  $  -3.99 \pm    0.02$ &  $  -4.33 \pm    0.03$  &    0.17 &   0.31 \\
  10.72 &   -9.91 &  $  -4.09 \pm    0.02$  &  $  -4.20 \pm    0.02$ &  $  -4.75 \pm    0.04$  &    0.35 &   0.22 \\
  10.72 &   -9.72 &  $  -4.52 \pm    0.03$  &  $  -4.62 \pm    0.03$ &  $  -5.21 \pm    0.06$  &    0.54 &   0.20 \\
  10.72 &   -9.53 &  $  -4.95 \pm    0.04$  &  $  -5.02 \pm    0.04$ &  $  -5.81 \pm    0.11$  &    0.73 &   0.14 \\
  10.72 &   -9.34 &  $  -5.70 \pm    0.09$  &  $  -5.81 \pm    0.10$ &  $  -6.50 \pm    0.18$  &    0.92 &   0.16 \\
  10.72 &   -9.16 &  $  -7.12 \pm    0.30$  &  $  -7.12 \pm    0.30$ & ... &    1.10 & ... \\
\hline
  11.03 &  -12.53 &  $  -6.04 \pm    0.22$  & ...&  $  -6.04 \pm    0.21$  &   -2.16 &   1.00 \\
  11.03 &  -12.34 &  $  -4.82 \pm    0.06$  & ...&  $  -4.84 \pm    0.06$  &   -1.97 &   0.96 \\
  11.03 &  -12.16 &  $  -4.34 \pm    0.03$  & ...&  $  -4.35 \pm    0.03$  &   -1.79 &   0.99 \\
  11.03 &  -11.97 &  $  -4.24 \pm    0.02$  &  $  -6.04 \pm    0.19$ &  $  -4.25 \pm    0.02$  &   -1.60 &   0.98 \\
  11.03 &  -11.78 &  $  -4.28 \pm    0.02$  &  $  -5.94 \pm    0.15$ &  $  -4.29 \pm    0.02$  &   -1.41 &   0.97 \\
  11.03 &  -11.59 &  $  -4.35 \pm    0.02$  &  $  -5.87 \pm    0.11$ &  $  -4.37 \pm    0.02$  &   -1.22 &   0.96 \\
  11.03 &  -11.41 &  $  -4.46 \pm    0.02$  &  $  -5.63 \pm    0.10$ &  $  -4.49 \pm    0.03$  &   -1.04 &   0.92 \\
  11.03 &  -11.22 &  $  -4.45 \pm    0.02$  &  $  -5.33 \pm    0.08$ &  $  -4.52 \pm    0.03$  &   -0.85 &   0.86 \\
  11.03 &  -11.03 &  $  -4.46 \pm    0.02$  &  $  -5.33 \pm    0.08$ &  $  -4.53 \pm    0.02$  &   -0.66 &   0.86 \\
  11.03 &  -10.84 &  $  -4.38 \pm    0.02$  &  $  -4.99 \pm    0.06$ &  $  -4.50 \pm    0.03$  &   -0.47 &   0.75 \\
  11.03 &  -10.66 &  $  -4.32 \pm    0.02$  &  $  -4.85 \pm    0.04$ &  $  -4.47 \pm    0.02$  &   -0.29 &   0.70 \\
  11.03 &  -10.47 &  $  -4.24 \pm    0.02$  &  $  -4.62 \pm    0.03$ &  $  -4.47 \pm    0.03$  &   -0.10 &   0.59 \\
  11.03 &  -10.28 &  $  -4.25 \pm    0.02$  &  $  -4.51 \pm    0.02$ &  $  -4.62 \pm    0.03$  &    0.09 &   0.43 \\
  11.03 &  -10.09 &  $  -4.40 \pm    0.02$  &  $  -4.64 \pm    0.03$ &  $  -4.79 \pm    0.03$  &    0.28 &   0.40 \\
  11.03 &   -9.91 &  $  -4.79 \pm    0.03$  &  $  -4.96 \pm    0.04$ &  $  -5.30 \pm    0.05$  &    0.46 &   0.31 \\
  11.03 &   -9.72 &  $  -5.36 \pm    0.05$  &  $  -5.50 \pm    0.07$ &  $  -5.95 \pm    0.09$  &    0.65 &   0.26 \\
  11.03 &   -9.53 &  $  -5.90 \pm    0.08$  &  $  -6.00 \pm    0.10$ &  $  -6.58 \pm    0.16$  &    0.84 &   0.21 \\
  11.03 &   -9.34 &  $  -6.31 \pm    0.15$  &  $  -6.42 \pm    0.16$ &  $  -6.99 \pm    0.30$  &    1.03 &   0.21 \\
  11.03 &   -9.16 &  $  -7.39 \pm    0.30$  &  $  -7.39 \pm    0.30$ & ... &    1.21 & ... \\
\hline
  11.34 &  -12.72 &  $  -6.95 \pm    0.30$  & ...&  $  -6.95 \pm    0.30$  &   -2.24 &   1.00 \\
  11.34 &  -12.53 &  $  -5.85 \pm    0.13$  & ...&  $  -5.85 \pm    0.12$  &   -2.05 &   1.00 \\
  11.34 &  -12.34 &  $  -5.08 \pm    0.05$  & ...&  $  -5.08 \pm    0.05$  &   -1.86 &   1.00 \\
  11.34 &  -12.16 &  $  -4.74 \pm    0.03$  & ...&  $  -4.75 \pm    0.03$  &   -1.67 &   0.97 \\
  11.34 &  -11.97 &  $  -4.68 \pm    0.03$  &  $  -6.48 \pm    0.19$ &  $  -4.69 \pm    0.03$  &   -1.49 &   0.97 \\
  11.34 &  -11.78 &  $  -4.82 \pm    0.03$  &  $  -6.57 \pm    0.21$ &  $  -4.83 \pm    0.03$  &   -1.30 &   0.98 \\
  11.34 &  -11.59 &  $  -4.98 \pm    0.03$  &  $  -6.31 \pm    0.15$ &  $  -5.00 \pm    0.03$  &   -1.11 &   0.95 \\
  11.34 &  -11.41 &  $  -5.00 \pm    0.03$  &  $  -6.12 \pm    0.10$ &  $  -5.03 \pm    0.04$  &   -0.92 &   0.92 \\
  11.34 &  -11.22 &  $  -4.95 \pm    0.03$  &  $  -5.89 \pm    0.10$ &  $  -5.01 \pm    0.04$  &   -0.74 &   0.88 \\
  11.34 &  -11.03 &  $  -5.15 \pm    0.04$  &  $  -6.07 \pm    0.11$ &  $  -5.21 \pm    0.04$  &   -0.55 &   0.86 \\
  11.34 &  -10.84 &  $  -5.08 \pm    0.03$  &  $  -5.76 \pm    0.08$ &  $  -5.18 \pm    0.04$  &   -0.36 &   0.79 \\
  11.34 &  -10.66 &  $  -5.00 \pm    0.04$  &  $  -5.56 \pm    0.08$ &  $  -5.15 \pm    0.04$  &   -0.17 &   0.72 \\
  11.34 &  -10.47 &  $  -5.05 \pm    0.04$  &  $  -5.52 \pm    0.05$ &  $  -5.24 \pm    0.04$  &    0.01 &   0.65 \\
  11.34 &  -10.28 &  $  -5.16 \pm    0.04$  &  $  -5.53 \pm    0.06$ &  $  -5.43 \pm    0.05$  &    0.20 &   0.54 \\
  11.34 &  -10.09 &  $  -5.56 \pm    0.06$  &  $  -5.77 \pm    0.07$ &  $  -5.98 \pm    0.08$  &    0.39 &   0.38 \\
  11.34 &   -9.91 &  $  -5.90 \pm    0.08$  &  $  -6.14 \pm    0.10$ &  $  -6.29 \pm    0.14$  &    0.58 &   0.40 \\
  11.34 &   -9.72 &  $  -6.50 \pm    0.13$  &  $  -6.86 \pm    0.18$ &  $  -6.75 \pm    0.16$  &    0.76 &   0.56 \\
  11.34 &   -9.53 &  $  -7.30 \pm    0.18$  &  $  -7.30 \pm    0.18$ & ... &    0.95 & ... \\
\hline
  11.65 &  -12.72 &  $  -7.25 \pm    0.30$  & ...&  $  -7.25 \pm    0.30$  &   -2.13 &   1.00 \\
  11.65 &  -12.53 &  $  -6.58 \pm    0.19$  & ...&  $  -6.58 \pm    0.18$  &   -1.94 &   1.00 \\
  11.65 &  -12.34 &  $  -5.73 \pm    0.07$  & ...&  $  -5.73 \pm    0.07$  &   -1.75 &   1.00 \\
  11.65 &  -12.16 &  $  -5.54 \pm    0.05$  & ...&  $  -5.54 \pm    0.05$  &   -1.56 &   1.00 \\
  11.65 &  -11.97 &  $  -5.70 \pm    0.05$  &  $  -7.29 \pm    0.18$ &  $  -5.71 \pm    0.05$  &   -1.38 &   0.96 \\
  11.65 &  -11.78 &  $  -5.87 \pm    0.07$  &  $  -7.13 \pm    0.20$ &  $  -5.89 \pm    0.07$  &   -1.19 &   0.95 \\
  11.65 &  -11.59 &  $  -6.16 \pm    0.10$  & ...&  $  -6.16 \pm    0.10$  &   -1.00 &   1.00 \\
  11.65 &  -11.41 &  $  -6.23 \pm    0.10$  &  $  -7.52 \pm    0.30$ &  $  -6.25 \pm    0.10$  &   -0.81 &   0.95 \\
  11.65 &  -11.22 &  $  -6.03 \pm    0.09$  &  $  -6.43 \pm    0.17$ &  $  -6.26 \pm    0.10$  &   -0.63 &   0.60 \\
  11.65 &  -11.03 &  $  -6.20 \pm    0.09$  &  $  -7.59 \pm    0.30$ &  $  -6.23 \pm    0.09$  &   -0.44 &   0.92 \\
  11.65 &  -10.84 &  $  -6.48 \pm    0.13$  &  $  -7.15 \pm    0.30$ &  $  -6.58 \pm    0.12$  &   -0.25 &   0.79 \\
  11.65 &  -10.66 &  $  -6.39 \pm    0.11$  &  $  -6.82 \pm    0.18$ &  $  -6.63 \pm    0.14$  &   -0.06 &   0.57 \\
  11.65 &  -10.47 &  $  -6.61 \pm    0.13$  &  $  -7.04 \pm    0.21$ &  $  -6.81 \pm    0.16$  &    0.12 &   0.63 \\
  11.65 &  -10.28 &  $  -7.16 \pm    0.23$  & ...&  $  -7.16 \pm    0.21$  &    0.31 &   1.00 \\
  11.65 &  -10.09 &  $  -6.92 \pm    0.19$  &  $  -7.29 \pm    0.18$ &  $  -7.56 \pm    0.30$  &    0.50 &   0.23 \\
  11.65 &   -9.91 &  $  -7.54 \pm    0.30$  &  $  -7.54 \pm    0.30$ & ... &    0.69 & ... \\
\hline
\enddata
\tablenotetext{a}{Units of $\phi(\ssfr, \Mstar)$ in Mpc$^{-3}$ bin$^{-1}$ where each bin is 0.3 dex wide in \Mstar~and 0.1875 dex wide in \ssfr.}
\label{table:ssfr_mass_cut}
\end{deluxetable}
}
}

\clearpage

 

\begin{thebibliography}{}
 
 %
 
 \bibitem[Adelman-McCarthy et al.(2006)]{Adelman-McCarthy2006} 
Adelman-McCarthy, J.~K., et al.\ 2006, \apjs, 162, 38 


\bibitem[Ag{\"u}eros et al.(2005)]{Agueros2005} Ag{\"u}eros, M.~A., et al.\ 
2005, \aj, 130, 1022 



\bibitem[Bell \& de Jong(2000)]{Bell2000} Bell, E.~F., \& de Jong, R.~S.\ 
2000, \mnras, 312, 497 



\bibitem[Bell(2002)]{Bell2002} Bell, E.~F.\ 2002, \apj, 577, 150 

\bibitem[Bell(2003a)]{Bell2003a} Bell, E.~F.\ 2003, \apj, 586, 794 

\bibitem[Bell et al.(2003b)]{Bell2003b} Bell, E.~F., McIntosh, D.~H., Katz, 
N., \& Weinberg, M.~D.\ 2003, \apjs, 149, 289 

\bibitem[Bell et al.(2005)]{Bell2005} Bell, E.~F., et al.\ 2005, \apj, 625, 
23 

\bibitem[Bell et al.(2006)]{Bell2006} Bell, E.~F., Phleps, S., Somerville, 
R.~S., Wolf, C., Borch, A., \& Meisenheimer, K.\ 2006, \apj, 652, 270 

\bibitem[Bell et al.(2007)]{Bell2007} Bell, E.~F., Zheng, X.~Z., Papovich, 
C., Borch, A., Wolf, C., \& Meisenheimer, K.\ 2007, \apj, 663, 834 


\bibitem[Bianchi et al.(2006)]{Bianchi2006} Bianchi, L., et al.\ 2006, 
ArXiv Astrophysics e-prints, arXiv:astro-ph/0611926 

\bibitem[Binney \& de Vaucouleurs(1981)]{Binney1981} Binney, J., \& de 
Vaucouleurs, G.\ 1981, \mnras, 194, 679 



\bibitem[Blanton et al.(2003a)]{Blanton2003a} Blanton, M.~R., et al.\ 2003, 
\aj, 125, 2348 

\bibitem[Blanton et al.(2003b)]{Blanton2003b} Blanton, M.~R., et al.\ 2003, 
\apj, 592, 819 

\bibitem[Blanton et al.(2003c)]{Blanton2003c} Blanton, M.~R., et al.\ 2003, 
\apj, 594, 186 

\bibitem[Blanton et al.(2005a)]{Blanton2005a} Blanton, M.~R., et al.\ 2005, 
\apj,  629, 143

\bibitem[Blanton et al.(2005b)]{Blanton2005b} Blanton, M.~R., et al.\ 2005, 
\aj, 129, 2562 


\bibitem[Blanton(2006)]{Blanton2006} Blanton, M.~R.\ 2006, \apj, 648, 268 

\bibitem[Blanton \& Roweis(2007)]{Blanton2007} Blanton, M.~R., \& Roweis, 
S.\ 2007, \aj, 133, 734 




\bibitem[Borch et al.(2006)]{Borch2006} Borch, A., et al.\ 2006, \aap, 453, 
869 


\bibitem[Boselli et al.(2001)]{Boselli2001} Boselli, A., Gavazzi, G., 
Donas, J., \& Scodeggio, M.\ 2001, \aj, 121, 753 


\bibitem[Brinchmann et al.(2004)]{Brinchmann2004} Brinchmann, J., Charlot, 
S., White, S.~D.~M., Tremonti, C., Kauffmann, G., Heckman, T., \& 
Brinkmann, J.\ 2004, \mnras, 351, 1151 


\bibitem[Bruzual \& Charlot(2003)]{Bruzual2003} Bruzual, G., \& Charlot, 
S.\ 2003, \mnras, 344, 1000 


\bibitem[Calzetti et al.(2000)]{Calzetti2000} Calzetti, D., Armus, L., 
Bohlin, R.~C., Kinney, A.~L., Koornneef, J., \& Storchi-Bergmann, T.\ 2000, 
\apj, 533, 682 

\bibitem[Cattaneo et al. (2007)]{Cattaneo2007} Cattaneo, A., et al.\ 2007, 
\mnras, 377, 63 

\bibitem[Dahlen et al.(2007)]{Dahlen2007} Dahlen, T., Mobasher, B., 
Dickinson, M., Ferguson, H.~C., Giavalisco, M., Kretchmer, C., \& 
Ravindranath, S.\ 2007, \apj, 654, 172 



\bibitem[Dale et al.(2007)]{Dale2007} Dale, D.~A., et al.\ 2007, \apj, 655, 
863 


\bibitem[De Lucia et al.(2006)]{De Lucia2006} De Lucia, G., Springel, V., 
White, S.~D.~M., Croton, D., \& Kauffmann, G.\ 2006, \mnras, 366, 499 


\bibitem[De Propris et al.(2005)]{De Propris2005} De Propris, R., Liske, 
J., Driver, S.~P., Allen, P.~D., \& Cross, N.~J.~G.\ 2005, \aj, 130, 1516 


\bibitem[de Zeeuw \& Franx(1991)]{de Zeeuw1991} de Zeeuw, T., \& Franx, M.\ 
1991, \araa, 29, 239 


\bibitem[Driver et al.(2006)]{Driver2006} Driver, S.~P., et al.\ 2006, 
\mnras, 368, 414 


\bibitem[Erb et al.(2006)]{Erb2006} Erb, D.~K., Steidel, C.~C., Shapley, 
A.~E., Pettini, M., Reddy, N.~A., \& Adelberger, K.~L.\ 2006, \apj, 646, 
107 


\bibitem[Faber et al.(2007)]{Faber2007} Faber, S.~M., et al.\ 2007, \apj, 665, 265


\bibitem[Fabian et al.(1982)]{Fabian1982} Fabian, A.~C., Nulsen, P.~E.~J., 
\& Canizares, C.~R.\ 1982, \mnras, 201, 933 


\bibitem[Feulner et al.(2006)]{Feulner2006} Feulner, G., Hopp, U., \& 
Botzler, C.~S.\ 2006, \aap, 451, L13 


\bibitem[Haynes \& Giovanelli(1984)]{Haynes1984} Haynes, M.~P., \& 
Giovanelli, R.\ 1984, \aj, 89, 758 


\bibitem[Heckman et al.(2005)]{Heckman2005} Heckman, T.~M., et al.\ 2005, 
\apjl, 619, L35 


\bibitem[Hoopes et al.(2006)]{Hoopes2006} Hoopes, C.~G., et al.\ 2006, 
ArXiv Astrophysics e-prints, arXiv:astro-ph/0609415 


\bibitem[Hopkins \& Beacom(2006)]{HopkinsA2006} Hopkins, A.~M., \& Beacom, 
J.~F.\ 2006, \apj, 651, 142 


\bibitem[Hopkins et al.(2006)]{Hopkins2006} Hopkins, P.~F., Hernquist, L., 
Cox, T.~J., Di Matteo, T., Robertson, B., \& Springel, V.\ 2006, \apjs, 
163, 1 


\bibitem[Jansen \& Kannappan(2001)]{Jansen2001} Jansen, R.~A., \& 
Kannappan, S.~J.\ 2001, \apss, 276, 1151 


\bibitem[Johnson et al.(2006)]{Johnson2006} Johnson, B.~D., et al.\ 2006, 
\apjl, 644, L109 

\bibitem[Johnson et al.(2007)]{Johnson2007} Johnson, B.~D., et al.\ 2007, this volume 


\bibitem[Jonsson et al.(2006)]{Jonsson2006} Jonsson, P., Cox, T.~J., 
Primack, J.~R., \& Somerville, R.~S.\ 2006, \apj, 637, 255 


\bibitem[Kannappan(2004)]{Kannappan2004} Kannappan, S.~J.\ 2004, \apjl, 
611, L89 

\bibitem[Kauffmann et al.(2003)]{Kauffmann2003c} Kauffmann, G., et al.\ 
2003, \mnras, 346, 1055 


\bibitem[Kauffmann et al.(1993)]{Kauffmann1993} Kauffmann, G., White, 
S.~D.~M., \& Guiderdoni, B.\ 1993, \mnras, 264, 201 


\bibitem[Kauffmann et al.(2003a)]{Kauffmann2003a} Kauffmann, G., et al.\ 
2003, \mnras, 341, 33 


\bibitem[Kauffmann et al.(2003b)]{Kauffmann2003b} Kauffmann, G., et al.\ 
2003, \mnras, 341, 54 

\bibitem[Kauffmann et al.(2006a)]{Kauffmann2006a} Kauffmann, G., Heckman, 
T.~M., De Lucia, G., Brinchmann, J., Charlot, S., Tremonti, C., White, 
S.~D.~M., \& Brinkmann, J.\ 2006, \mnras, 367, 1394 

\bibitem[Kauffmann et al.(2006b)]{Kauffmann2006b} Kauffmann, G., et al.\ 
2006, ArXiv Astrophysics e-prints, arXiv:astro-ph/0609436 


\bibitem[Kennicutt(1998a)]{Kennicutt1998a} Kennicutt, R.~C., Jr.\ 1998, \apj, 
498, 541 


\bibitem[Kennicutt(1998b)]{Kennicutt1998b} Kennicutt, R.~C., Jr.\ 1998, 
\araa, 36, 189 


\bibitem[Kere{\v s} et al.(2005)]{Keres2005} Kere{\v s}, D., Katz, N., 
Weinberg, D.~H., \& Dav{\'e}, R.\ 2005, \mnras, 363, 2 


\bibitem[Knapp et al.(1992)]{Knapp1992} Knapp, G.~R., Gunn, J.~E., \& 
Wynn-Williams, C.~G.\ 1992, \apj, 399, 76 

\bibitem[Kong et al.(2004)]{Kong2004} Kong, X., Charlot, S., Brinchmann, 
J., \& Fall, S.~M.\ 2004, \mnras, 349, 769 


\bibitem[Labbe et al.(2007)]{Labbe2007} Labbe, I., et al.\ 2007, ArXiv 
e-prints, 705, arXiv:0705.3325 


\bibitem[Larson \& Tinsley(1978)]{Larson1978} Larson, R.~B., \& Tinsley, 
B.~M.\ 1978, \apj, 219, 46 


\bibitem[Maller et al.(2006)]{Maller2006} Maller, A.~H., Katz, N., Kere{\v 
s}, D., Dav{\'e}, R., \& Weinberg, D.~H.\ 2006, \apj, 647, 763 


\bibitem[Martin \& Kennicutt(2001)]{Martin2001} Martin, C.~L., \& 
Kennicutt, R.~C., Jr.\ 2001, \apj, 555, 301 


\bibitem[Martin et al.(2005)]{Martin2005} Martin, D.~C., et al.\ 2005, 
\apjl, 619, L1 


\bibitem[Martin et al.(2007)]{Martin2007} Martin, D.~C., et al.\ 2007, 
ArXiv Astrophysics e-prints, arXiv:astro-ph/0703281 


\bibitem[Masjedi et al.(2006)]{Masjedi2006} Masjedi, M., et al.\ 2006, 
\apj, 644, 54 


\bibitem[Mathews \& Brighenti(2003)]{Mathews2003} Mathews, W.~G., \& 
Brighenti, F.\ 2003, \araa, 41, 191 

\bibitem[Melbourne et al.(2005)]{Melbourne2005} Melbourne, J., Koo, D.~C., 
\& Le Floc'h, E.\ 2005, \apjl, 632, L65 


\bibitem[Menanteau et al.(2005)]{Menanteau2005} Menanteau, F., et al.\ 
2005, \apj, 620, 697 

\bibitem[Meurer et al.(1999)]{Meurer1999} Meurer, G.~R., Heckman, T.~M., \& 
Calzetti, D.\ 1999, \apj, 521, 64 


\bibitem[Morrissey et al.(2005)]{Morrissey2005} Morrissey, P., et al.\ 
2005, \apjl, 619, L7 


\bibitem[Morrissey et al.(2007)]{Morrissey2007} Morrissey, P., et al.\ 
2007, ArXiv e-prints, 706, arXiv:0706.0755 


\bibitem[Noeske et al.(2007a)]{Noeske2007a} Noeske, K.~G., et al.\ 2007, 
\apjl, 660, L43 

\bibitem[Noeske et al.(2007b)]{Noeske2007b} Noeske, K.~G., et al.\ 2007, 
\apjl, 660, L47 

\bibitem[Pierini et al.(2004)]{Pierini2004} Pierini, D., Gordon, K.~D., 
Witt, A.~N., \& Madsen, G.~J.\ 2004, \apj, 617, 1022 




\bibitem[Reddy et al.(2006)]{Reddy2006} Reddy, N.~A., Steidel, C.~C., 
Fadda, D., Yan, L., Pettini, M., Shapley, A.~E., Erb, D.~K., \& Adelberger, 
K.~L.\ 2006, \apj, 644, 792 


\bibitem[Rich et al.(2005)]{Rich2005} Rich, R.~M., et al.\ 2005, \apjl, 
619, L107 


\bibitem[Roberts \& Haynes(1994)]{Roberts1994} Roberts, M.~S., \& Haynes, 
M.~P.\ 1994, \araa, 32, 115 


\bibitem[Salim et al.(2005)]{Salim2005} Salim, S., et al.\ 2005, \apjl, 
619, L39 


\bibitem[Salim et al.(2007)]{Salim2007} Salim, S., et al.\ 2007, ArXiv 
e-prints, 704, arXiv:0704.3611 


\bibitem[Schmidt(1959)]{Schmidt1959} Schmidt, M.\ 1959, \apj, 129, 243 


\bibitem[Searle et al.(1973)]{Searle1973} Searle, L., Sargent, W.~L.~W., \& 
Bagnuolo, W.~G.\ 1973, \apj, 179, 427 


\bibitem[Seibert et al.(2005)]{Seibert2005} Seibert, M., et al.\ 2005, 
\apjl, 619, L55 


\bibitem[Shen et al.(2003)]{Shen2003} Shen, S., Mo, H.~J., White, S.~D.~M., 
Blanton, M.~R., Kauffmann, G., Voges, W., Brinkmann, J., \& Csabai, I.\ 
2003, \mnras, 343, 978 


\bibitem[Solanes et al.(2001)]{Solanes2001} Solanes, J.~M., Manrique, A., 
Garc{\'{\i}}a-G{\'o}mez, C., Gonz{\'a}lez-Casado, G., Giovanelli, R., \& 
Haynes, M.~P.\ 2001, \apj, 548, 97 


\bibitem[Somerville et al.(2006)]{Somerville2006} Somerville, R.~S., et 
al.\ 2006, ArXiv Astrophysics e-prints, arXiv:astro-ph/0612428 


\bibitem[Stringer \& Benson(2007)]{Stringer2007} Stringer, M.~J., \& 
Benson, A.~J.\ 2007, ArXiv Astrophysics e-prints, arXiv:astro-ph/0703380 


\bibitem[Tinsley(1968)]{Tinsley1968} Tinsley, B.~M.\ 1968, \apj, 151, 547 


\bibitem[Tremonti et al.(2004)]{Tremonti2004} Tremonti, C.~A., et al.\ 
2004, \apj, 613, 898 

\bibitem[Tremonti et al.(2007)]{Tremonti2007} Tremonti, C.~A., Moustakas, 
J., \& Diamond-Stanic, A.~M.\ 2007, \apjl, 663, L77 



\bibitem[Treyer et al.(2005)]{Treyer2005} Treyer, M., et al.\ 2005, \apjl, 
619, L19 


\bibitem[Vincent \& Ryden(2005)]{Vincent2005} Vincent, R.~A., \& Ryden, 
B.~S.\ 2005, \apj, 623, 137 


\bibitem[Wyder et al.(2005)]{Wyder2005} Wyder, T.~K., et al.\ 2005, \apjl, 
619, L15 

\bibitem[Wyder et al.(2007)]{Wyder2007} Wyder, T.~K., et al.\ 2007, ArXiv 
e-prints, 706, arXiv:0706.3938 



\bibitem[Yi et al.(2005)]{Yi2005} Yi, S.~K., et al.\ 2005, \apjl, 619, L111 


\bibitem[Zheng et al.(2007)]{Zheng2007} Zheng, X.~Z., Dole, H., Bell, 
E.~F., Le Floc'h, E., Rieke, G.~H., Rix, H.-W., \& Schiminovich, D.\ 2007, 
ArXiv e-prints, 706, arXiv:0706.0003 


 
 
 

\end{thebibliography}
 \end{document}